\documentclass[12pt]{article}

\usepackage[top=0.7in, bottom=1in, left=1in, right=1in]{geometry}
\usepackage{amsmath, amssymb, amsfonts}
\usepackage{graphicx}
\usepackage{booktabs}
\usepackage{multirow}
\usepackage{hyperref}
\usepackage{url}
\usepackage[numbers,sort&compress]{natbib}
\usepackage{setspace}
\usepackage{authblk}
\usepackage[font=small,labelfont=bf]{caption}
\usepackage{subcaption}
\usepackage{xcolor}

\hypersetup{
colorlinks=true,
linkcolor=blue,
citecolor=blue,
urlcolor=blue
}

\title{\textbf{{\it Ab initio} insights into plasmonic and strong-field contributions to H$_2$ dissociation on silver nanoshells}}

\author[a]{Natalia E. Koval\thanks{Corresponding author. Email: \href{mailto:natalia.koval@ehu.eus}{natalia.koval@ehu.eus}}}
\author[b,a,c]{J. I\~{n}aki Juaristi}
\author[a,c]{Maite Alducin\thanks{Corresponding author. Email: \href{mailto:maite.alducin@ehu.eus}{maite.alducin@ehu.eus}}}

\affil[a]{\small Centro de F\'{\i}sica de Materiales CFM/MPC (CSIC-EHU), Paseo Manuel de Lardizabal 5, 20018 Donostia-San Sebasti\'an, Spain}
\affil[b]{\small Departamento de Pol\'{i}meros y Materiales Avanzados: F\'{i}sica, Qu\'{i}mica y Tecnolog\'{i}a, Facultad de Qu\'{i}mica (EHU), Apartado 1072, 20080 Donostia-San Sebasti\'{a}n, Spain}
\affil[c]{\small Donostia International Physics Center (DIPC),
Paseo Manuel de Lardizabal 4, 20018 Donostia-San Sebasti\'an, Spain}

\date{}

\begin{document}

\maketitle

\begin{abstract}
Modeling plasmonic catalysis by applying femtosecond laser pulses of high intensity ($10^{13}-10^{15}$ W~cm$^{-2}$), although justified by the time-dependent density functional theory (TDDFT) time-scale limitations, can lead to a dissociation mechanism that is completely unrelated to the plasmon excitation created under low-intensity continuous light in experiments (on the order of 1 W~cm$^{-2}$).
In this study, we examine the dissociation of H$_2$ on a large octahedral Ag nanoshell under varying field intensity, frequency, and duration, and we explore the possibility of identifying optimal modeling conditions accessible with current TDDFT simulations. We show that using this large nanoshell that consists in the outer layer of the Ag$_{231}$ cluster, it is still possible to disentangle the role of the plasmon from strong-field effects at applied field intensities as high as $(2-8) \times 10^{13}$~W~cm$^{-2}$. In particular, although strong-field effects are always present at these intensities, we find that the excited plasmon dominates the dissociation process at the lowest applied intensity of $2 \times 10^{13}$~W~cm$^{-2}$. Furthermore, at the highest intensity, at which strong-field effects become dominant, the plasmon contributes to accelerating the dissociation of the molecule. Overall, our simulations pave the way to bridge the intensity gap between TDDFT modeling and experiments in plasmonic catalysis.
\end{abstract}

\section{Introduction}

Plasmonic catalysis is a promising field of research which explores the use of localized surface plasmon resonance (LSPR) in metal nanoparticles to accelerate and enhance chemical reaction
rates under the influence of light \cite{Linic2011,Christopher2012,Linic2015,acswc,Zhang_2013,Wei2018,Zhang2019,Cortes2020,SivanDubi2020,Kumar2021,Newmeyer2022,Jain2022,pubs.acs.org/acscatalysis2023,Dong2023,Amirjani2023,ConstantinosMoularas2024}. 
Experimentally, there are multiple studies showing that the rate of molecular dissociation at metal nanoparticle surfaces increases when the system is illuminated with a light of the same frequency as the plasmon resonance of the nanoparticle~\cite{acswc,Christopher2011,Mukherjee2012,Brongersma2015,Boerigter2016a,CheruvathoorPoulose2022}. The effect is often explained by the plasmon decay that creates energetic charge carriers ("hot electrons") in the nanoparticle that can transfer to the adsorbed molecule leading to the weakening of its bonds \cite{Christopher2011,Mukherjee2012,CheruvathoorPoulose2022}. For silver nanoparticles, from a 4-fold \cite{Christopher2011} to an 8-fold \cite{Miralles2024} acceleration of the chemical reaction was observed when illuminated with resonant light. However, the precise mechanism of the plasmon-induced molecular dissociation is not possible to understand from experiments alone due to the short time of plasmon excitation and decay (on the order of femtoseconds, fs) that cannot be resolved experimentally. On top of that, the mechanism can be system-specific and is still a highly-debated issue \cite{Brosseau2024,VermaNatCom2024,RebecaMiyar2025}. Therefore, \emph{ab initio} calculations are crucial for getting insights into the interplay between plasmon excitation in a metal nanoparticle and dissociation of molecules adsorbed on its surface. 

Time-dependent density functional theory (TDDFT) \cite{Runge1984,2006-book-tddft,ullrich2011time} combined with Ehrenfest dynamics (ED) is a state-of-the-art methodology able to describe electronic and atomic excitations at the nanoscale and on the femtosecond time scale \cite{Herring2023a}. Real-time TDDFT allows for an accurate (formally exact, but practically limited by the choice of the exchange-correlation functional) description of the plasmon excitation and decay. TDDFT-ED can be used to describe the atomic motion following the plasmon decay.
Since ED is a mean-field approach, the dynamics follows a single average trajectory, so that it cannot account for different pathways associated with different electronic quantum states~\cite{Tully2023},
which may be relevant in the dissociation processes. However, TDDFT-ED is one of the very few methods capable of handling complex light-matter interactions and the electron and nuclear dynamics following a plasmon decay. Thus, despite its limitations, this approach is able to provide valuable insights into the mechanisms of 
plasmonic catalysis. 

There is an increasing number of studies using TDDFT-ED that report plasmon-assisted dissociation of molecules on
plasmonic nanoparticles of different sizes and shapes \cite{Yan2015,YuZhang2018,Hull2020,YiminZhang2021,Huang2021,KudaSingappulige2023,Zhang2021b,Herring2023,Li2023c,Wang2023c,Wang2024a,Herring2025,Verma2025a,Kar2025,Zhang2025}. Due to the high computational cost of such calculations, these studies, aimed to model plasmon-induced reactions, are commonly performed for small systems irradiated with high-intensity laser pulses (10$^{13}$--10$^{15}$ W cm$^{-2}$).  
These conditions accelerate the dissociation process, but can bring about strong-field effects \cite{Fennel2010} that are rarely discussed. 
Such discrepancy in intensities makes it difficult to extrapolate \emph{ab initio} results to experimental conditions (intensities of the order of 1 W cm$^{-2}$) because strong-field effects can suppress the role of plasmon excitation and change the mechanism of molecular dissociation, as demonstrated recently in small ($\approx 1$ nm) nanoparticles \cite{Koval2024}. In that work \cite{Koval2024}, we studied H$_2$ dissociation on a silver nanoshell (a hollow shell consisting in the outer layer of the icosahedral Ag$_{55}$) and compared the results obtained when applying both resonant and off-resonant (with the nanoparticle plasmon frequency) external fields of high intensities. Our TDDFT-ED calculations showed that both the metal nanoshell and H$_2$ become ionized at these high intensities because of multiphoton absorption. 
As a result, H$_2$ dissociates at both resonant and off-resonant pulse frequencies, making it difficult to determine the contribution of plasmon excitation to the process.  

Motivated by those findings that emphasized the need to reconsider the use of strong external fields in plasmonic catalysis simulations, the aim of this work is twofold. First, we assess the possibility of reducing nonlinear effects while using external field intensities that allow us to access the reaction timescale with current TDDFT-ED simulations. Second, we conduct a thorough analysis of the dissociation mechanism to identify the role of plasmon and its interplay with strong-field nonlinear effects.

One possible way of reducing strong-field effects without reducing the intensity of the external pulse is to use large nanoparticles. It has been shown that the threshold intensity for dissociation decreases with increasing nanoparticle size \cite{Giri2023}. Therefore, by modeling larger systems, it is possible to determine the optimal external field conditions that would generate strong plasmonic excitation, while reducing nonlinear effects without significantly increasing the dissociation timescale. 
In particular, we present a comprehensive analysis of the plasmon effect on an H$_2$ molecule adsorbed at the vertex of an octahedral silver nanoshell \cite{Huang2016} under various external field conditions. The choice of the octahedron is motivated by the stronger field enhancement that the plasmon creates on its vertex compared to other shapes \cite{Agrawal2015,Moon2016,MontanoPriede2019,Kang2024}. This property can additionally increase the effect on the molecule at a lower external field intensity.
By comparing different field duration, frequency (resonant vs off-resonant), and peak intensity, we analyze the correlation between plasmonic and strong-field effects in detail. 

\section{Computational methodology}

\subsection{Geometry optimization}
The first step in our computational approach involves optimization of the system's geometry using density functional theory (DFT). We used the Perdew-Burke-Ernzerhof (PBE) functional \cite{Perdew1996} within the CP2K software package \cite{Hutter2013,Kuehne2020,cp2k}, which implements the Gaussian plane wave (GPW) method \cite{LIPPERT1997,VandeVondele2005}. We used DZVP basis sets including 1 and 11 electrons for H and Ag, respectively, and a cutoff of 600 Ry for the plane wave auxiliary basis set. 
Norm-conserving Goedecker-Teter-Hutter (GTH) pseudopotentials \cite{Goedecker1996} were used to represent the interaction of valence electrons with atomic cores. The initial coordinates of the octahedral Ag$_{231}$ cluster (edge length 7 atoms) were obtained using the Atomic Simulation Environment (ASE) builder (function ase.cluster.Octahedron) \cite{HjorthLarsen2017}. The nanoshell (denoted hereafter Ag$_{231}^{\mathrm{L1}}$) was next constructed by taking only the outer layer of the unrelaxed Ag$_{231}$ cluster, which is formed by 146 atoms.  The H$_2$ molecule was placed at one nanoshell vertex with its molecular axis oriented parallel to the octahedron base (see Fig.~\ref{fig:abs_spect}a). The $z-$axis is defined along the inner diagonal that contains the vertex at which H$_2$ adsorbs. Note that the nanoshell with an inner diagonal of 2.45~nm before relaxation is smaller than those commonly employed in experiments. However, although less often, nanoparticles as small as a few nm have been studied experimentally as well \cite{Campos2018,Deka2024}.

Using a non-periodic simulation cell of $32\times32\times32$ \AA$^3$, the Ag$_{231}^{\mathrm{L1}}$+H$_2$ geometry was optimized until all atomic forces were below 0.001 Ha Bohr$^{-1}$. After relaxation, the edges of the nanoshell (colored in green in Fig.~\ref{fig:abs_spect}a) slightly bent towards the inner empty space of the hollow shell, whereas the facets (blue atoms in Fig.~\ref{fig:abs_spect}a) bent slightly outwards. Adsorption of H$_2$ causes slight distortion of the nearest Ag atoms with the underneath Ag vertex being displaced towards H$_2$. The optimized distance between the Ag vertex and the H$_2$ center of mass (CM) is 2.275~{\AA}. The adsorbed H$_2$ remains parallel to the octahedron base with an equilibrium bond length of 0.763 {\AA}. The adsorption energy $E_{\mathrm{ads}}=-0.17$~eV is calculated from the ground state DFT+PBE energies as, $E_{\mathrm{ads}} = E_{\mathrm{Ag_{231}^{L1}+H_2}} - E_{\mathrm{Ag_{231}^{L1}}} - E_{\mathrm{H_2}}$. Similarly, the Ag$_{231}^{\mathrm{L1}}$+H$_2$ ionization potential $I_p=4.6$~eV is obtained as the ground state energy difference between the positively charged and neutral systems, i.e., $I_p =E_\mathrm{(Ag_{231}^{\mathrm{L1}}+H_2)^+}- E_\mathrm{(Ag_{231}^{\mathrm{L1}}+H_2)^0}$. 
 
\subsection{Real-time time-dependent density functional theory calculations of the absorption spectrum}

\begin{figure}[h]
\centering
    \includegraphics[width=0.5\linewidth]{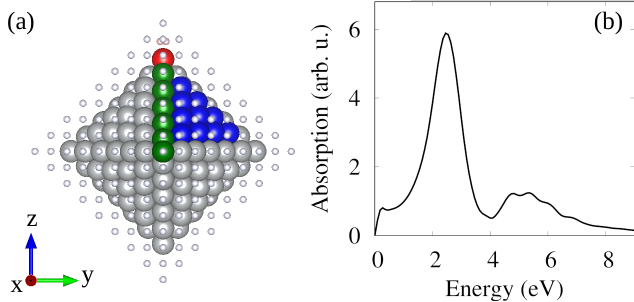}
    \caption{\label{fig:abs_spect} (a) Relaxed structure of the Ag$_{231}^{\mathrm{L1}}$ nanoshell with H$_2$ (light pink spheres) adsorbed at 2.275~{\AA} from the vertex (red sphere) surrounded by a layer of 258 ghost Ag atoms (small gray spheres). To facilitate visualization, green and blue spheres depict the nanoshell edge and facet Ag atoms, respectively. (b) Absorption spectrum of Ag$_{231}^{\mathrm{L1}}$+H$_2$ calculated with RT-TDDFT, showing the plasmon resonance at $\hbar \omega_{\mathrm{p}} = 2.48$~eV.} 
\end{figure}

The Ag$_{231}^{\mathrm{L1}}$+H$_2$ absorption spectrum was calculated within the real-time time-dependent density functional theory (RT-TDDFT) approach implemented in the CP2K software package \cite{Hutter2013,Kunert2003,Andermatt2016,Kuehne2020,cp2k,LIPPERT1997,VandeVondele2005} by perturbing the system with a weak broadband electric field, $E(t)=E_0 \delta(t) \mathbf{\hat{e}}$~\cite{Yabana1996}. In particular, we applied a $z$-polarized $\delta$-kick field of strength $E_0$=0.001~a.u.~$\approx 0.5$ V~nm$^{-1}$ at $t = 0$ and let the system orbitals evolve in time during 25~fs using the enforced time reversible symmetry (ETRS) real-time propagation scheme with a time step $\Delta t\!=\!0.005$~fs. 
Calculations were performed in the same non-periodic cell of $32\times32\times32$~{\AA}$^3$ employed for geometry optimization. The absorption spectrum in the frequency domain was next computed by applying a discrete Fourier transform to the (induced) time-dependent dipole moment. For completeness, we also calculated the optical absorption spectra for $x-$ and $y-$polarized electric fields 
and found them almost identical because of the symmetry of the nanoshell. 
The absorption spectrum for the $z$-polarized field is shown in Fig.~\ref{fig:abs_spect}b. It features an intense and well-defined plasmon peak at $\hbar\omega_{\mathrm{p}}=2.48$~eV and a less intense signal that extends from 4 to 7~eV, approximately. A similar optical absorption spectrum was obtained with RT-TDDFTB for the closely related system Ag$_{231}$+H$_2$, showing the plasmon peak at around 2.9~eV and a high-energy broad signal above 3~eV that is attributed to $d$-$sp$ intraband transitions~\cite{Giri2023}. Experimentally, localized-surface plasmon resonances of 2.75--3.0~eV were measured in synthesized Ag octahedra with edge lengths varying in the range of 20-70~nm~\cite{Wang2013}. 

\subsection{Ehrenfest molecular dynamics simulations}

The response of the electrons and nuclei to an external field was studied with the real-time TDDFT simulations combined with Ehrenfest molecular dynamics (ED) as implemented in the CP2K software package \cite{Hutter2013,Kunert2003,Andermatt2016,Kuehne2020,cp2k,LIPPERT1997,VandeVondele2005}.  
The external field was modeled by a Gaussian envelope:
\begin{equation}
    \mathbf{E}(\omega,t) = E_0~ \mathrm{exp}\left[-\frac{(t-t_0)^2}{2\sigma^2} \right] \mathrm{cos}[\omega (t-t_0)]\, \hat{\mathbf{k}} \, ,
\end{equation}
where $\sigma$ is the standard deviation and $t_0$ is the center of the envelope. The electric field was polarized in the $z-$direction. The maximum intensity is given by $I_\mathrm{max}=c \epsilon_0 E_0^2$, where $c$ is the speed of light, $\epsilon_0$ is the permittivity in vacuum, and $E_0$ is the maximum field strength. In order to determine those field conditions that allow us to distinguish between plasmon and strong-field contributions when studying the catalytic properties of plasmonic nanoparticles, we have applied four different pulses, resonant ($\hbar \omega = \hbar \omega_{\mathrm{p}}= 2.48$~eV) and off-resonant ($\hbar \omega = 8$~eV), of varying intensity and duration to the Ag$_{231}^{\mathrm{L1}}+$ H$_2$ system. 
The intensity and duration of each pulse are summarized in Table~\ref{table:pulses} and depicted in Fig.~S1. 
\begin{table}[b]
\small
\centering
  \caption{\ Gaussian field pulses applied to Ag$_{231}^{\mathrm{L1}}+$ H$_2$ in TDDFT-ED}
  \label{table:pulses}
  \begin{tabular*}{0.68\textwidth}{@{\extracolsep{\fill}}ccccc}
    \hline
    & pulse-1 & pulse-2 & pulse-3 &pulse-4 \\
    \hline
    $I_{\mathrm{max}}$( W~cm$^{-2}$) & $ 2 \times 10^{13}$ & $ 2 \times 10^{13}$ & $ 2 \times 10^{13}$ &$ 8 \times 10^{13}$\\
    $t_0$(fs)& 18 & 35 & 55 & 9\\
    $\sigma$(fs) & 5 & 10 & 15 & 2.5\\
     
    \hline
  \end{tabular*}
\end{table}

For each applied field, the TDDFT-ED simulations were initiated from the optimized geometry of the H$_2$ molecule adsorbed on the nanoshell, while the initial atomic velocities correspond to an initial temperature of 300~K.
All atoms were allowed to move freely without any geometry constraints during the dynamics (i.e., no frozen atoms). We used a converged time step $\Delta t\!=\!0.002$~fs and a non-periodic unit cell of size $40\times40\times45$~{\AA}$^3$. Since CP2K is based on atom-centered basis sets, this large cell allows us to account for possible electron emission by adding a layer of ghost atoms around the nanoshell, as done in ref.~\citenum{Koval2024}. In the present work, the ghost layer corresponds to the outer layer of an octahedral nanocluster with an edge length of 9 atoms (see Fig.~\ref{fig:abs_spect}a). The ghost atoms Ag$_g$ are represented by the Ag DZVP basis set and the GTH pseudopotentials.

Data analysis and visualization were performed using NumPy~\cite{Harris2020}, Matplotlib~\cite{Hunter2007}, VESTA~\cite{Momma2011}, and Gnuplot~\cite{Gnuplot_5.4}.

\section{Results and Discussion}

At an external field frequency matching the plasmon resonance, the metal nanoparticle exhibits a distinct strong response in the linear regime. The amplitude of the dipole moment induced in the system is much larger at the plasmon frequency than at any other frequency. In a strong field, a clear manifestation of the nonlinearity of the perturbation is the absence of such a resonant behavior. This is what we found for a small icosahedral nanoshell of the Ag$_{55}$ cluster  (Ag$_{55}^{\mathrm{L1}}$) \cite{Koval2024}, as mentioned in the Introduction. At the usual intense fields employed in TDDFT-ED studies ($I_\mathrm{max}\geq 10^{13}$~W~cm$^{-2}$), there was no difference in the amplitude of the time-dependent dipole moments that were induced by fields in resonance or off-resonance with the Ag$_{55}^{\mathrm{L1}}$ plasmon mode. Therefore, our first step has been to assess whether field intensities in the order of $10^{13}$~W~cm$^{-2}$ fall within the linear response regime for the larger octahedral nanoshell studied here. In particular, we start by applying to the Ag$_{231}^{\mathrm{L1}}$ nanoshell with the H$_2$ adsorbate the Gaussian pulse denoted pulse-1 in Table~\ref{table:pulses}, for which plasmon effects were completely masked in Ag$_{55}^{\mathrm{L1}}$+H$_2$~\cite{Koval2024}. 

\begin{figure}[h]
\centering
    \includegraphics[width=0.5\linewidth]{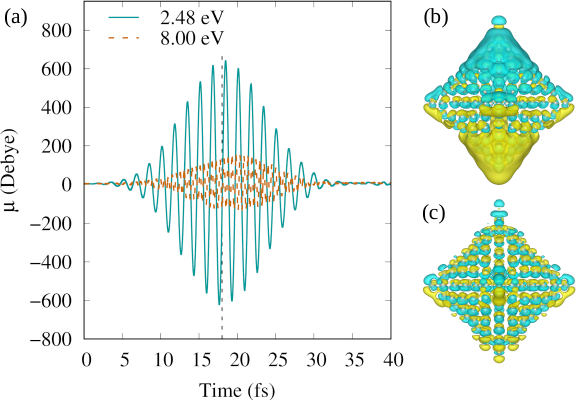}
    \caption{\label{fig:dipole} (a) Transient dipole moment $\mu(t)$ of the Ag$_{231}^{\mathrm{L1}}$+H$_2$ system induced by the Gaussian external field denoted pulse-1 in Table~\ref{table:pulses} with a frequency of 2.48~eV (resonant, cyan curve) and 8~eV (off-resonant, orange curve). Corresponding Ag$_{231}^{\mathrm{L1}}$+H$_2$ induced density created at time $t = 10$~fs (i.e., near the induced dipole moment maxima) by (b) the resonant pulse-1 (isosurface value 0.00027588 e$^-$~bohr$^{-3}$) and (c) the off-resonant pulse-1 (isosurface value 0.000148458 e$^-$~bohr$^{-3}$). Vertical dashed line marks the instant $t_0 = 18$~fs of the pulse maximum.}
\end{figure}

The induced dipole moments for resonant and off-resonant external field frequencies are compared in Fig.~\ref{fig:dipole}a. The dipole moment at the plasmon frequency of 2.48~eV has a significantly larger amplitude than at the higher, off-resonant frequency of 8~eV (see the absorption spectrum in Fig.~\ref{fig:abs_spect}b). However, in the linear-response regime, the created plasmonic oscillations of the dipole moment persist long after the pulse is off (see the evolution of the dipole moment for a much lower intensity in Fig.~S2)~\cite{Nishizawa2024, Verma2025}. Hence, the rapid decay of the dipole moment shown in Fig.~\ref{fig:dipole}a, which becomes almost zero immediately after the pulse is off, still manifests the nonlinear response of the system and it remarks that the applied perturbation is still strong. In spite of it, analysis of the density changes induced by pulse-1 at the resonant frequency allows us to identify the distinct collective nature of the plasmon excitation, confirming that the plasmon is not suppressed. Figure~\ref{fig:dipole}b and c show selected snapshots of the induced density created when applying the resonant and off-resonant pulse-1 to the Ag$_{231}^{\mathrm{L1}}$-H$_2$ system. Each snapshot corresponds to an instant at which the value of the induced dipole (Fig.~\ref{fig:dipole}a) is close to one of its local maxima. When pulse-1 is resonant with $\hbar\omega_{\mathrm{p}}$, we clearly observe the collective electron displacement that is characteristic of plasmon excitation (Fig.~\ref{fig:dipole}b). In contrast, when the pulse frequency is far from the resonance, at 8~eV, the distribution of the induced density suggests that only single- and multi-pair excitations are created (Fig.~\ref{fig:dipole}c). Altogether, the observed stronger dipolar response and corresponding induced density distribution under the resonant pulse-1 suggest that by applying an intense external field to the large nanoshell Ag$_{231}^{\mathrm{L1}}$, we can keep the conditions closer to the linear regime, with a potential for distinguishing the role of 
plasmon in catalysis, as we analyze next. 

\begin{figure}[h!]
\centering
    \includegraphics[width=0.5\linewidth]{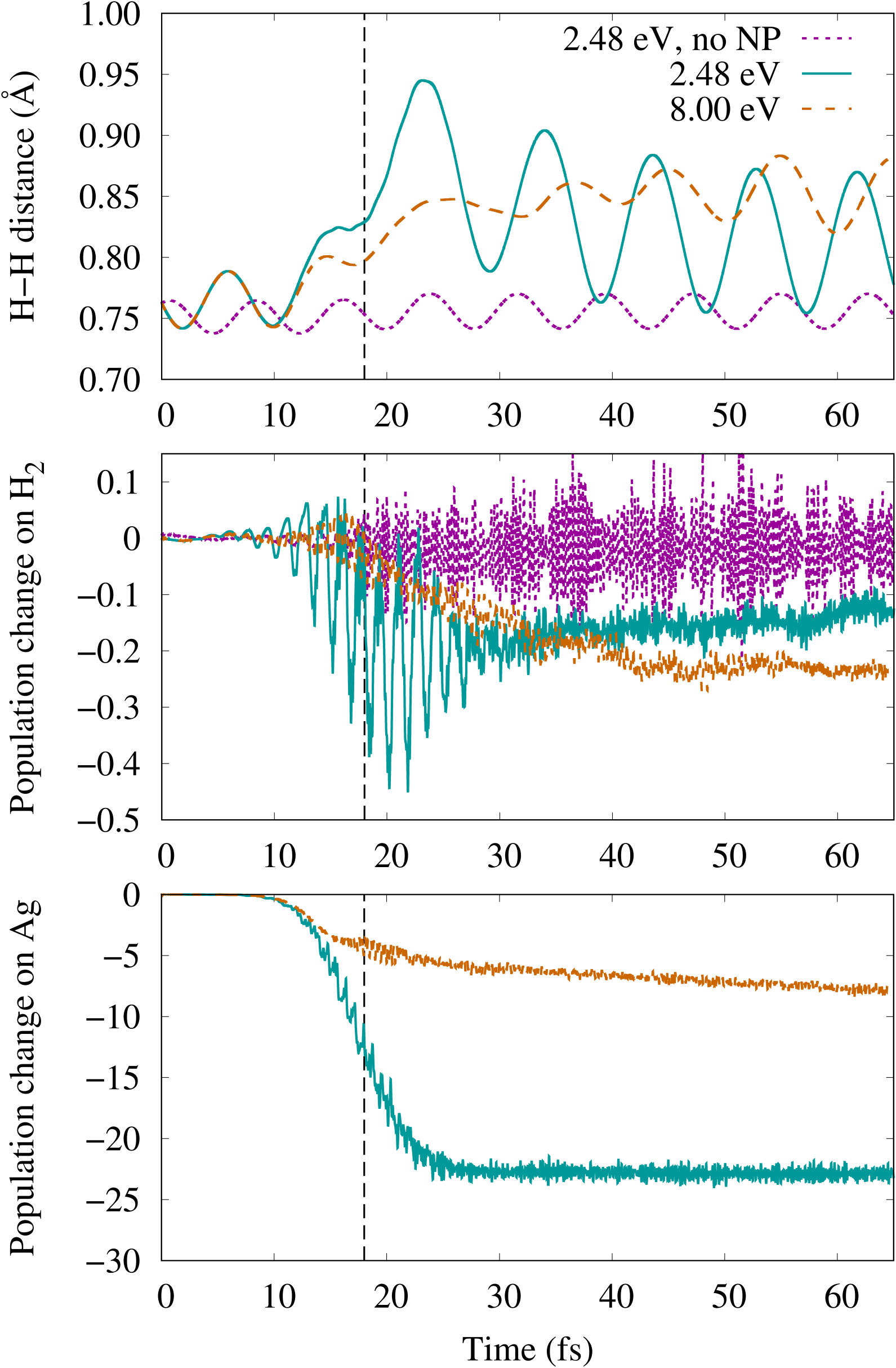}
    \caption{\label{fig:hh_1pulse} (a) H$_2$ bond length and (b) and (c) Mulliken population change $\left[\Delta N_e = N_e(t)-N_e(t=0)\right]$ on H$_2$ and Ag$_{231}^{\mathrm{L1}}$, respectively, as a function of time. Results obtained from TDDFT-ED simulations of H$_2$ adsorbed on Ag$_{231}^{\mathrm{L1}}$ using  as external field, pulse-1 (see Table~\ref{table:pulses}) at resonant (2.48~eV, cyan solid lines) and off-resonant (8~eV, orange dashed lines) frequencies. For comparison, the results obtained for a gas-phase H$_2$ when applying the resonant pulse-1 are also shown by magenta dotted lines in (a) and (b). Vertical dashed line marks the instant $t_0 = 18$~fs of the pulse maximum. }
\end{figure}

In order to determine the plasmon contribution to the dynamics and possible dissociation of H$_2$ on Ag$_{231}^{\mathrm{L1}}$, we have performed TDDFT-ED calculations by applying pulse-1 (Table~\ref{table:pulses}) with frequencies $\hbar \omega =$ 2.48~eV (resonant) and 8~eV (off-resonant). Figure~\ref{fig:hh_1pulse}a shows the time evolution of the H-H bond length for the two field frequencies. For comparison, it also includes the result for H$_2$ without the nanoshell (14 ghost atoms surround the molecule), in which case the molecule oscillates  around its equilibrium bond length. When the molecule is adsorbed on the nanoshell, the H-H bond is visibly stretched at both field frequencies during the interval around 10--30~fs (i.e., while the pulse is more active), but the resonant pulse at $\hbar \omega = 2.48$~eV produces the largest elongation. Once the pulse is extinguished, the molecule remains oscillating profoundly around a bound length of 0.8~{\AA} in the case of the resonant pulse and around 0.85~{\AA} but with little variations in the off-resonant case. Importantly, the molecule desorbs in both cases but without dissociating.

The observed changes in the H$_2$ bond length are associated to changes in the electron population. Figure~\ref{fig:hh_1pulse} shows the transient electron population change [$\Delta N_e = N_e(t)-N_e(t=0)$] experienced by the molecule (panel b) and the nanoshell (panel c). In each case, $N_e$ is the number of electrons (not the charge) obtained from a Mulliken population analysis, with negative values of $\Delta N_e$ meaning a reduction in the
number of electrons. Excitation of the plasmon mode in the case of the resonant pulse (2.48~eV), which is characterized by an oscillatory displacement of electrons between the adsorbate and the nanoshell, is reflected in the large-amplitude oscillations of $\Delta N_e$ that take place in the molecule while the pulse is on. Together with these oscillations, there is an underlying minor decrease of up to 0.1--0.2 electrons that persists once the external electric field is off. Inspection of the system orbitals remarks that many of them involve both the molecule and the nanoshell, as illustrated in Fig.~S3. Therefore, this small ionization of the molecule is possibly related to the ionization of the nanoshell that loses more than 20 electrons according to the Mulliken analysis of the electron population change (Fig.~\ref{fig:hh_1pulse}c, cyan curve). Importantly, since the resonant frequency is smaller than the ionization potential of the system (4.6 eV), the existence of electron emission is a fingerprint of the strong-field effects associated to the intense pulse-1.

The absence of plasmon excitation in the case of the off-resonant pulse ($\hbar\omega = 8$~eV) leads to much smaller oscillations of $\Delta N_e$ in the molecule (see orange curve in Fig.~\ref{fig:hh_1pulse}b) and to minimal ionization of the nanoshell despite the fact that 8~eV is larger than the ionization potential and both single-photon and multi-photon absorption are possible (see orange curve in Fig.~\ref{fig:hh_1pulse}c). As a result, the initial perturbation (10--30~fs) created by pulse-1 in the H-H bond is weaker in the off-resonant case, as shown in Fig.~\ref{fig:hh_1pulse}a. Notably, the isolated molecule does not lose any charge, supporting the fact that the electron population change on H$_2$ is mediated by the nanoshell. Altogether, the comparison of the results obtained with both field frequencies allows us to distinguish the plasmon from strong-field contributions. The large electron density fluctuations associated to the plasmon are responsible for a larger vibrational excitation of the molecule (larger oscillations of the H-H bond length). Furthermore, even though the plasmon energy is smaller than the system ionization potential and the off-resonant frequency, the excited plasmon also contributes to a higher ionization of the nanoshell by enhancing the coupling to the strong external field.

The observed rapid increase of the H-H bond length occurring when the resonant pulse-1 is more intense (18$\pm$5~fs) suggests that the duration of both the pulse and the excited plasmon, which decays rapidly once the pulse is off (Fig.~\ref{fig:dipole}a), was too short to cause dissociation.  Therefore, to assess the effect of a longer perturbation, we have performed TDDFT-ED simulations for two additional resonant pulses, pulse-2 (Fig.~S1b) and pulse-3 (Fig.~S1c) in Table~\ref{table:pulses}, that share $I_\textrm{max}$ but have durations two and three times longer, respectively, than that of pulse-1.

\begin{figure}[h]
\centering
 \includegraphics[width=0.5\linewidth]{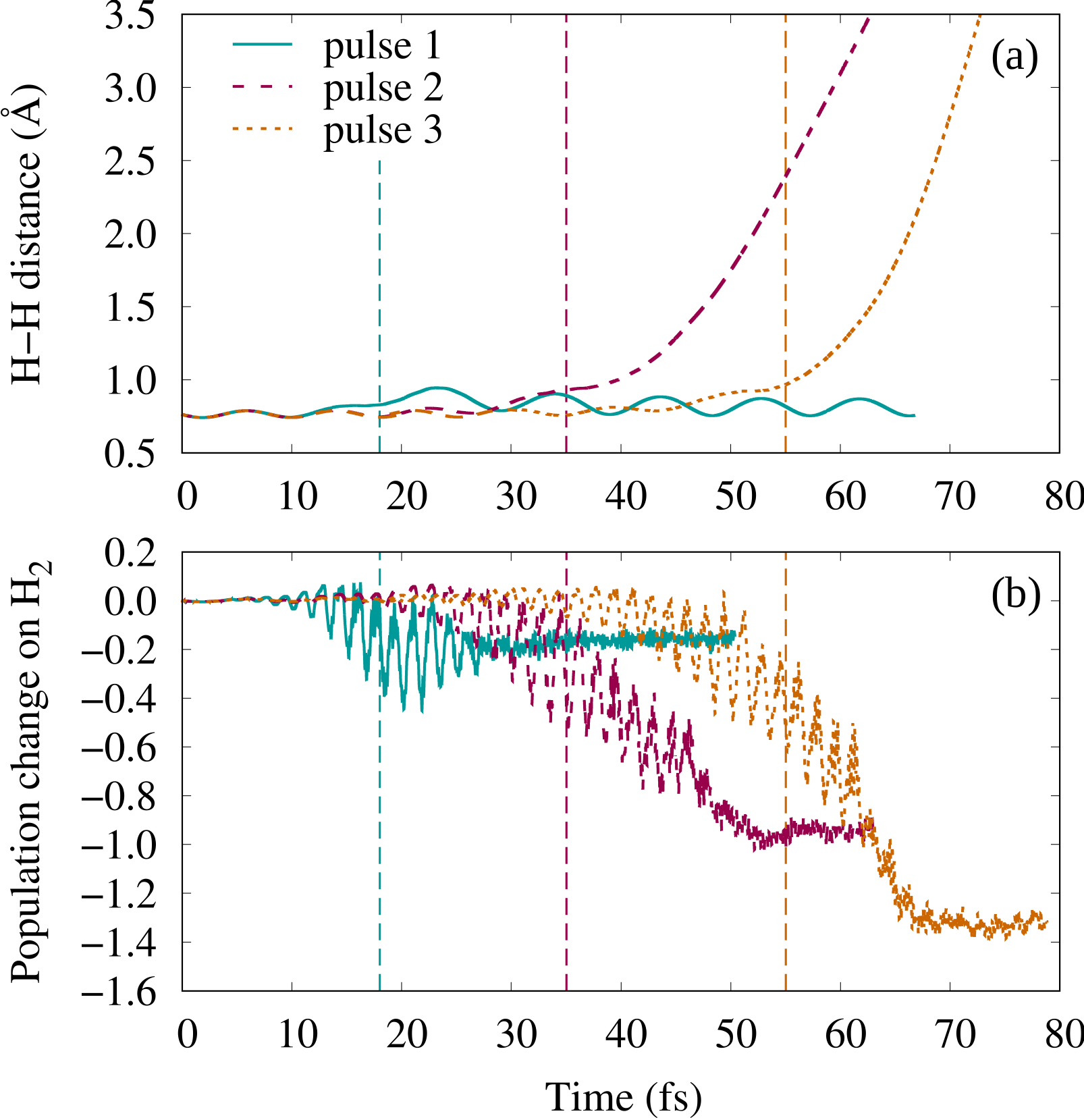}
    \caption{\label{fig:H-H_3pulses} (a) Time evolution of the H-H internuclear distance and 
    (b) Mulliken population change $\left[\Delta N_e = N_e(t)-N_e(t=0)\right]$ on H$_2$. Results obtained from TDDFT-ED simulations for Ag$_{231}^{\mathrm{L1}}+$H$_2$ using pulse-1 (cyan), pulse-2 (magenta), and pulse-3 (orange) as resonant (2.48~eV) external fields (see Table~\ref{table:pulses} for pulse details). Peak intensity instants $t_0$ are marked by vertical dashed lines following the same color code.} 
\end{figure}

Figure~\ref{fig:H-H_3pulses}a shows the H-H bond length as a function of time for the three pulses. Contrary to the case of the shortest pulse-1, both pulse-2 and pulse-3 induce molecular dissociation and desorption of the nascent H atoms. The H-H distance of 3.5~{\AA} that is reached with the two pulses during the time span of the simulations assures that H$_2$ is dissociated (see the potential energy as a function of the internuclear distance for isolated and adsorbed H$_2$ in Fig~S4). The electron population changes in Fig.~\ref{fig:H-H_3pulses}b show that the longer duration of pulse-2 and pulse-3 facilitates the ionization of the molecule, which subsequently dissociates.
In both cases, we recognize the large-amplitude oscillations in $\Delta N_e$ that are associated to plasmon excitation and the underlying decrease in the electron population that were observed with pulse-1. The difference is the higher degree of H$_2$ ionization that is reached because the new pulses are longer. As noted previously, the H$_2$ ionization is linked to the ionization of the nanoshell that loses 35 (pulse-2) and 45 (pulse-3) electrons (see Fig.~S5). 

In the previous analysis, not only the duration of the pulses is different but also their total energy. Therefore, the obtained H$_2$ dissociation can simply be a consequence of increasing the pulse energy when the maximum intensity is kept constant. 
In the following, we compare the H$_2$ dynamics on the Ag$_{231}^{\mathrm{L1}}$ nanoshell induced by pulses with the same total energy, but different maximum intensity and duration. In particular, we have performed TDDFT-ED simulations for both resonant and off-resonant field frequencies using pulse-4, which is four times shorter than pulse-2 and has a four times higher maximum intensity ($I_{\mathrm{max}} = 8 \times 10^{13}$ W cm$^{-2}$) than the previous pulses (see Table~\ref{table:pulses} and Fig.~S1d). Despite its high intensity, the plasmon excitation created by pulse-4 is still visible. This can be seen from the larger transient dipole moment amplitude induced by pulse-4 at the resonant frequency compared to the off-resonant frequency (see Fig.~S6).

\begin{figure}[h]
\centering
    \includegraphics[width=0.5\linewidth]{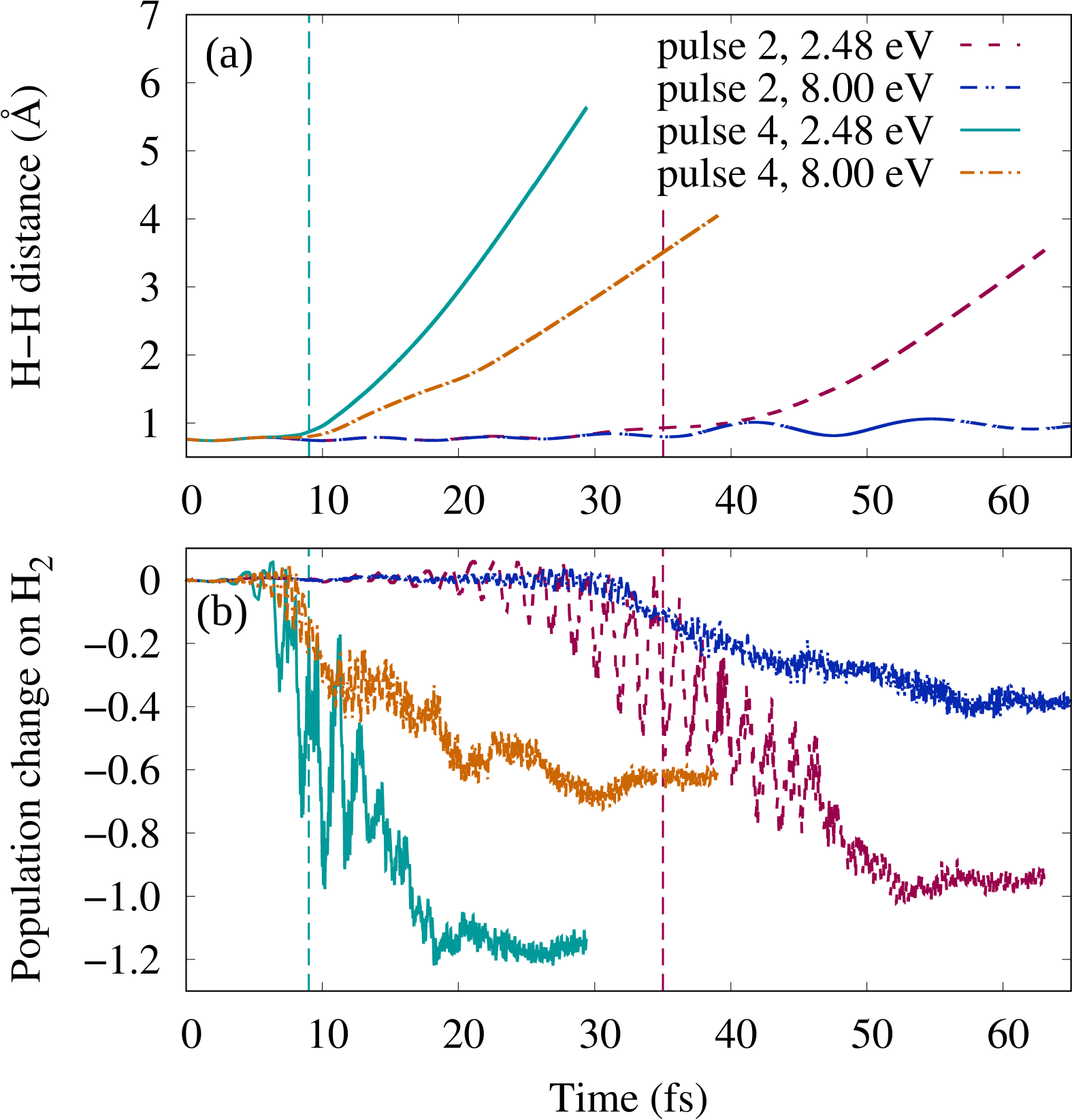}
    \caption{\label{fig:2pulses} (a) H-H internuclear distance and (b) Mulliken population change $\left[\Delta N_e = N_e(t)-N_e(t=0)\right]$ on H$_2$ as a function of time for pulse-2 and pulse-4 (see pulses details in Table~\ref{table:pulses}) at resonant (2.48~eV) and off-resonant (8~eV) field frequency.  Results obtained from TDDFT-ED simulations for Ag$_{231}^{\mathrm{L1}}+$H$_2$. Vertical cyan (pulse-4) and magenta (pulse-2) lines show the instants of each pulse maximum.}
\end{figure}

The time evolution of the H-H internuclear distance under pulse-2 and pulse-4 are compared in Fig.~\ref{fig:2pulses}a. Both pulses induce H$_2$ dissociation at the resonance frequency, but the shorter and more intense pulse-4 also does so under off-resonant conditions. Therefore, pulse-2 provides the most explicit demonstration of plasmon effects, as the excitation of the plasmon is key to inducing H$_2$ dissociation. However, it should be noted that pulse-4 causes the molecule to dissociate much faster at the resonance frequency, demonstrating the importance of the plasmon effect even for this intense pulse. Note in passing that the molecule ends up desorbing regardless of whether it dissociates or not. Although direct comparison to experiments is still questionable due to differences in field intensity and timescale between \textit{ab initio} simulations and experiments, our results are consistent with the observed acceleration of chemical reactions under resonant light~\cite{Miralles2024}. 

The results regarding the evolution of the H-H bond are consonant with the strength and nature of the H$_2$ ionization induced by each pulse and field frequency. This is observed in Fig.~\ref{fig:2pulses}b, in which we show the Mulliken population change on H$_2$. As found previously with pulse-1, the plasmon excitation is demonstrated by the large-amplitude charge oscillations that are obtained for both pulses at 2.48~eV, but not at the high off-resonant frequency of 8~eV. Remarkably, the degree of ionization of the molecule (around one electron), which is inherently a strong-field effect, is also larger with the resonant pulses. The same is observed when comparing the electron population change on the nanoshell (Fig.~S7). This is a clear indication that the excitation of the plasmon facilitates the electron emission that, ultimately, causes the dissociation of the molecule.  Strong-field plasmonic (multiphoton) photoemission from silver nanoparticles has also been observed experimentally under similar laser intensities (10$^{12}-10^{14}$ W cm$^{-2}$) due to a stronger field enhancement at LSPR~\cite{Evers2005, Bionta2016}. 

The fact that pulse-4 is more intense than pulse-2 is reflected in the larger H$_2$ ionization that occurs at the same field frequencies. As a consequence, at this field strength, the molecule also dissociates with the off-resonant pulse-4, despite the absence of plasmon excitation. In this case, although ionization amounts to just $\approx 0.6$ electrons at the end of the dynamics, it is sufficient to induce dissociation of the molecule. The chemical reaction in this particular case is therefore a purely strong-field effect in which plasmon excitation does not play any role.

\begin{figure}[h!]
\centering
    \includegraphics[width=0.5\linewidth]{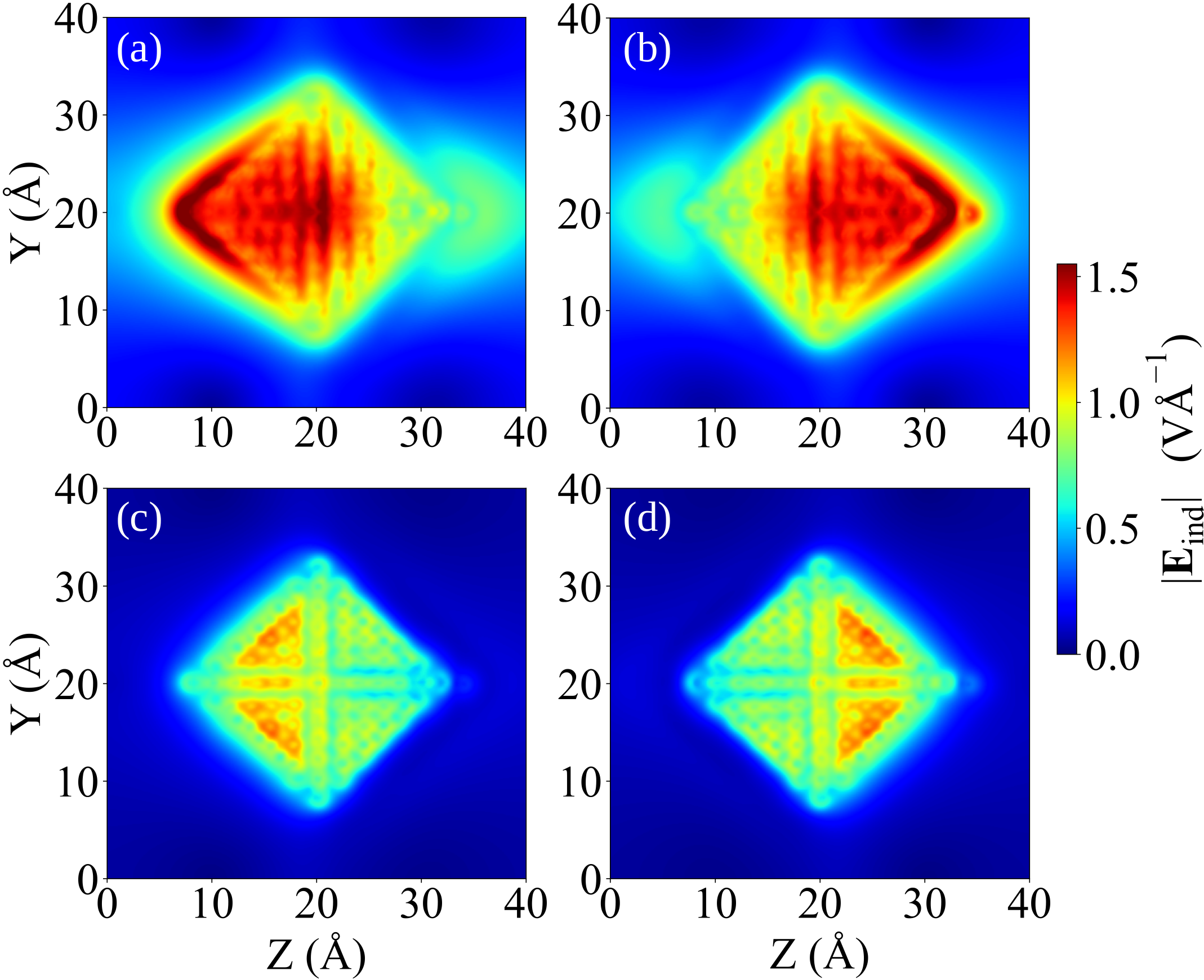}
    \caption{\label{fig:ind_field-pulse2} $(z,y )$-distribution of the $x$-averaged electric field induced on Ag$^{L1}_{231}+$H$_2$ by the external pulse-2 at different instants $t$, 
    $\langle\|\mathbf{E}_{\mathrm{ind}}(y,z;t)\|\rangle_x$. Selected $t$ correspond to instants at which the induced dipole moment $\mu_z(t)$  (see Fig.~S9) is close to a local minimum (left panels) and maximum (right panels). (a) and (b) Snapshots for the resonant field frequency 2.48~eV at instants $t=34.25$ and 35~fs, with actual maximum value of the induced electric field 1.53 and 1.65~V~{\AA}$^{-1}$, respectively. (c) and (d) Snapshots for the off-resonant field frequency 8~eV at instants $t=34$ and 34.25~fs, with actual maximum values of 1.16 and 1.31~V~{\AA}$^{-1}$, respectively. } 
\end{figure}

\begin{figure}[h!]
\centering
    \includegraphics[width=0.5\linewidth]{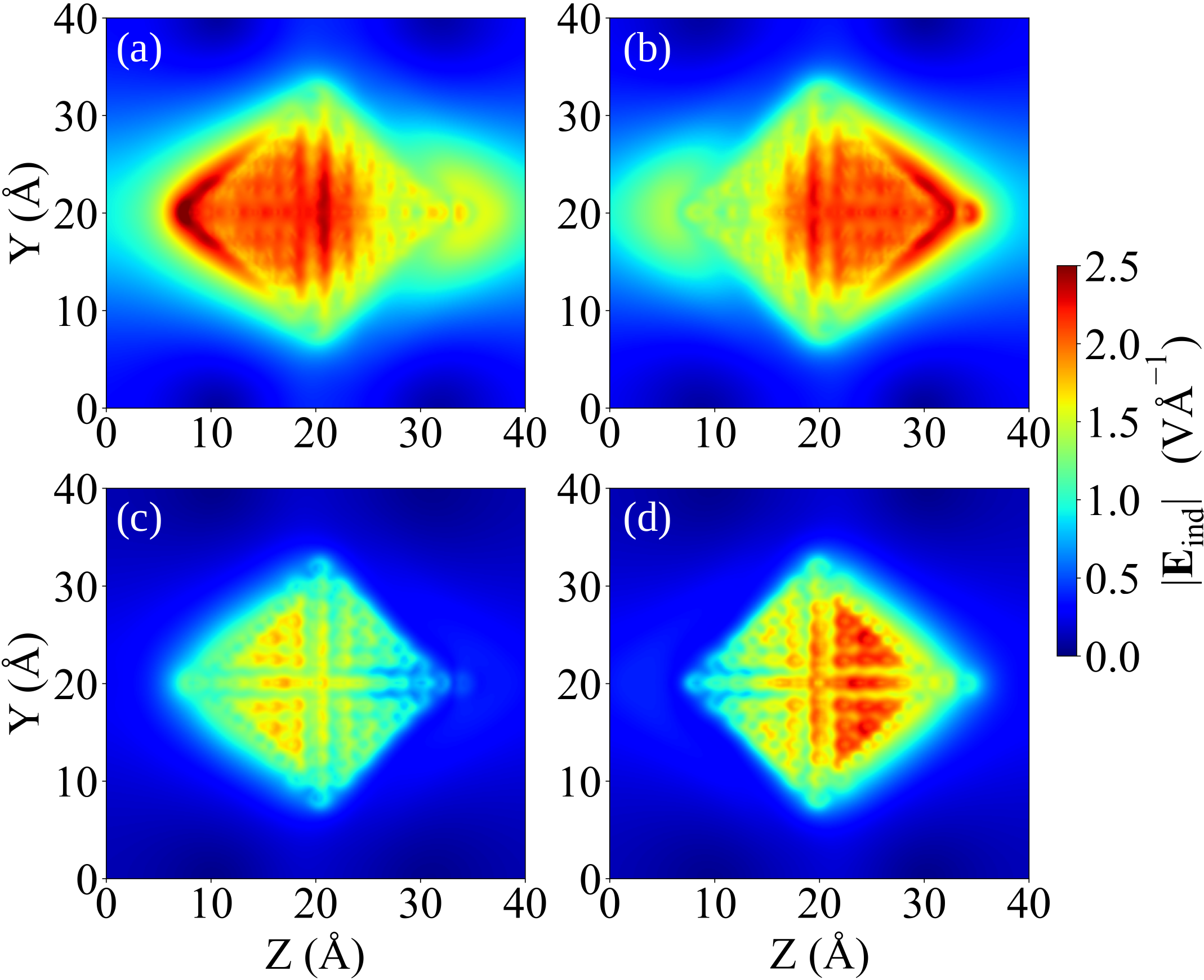}
    \caption{\label{fig:ind_field-pulse4} Same as Fig.\ref{fig:ind_field-pulse2} but for pulse-4. (a) and (b) Snapshots for the resonant field frequency 2.48~eV at instants $t=9.24$ and 10~fs (see Fig.~S10), with actual maximum values of the induced electric field 2.69 and 2.37~V~{\AA}$^{-1}$, respectively. (c) and (d) Snapshots for the off-resonant field frequency 8~eV at instants $t=9.24$ and 10~fs, with actual maximum values of the induced electric field 1.84 and 2.32~V~{\AA}$^{-1}$, respectively. } 
\end{figure}

The comparative analysis of the induced electric field $\mathbf{E}_\mathrm{ind}$ provides direct evidence on the different perturbation that each pulse creates in the Ag$_{231}^{\mathrm{L1}}+$ H$_2$ system and it can allow us to discern how the plasmon excitation, on the one hand, and the strong-field effects, on the other hand, contribute
to the observed dissociation. Figures~\ref{fig:ind_field-pulse2} and \ref{fig:ind_field-pulse4} show selected snapshots of the strength of the induced electric field created by pulse-2 and pulse-4, respectively, at resonant (panels a and b) and off-resonant (panels c and d) frequencies. The induced field is computed from the induced electron density as explained in SI. The two snapshots shown for each pulse and frequency correspond to instants at which the modulus of the induced dipole moment is close to a local maximum (see Figs. S9 and S10) and oriented either parallel or opposite to the $z-$axis. Each snapshot shows the instantaneous $(z,y)$-distribution of the $x$-averaged modulus of the electric field induced on Ag$^{L1}_{231}+$H$_2$ by the external pulse, $\langle\|\mathbf{E}_{\mathrm{ind}}(y,z;t)\|\rangle_x$. 
 Starting with pulse-2, Fig.~\ref{fig:ind_field-pulse2}a and b shows that excitation of the plasmon mode at the resonant frequency leads to a high induced field that oscillates in time between the two opposite vertices located along the $z$-direction, extending well outside the nanoshell. Most of the perturbation is thus transiently localized on the molecule, with the highest values varying in the range 1.1-1.3~V~{\AA}$^{-1}$ (see Figs.~S11-S12). In contrast, the off-resonant pulse-2 creates a much weaker induced field that, in addition, is mostly localized on the nanoshell and not on the molecule, as shown in Fig.~\ref{fig:ind_field-pulse2}c and d. Thus, the observed dissociation at the resonant pulse-2 is a consequence of the plasmon-enhanced high induced electric field that is created on the molecule. 


In the case of pulse-4, the induced electric field at both the resonant (Fig.~\ref{fig:ind_field-pulse4}a and b) and off-resonant (Fig.~\ref{fig:ind_field-pulse4}c and d) frequencies behaves similarly to that of pulse-2. The difference is that the induced electric field created by the more intense pulse-4 is also stronger. Thus, even if the regions where $E_\mathrm{ind}$ is more intense with the off-resonant pulse are localized at the nanoshell faces, the values at the vertex where the molecule is adsorbed are still transiently high (0.9-1.0~V~{\AA}$^{-1}$, see Fig.~S18) and comparable to those created by the resonant pulse-2. As a result, the molecule also dissociates with the off-resonant pulse-4. 
Altogether, the results obtained with pulse-2 and pulse-4 remark that the amount of energy provided by the external field is not the only defining parameter, as it is utilized much less effectively in the off-resonant case. At resonance, plasmonic field enhancement concentrates the energy near the molecule, enabling efficient electron emission and bond dissociation. In contrast, off-resonant excitation results in a weaker effect on the molecule.

\section{Conclusions and outlook}

In summary, we have analyzed whether using the high-intensity laser pulses (10$^{13}$--10$^{14}$ W~cm$^{-2}$) required to perform TDDFT-ED simulations, plasmonic effects can still be observed in photoinduced chemical reactions on nanoparticles. Our simulations show that by increasing the size of the nanoparticle we can indeed distinguish plasmonic effects that were completely suppressed by strong-field effects in the small nanoparticle studied in our previous work~\cite{Koval2024}.
Still, for the nanoparticle size and field intensities employed, nonlinear effects are manifested. The obtained dissociation of the H$_2$ molecule requires that it loses one electron. Moreover, it can take place even under off-resonant conditions when applying a short (2.5 fs) but very intense ($8 \times 10^{13}$ W~cm$^{-2}$) external pulse. 
However, the effect of the plasmon can be clearly distinguished in all cases when comparing the response of the system under resonant and off-resonant pulses. The plasmon excitation generated by the former induces much larger charge oscillations on the molecule. The induced electric field is also stronger and is more localized on the molecule. As a result, depending on the intensity of the external field, the plasmon excitation can be decisive to provoke the chemical reaction or to accelerate it.

Actual experiments are performed in the linear regime with field intensities of the order of 1 W cm$^{-2}$, much lower than the ones used here. As discussed above, this means that nonlinear effects are still playing a role in these simulations. As a consequence, the extrapolation of the results to disentangle the mechanisms governing the experimental observations is still questionable. Nevertheless, the observation that already at the utilized field intensities and nanoparticle sizes the TDDFT-ED approach can account for plasmon induced photochemical processes constitutes a promising fact for future research. Further reducing the field strength while increasing its duration,  and further increasing the nanoparticle size to the extent allowed by future computational capabilities may permit to reach a regime that finally can be safely extrapolated to the experimental conditions and allow for a better  understanding of the physics behind the measured photo-induced chemical processes.

\section*{Author contributions}
Conceptualization: NEK, MA, JIJ; Methodology: NEK, MA; Calculations: NEK;
Data analysis and validation: NEK, MA, JIJ; Visualization: NEK; Writing – original draft preparation: NEK; Writing – review and editing: NEK, MA, JIJ; Funding acquisition: MA, JIJ. All authors have accepted responsibility for the entire content of this manuscript and approved its submission.

\section*{Conflicts of interest}
There are no conflicts to declare.

\section*{Data availability}

The data generated during the current study are presented in the article and in the Supplementary Information (SI). SI contains additional figures and theoretical details on the calculation of the induced fields.

\section*{Acknowledgements}

Financial support was provided by the
Spanish MCIN$/$AEI$/$10.13039$/$501100011033$/$, FEDER Una manera de hacer Europa (Grant No.~PID2022-140163NB-I00), Gobierno Vasco-EHU (Project No.~IT1569-22), and the Basque Government Education Departments’ IKUR program, also co-funded by the European NextGenerationEU action through the Spanish Plan de Recuperación, Transformación y Resiliencia (PRTR). The authors thankfully acknowledge the computer resources at MareNostrum and the technical support provided by Barcelona Supercomputing Center (Projects No. RES-FI-2024-2-0022 and No. RES-FI-2025-1-0006) as well as the HPC resources provided by the Donostia International Physics Center (DIPC) Supercomputing Center.


\bibliographystyle{unsrt}
\bibliography{plasmon-induced-hot-carriers}

\end{document}


\maketitle

\section{Gaussian pulses used in TDDFT-ED simulations}

\begin{figure}[ht!]
\centering
    \includegraphics[width=0.6\linewidth]{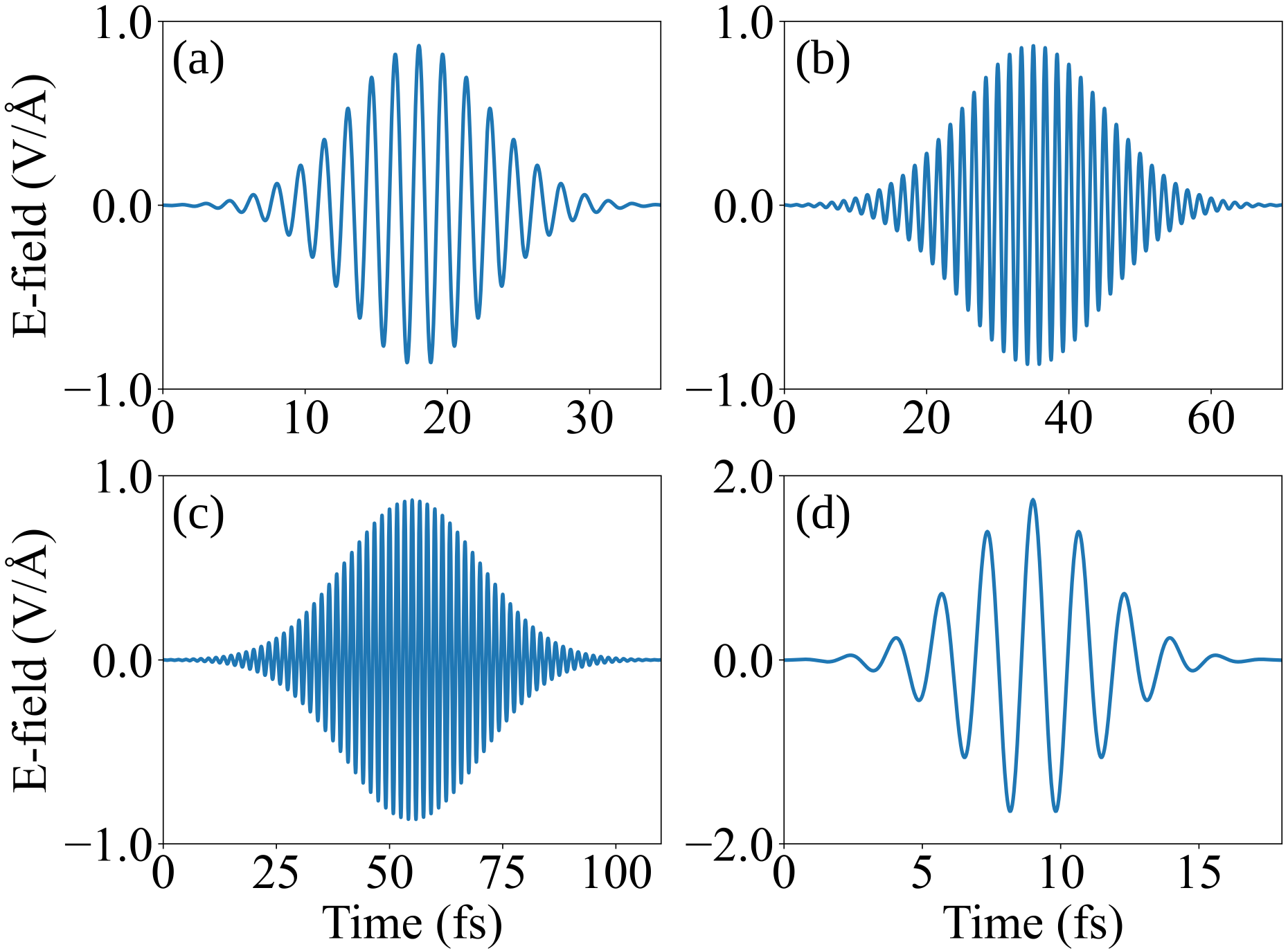}
    \caption{\label{fig:4pulses} Electric field strength as a function of time for the four pulses used in this work (all four shown at plasmon frequency $\hbar \omega = \hbar \omega_p = 2.48$ eV): (a) pulse-1: $I_{\mathrm{max}} = 2 \times 10^{13}$ W~cm$^{-2}$ ($E_0 = 0.87$ V~{\AA}$^{-1}$), $t_0=18$ fs, $\sigma=5$ fs, (b) pulse-2: $I_{\mathrm{max}} = 2 \times 10^{13}$ W~cm$^{-2}$, $t_0=35$ fs, $ \sigma=10$ fs, (c) pulse-3: $I_{\mathrm{max}} = 2 \times 10^{13}$ W~cm$^{-2}$, $t_0=55$ fs, $ \sigma=15$ fs, (d) pulse-4: $I_{\mathrm{max}} = 8 \times 10^{13}$ W~cm$^{-2}$ ($E_0 = 1.74$ V~{\AA}$^{-1}$), $t_0=9$ fs, $ \sigma=2.5$ fs.}
\end{figure}

\clearpage

\section{Induced dipole moment at low intensity}

\begin{figure}[h!]
 \centering
		\includegraphics[width=0.7\linewidth]{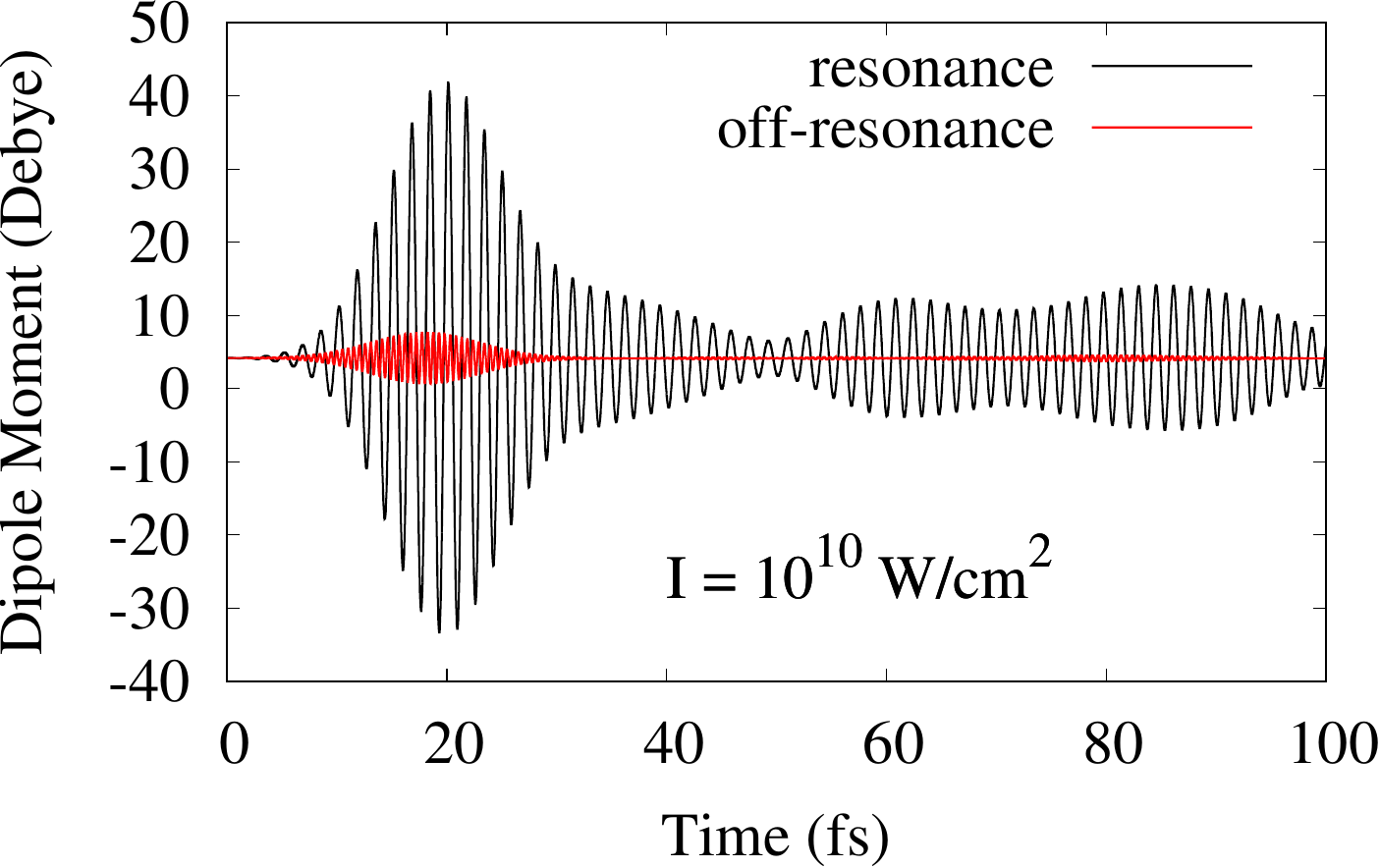}
		\caption{Time-dependent dipole moment induced in the Ag$_{231}^{L1}$ nanoshell with H$_2$ adsorbate by the external pulse with intensity $I_{\mathrm{max}}=2\times 10^{10}$ W cm$^{-2}$ and frequencies 2.48 eV (resonant condition) and 8 eV (off-resonance). The Gaussian pulse parameters are: $t_0=18$ fs, $\sigma=5$ fs.}
		\label{fig:ind_dip_Ag55}
	\end{figure}

\FloatBarrier

\section{Selected molecular orbitals}

\begin{figure}[h!]
 \centering
 \includegraphics[width=0.99\linewidth]{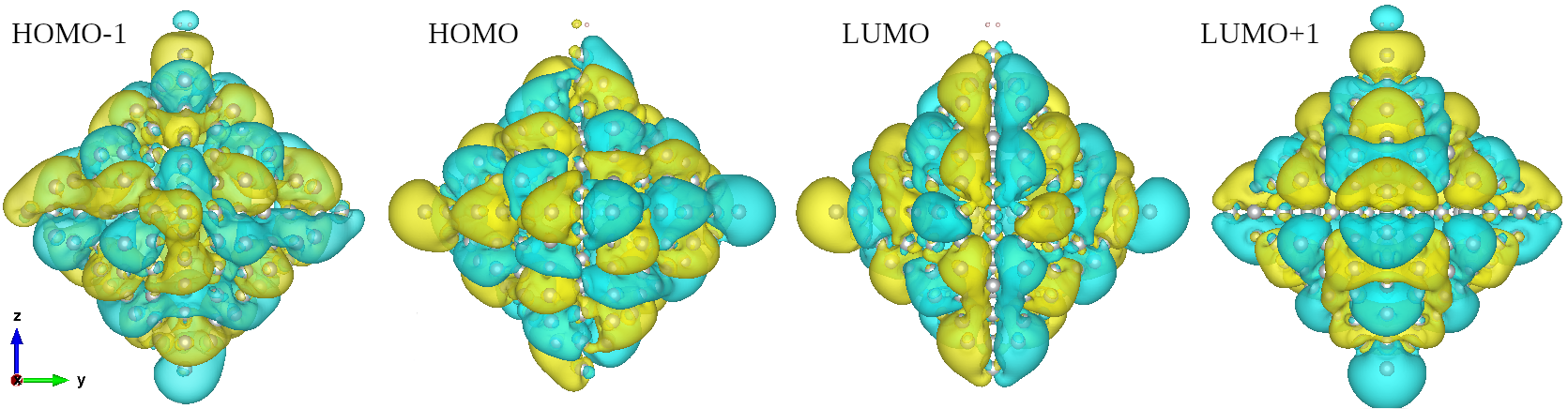}
		\caption{ Ground-state molecular orbitals for the Ag$_{231}^{L1}$ nanoshell with H$_2$ adsorbate.
        }
		\label{fig:orb}
	\end{figure}

\clearpage

\section{H$_2$ dissociation energy: gas-phase vs adsorbed}

\begin{figure}[h!]
 \centering
    \includegraphics[width=0.7\linewidth]{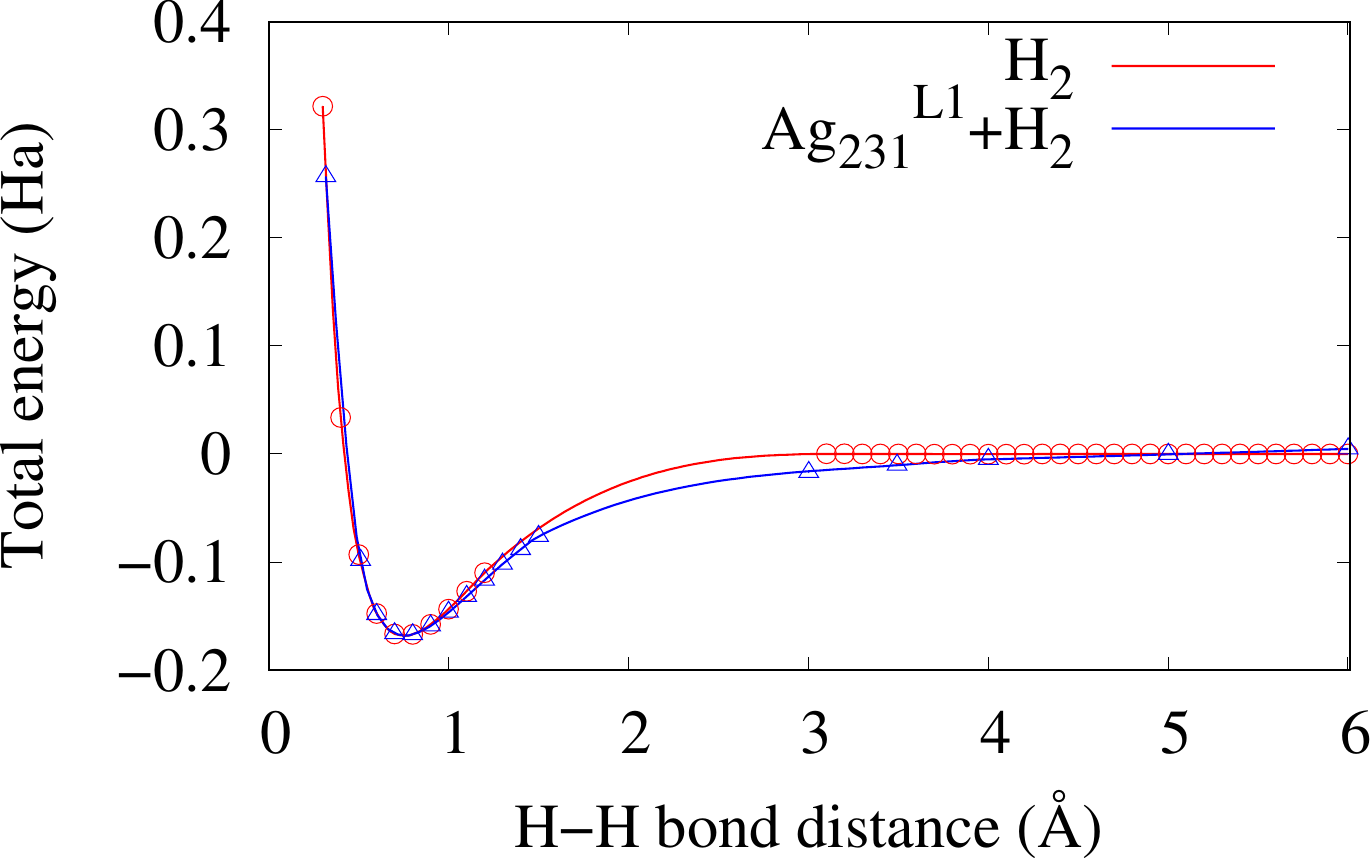}
		\caption{Potential energy (in Hartree units) as a function of the H-H internuclear distance for H$_2$ in gas-phase (red circles) and adsorbed on Ag$_{231}^{L1}$ (blue triangles). Symbols show the DFT+PBE energies and lines represent the extrapolation of the calculated data.}
		\label{fig:H2-dist-vs-en}
	\end{figure}

Figure \ref{fig:H2-dist-vs-en} shows the potential energy as a function of the H-H internuclear distance for H$_2$ in gas-phase and H$_2$ adsorbed on the Ag$_{231}^{L1}$ nanoshell. The energy is calculated using density functional theory (DFT) and the PBE exchange correlation functional with the CP2K software package. The calculated results are shown as symbols, while the extrapolated data is shown as lines. The potential energy at small (large) internuclear distances were calculated for a singlet (triplet) spin state of H$_2$ using unrestricted Kohn-Sham DFT with a spin multiplicity of 1 (3). For the triplet state of the adsorbed molecule, constrained DFT was employed to enforce the localization of spin-polarized electrons on the hydrogen atoms, ensuring a proper triplet configuration at large internuclear distances. The transition between the two spin states was captured by interpolating the energy using Akima spline extrapolation, providing a smooth transition between the singlet and triplet states.

\clearpage

\section{Population change on Ag$_{231}^{L1}$ under pulses-1,-2,-3}

\begin{figure}[h!]
 \centering
		\includegraphics[width=0.7\linewidth]{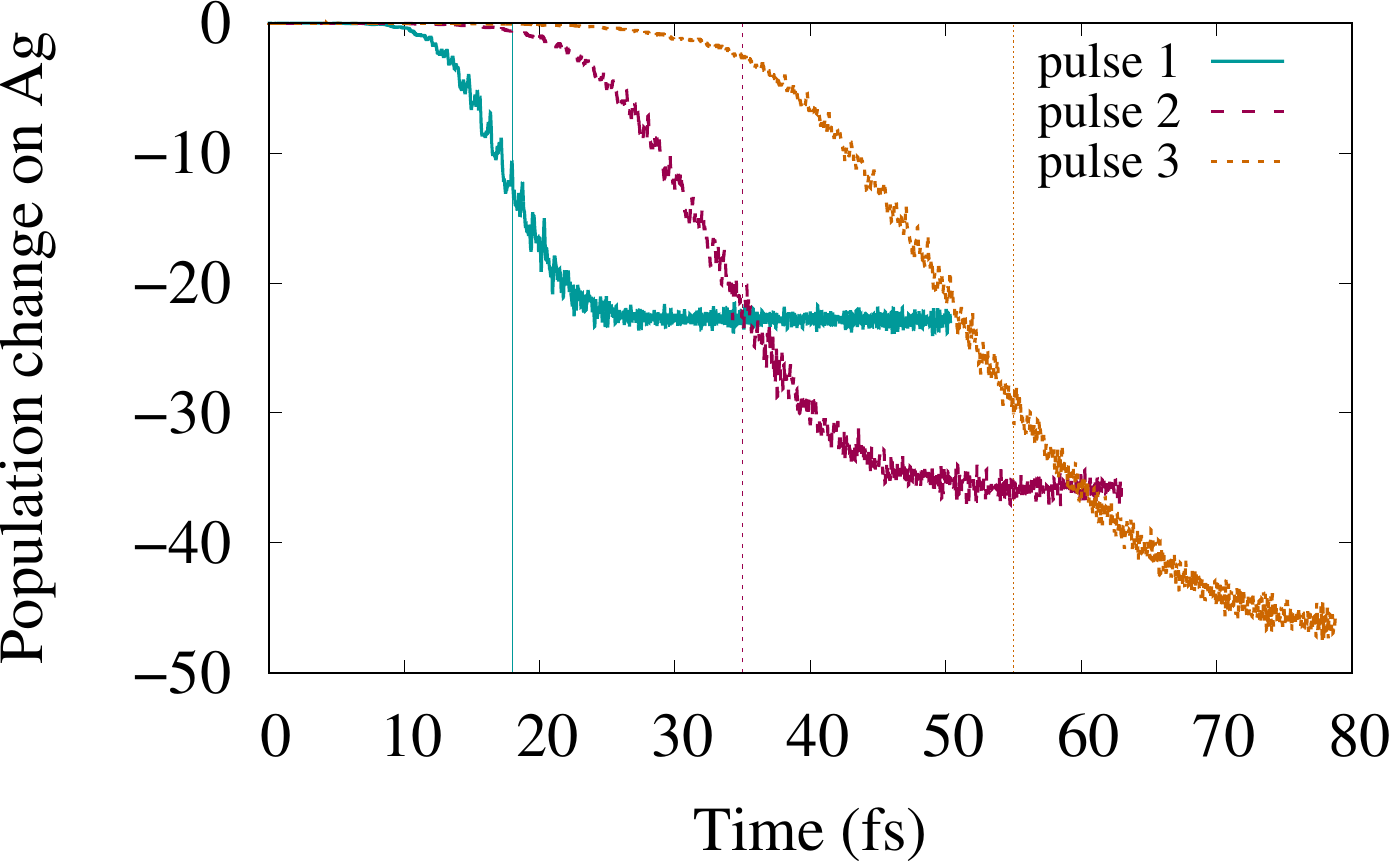}
      \caption{Mulliken population change $\left[\Delta N_e = N_e(t)-N_e(t=0)\right]$ on Ag$_{231}^{L1}$ nanoshell as a function of time for pulse-1, pulse-2, and pulse-3 (see pulses details in Fig.~\ref{fig:4pulses}a-c) at frequency 2.48 eV.}
\label{fig:Ag_populations_pulses1_3}
	\end{figure}

\section{Dipole moment and population change on Ag$_{231}^{L1}$ under pulse-4}

\begin{figure}[h!]
 \centering
		\includegraphics[width=0.6\linewidth]{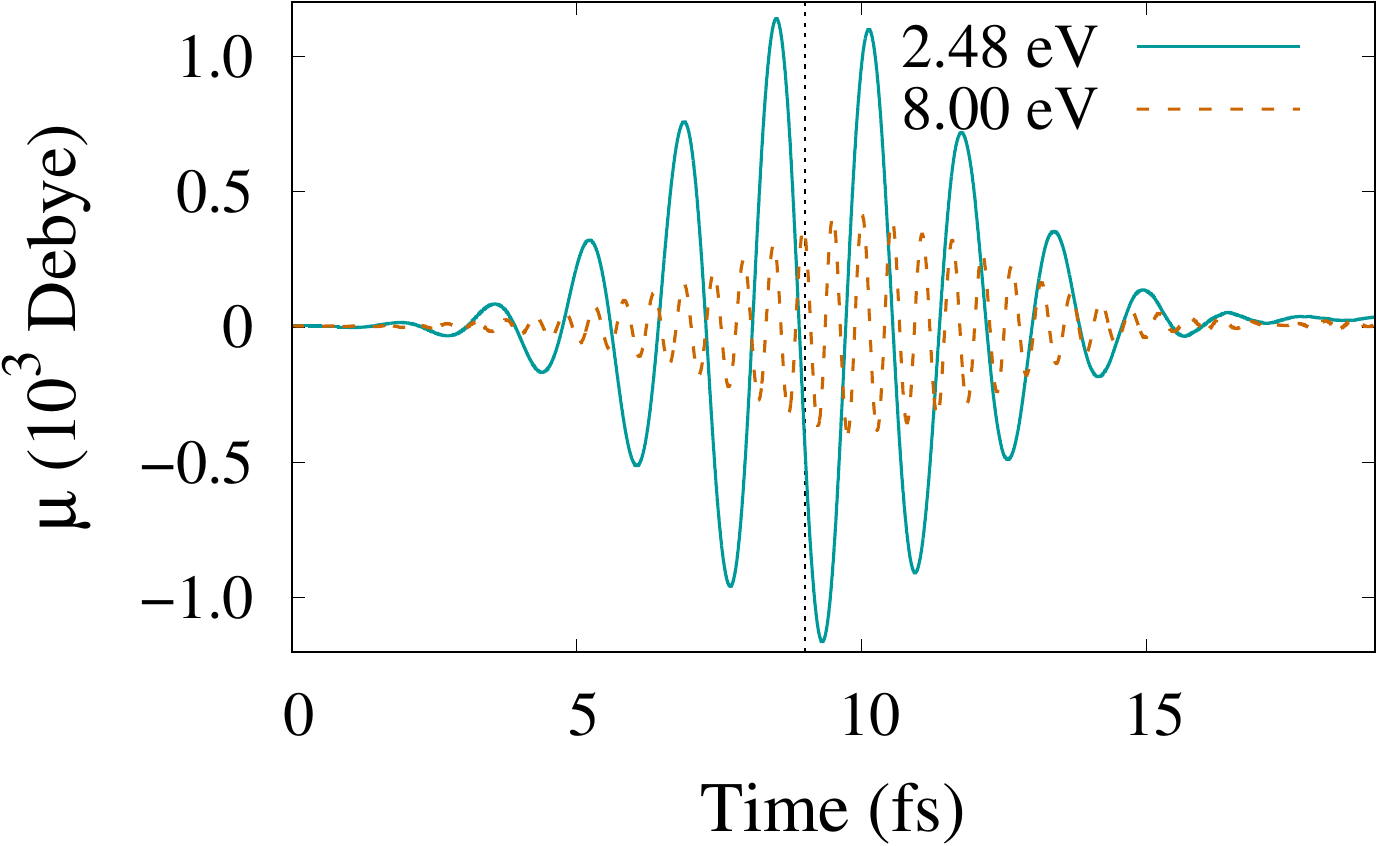}
		\caption{Time-dependent dipole moment for the Ag$_{231}^{L1}$ nanoshell with H$_2$ adsorbate under the external pulse-4 with intensity $I_{\mathrm{max}}=8\times 10^{13}$ W cm$^{-2}$ and both frequencies. The vertical line shows the pulse maximum.}
		\label{fig:dipole_pulse4}
	\end{figure}

\begin{figure}[h!]
 \centering
		\includegraphics[width=0.7\linewidth]{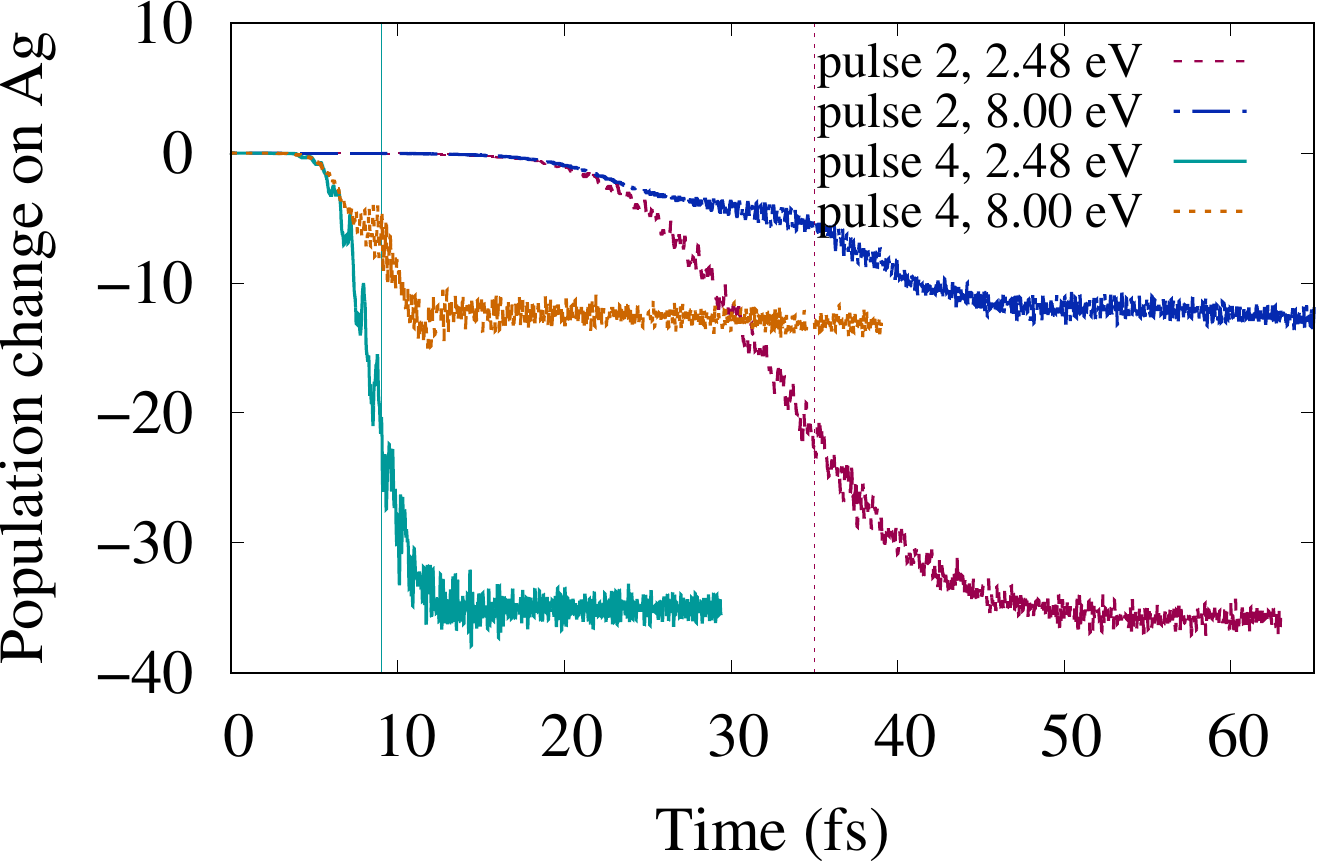}
      \caption{Mulliken population change $\left[\Delta N_e = N_e(t)-N_e(t=0)\right]$ on Ag$_{231}^{L1}$ nanoshell as a function of time for pulse-4 compared to pulse-2 (see pulses details in Fig.~\ref{fig:4pulses}b,d) at frequencies $\hbar \omega = \hbar \omega_{\mathrm{p}} = 2.48$ eV and $\hbar \omega =8$ eV.}
\label{fig:Ag_populations_pulses2_4}
	\end{figure}

\FloatBarrier 

\section{Calculation of the induced electric field on Ag$_{231}^{L1}+$H$_2$}

The induced charge density $\rho_{\text{ind}}(\mathbf{r})$ generates an electrostatic potential $\Phi_{\text{ind}}(\mathbf{r})$ governed by Poisson’s equation:
\begin{equation}
\nabla^2 \Phi_{\text{ind}}(\mathbf{r}) = - 4   \pi \rho_{\text{ind}}(\mathbf{r}),
\end{equation}
where $\epsilon_0$ is the vacuum permittivity. To solve this equation efficiently, we transform it into Fourier space. Taking the Fourier transform:
\begin{equation}
\mathcal{F} \left[ \nabla^2 \Phi_{\text{ind}}(\mathbf{r}) \right] = -4   \pi \mathcal{F} \left[ \rho_{\text{ind}}(\mathbf{r}) \right].
\end{equation}

Since the Fourier transform of the Laplacian is $-\mathbf{k}^2$, we obtain:
\begin{equation}
\tilde{\Phi}_{\text{ind}}(\mathbf{k}) = \frac{4 \pi \tilde{\rho}_{\text{ind}}(\mathbf{k})}{k^2},
\end{equation}
where $\mathbf{k} = (k_x, k_y, k_z)$ is the wavevector, and $k^2 = k_x^2 + k_y^2 + k_z^2$.

The induced electric field is then given by:
\begin{equation}
\mathbf{E}_{\text{ind}}(\mathbf{r}) = -\nabla \Phi_{\text{ind}}(\mathbf{r}),
\end{equation}
which, in Fourier space, becomes:
\begin{equation}
\tilde{\mathbf{E}}_{\text{ind}}(\mathbf{k}) = -i \mathbf{k} \tilde{\Phi}_{\text{ind}}(\mathbf{k}).
\end{equation}

Substituting $\tilde{\Phi}_{\text{ind}}(\mathbf{k})$:
\begin{equation}
\tilde{\mathbf{E}}_{\text{ind}}(\mathbf{k}) = -i \mathbf{k} \frac{4 \pi \tilde{\rho}_{\text{ind}}(\mathbf{k})}{ k^2}.
\end{equation}
To avoid division by zero at $\mathbf{k} = 0$, we set $k^2 = 1$ at that point.

Finally, the induced field in real space is obtained via the inverse Fourier transform:
\begin{equation}
\mathbf{E}_{\text{ind}}(\mathbf{r}) = \mathcal{F}^{-1} \left[ \tilde{\mathbf{E}}_{\text{ind}}(\mathbf{k}) \right].
\end{equation}

\begin{figure}[h!]
 \centering
		\includegraphics[width=0.46\linewidth]{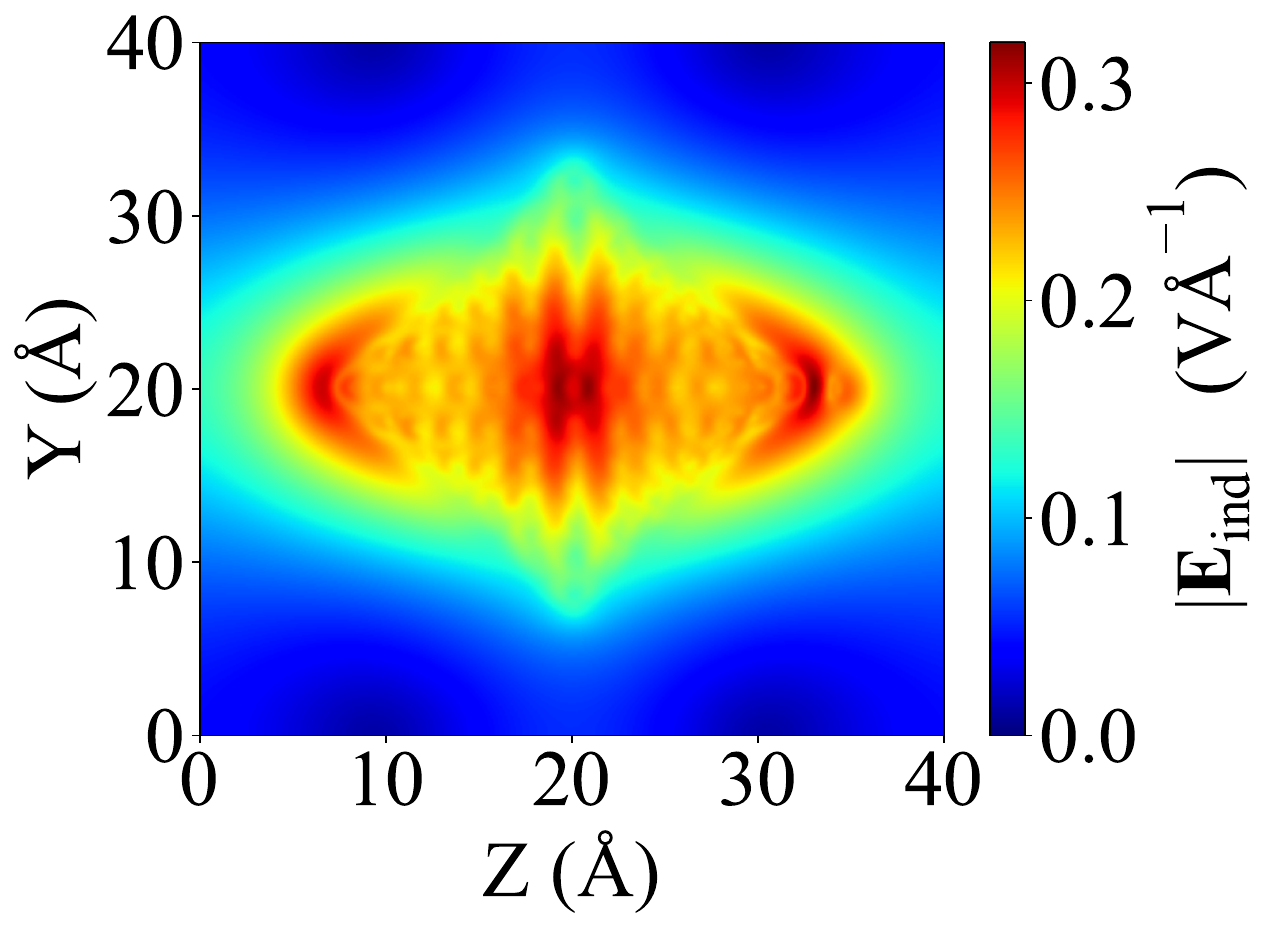}\hfill
        \includegraphics[width=0.49\linewidth]{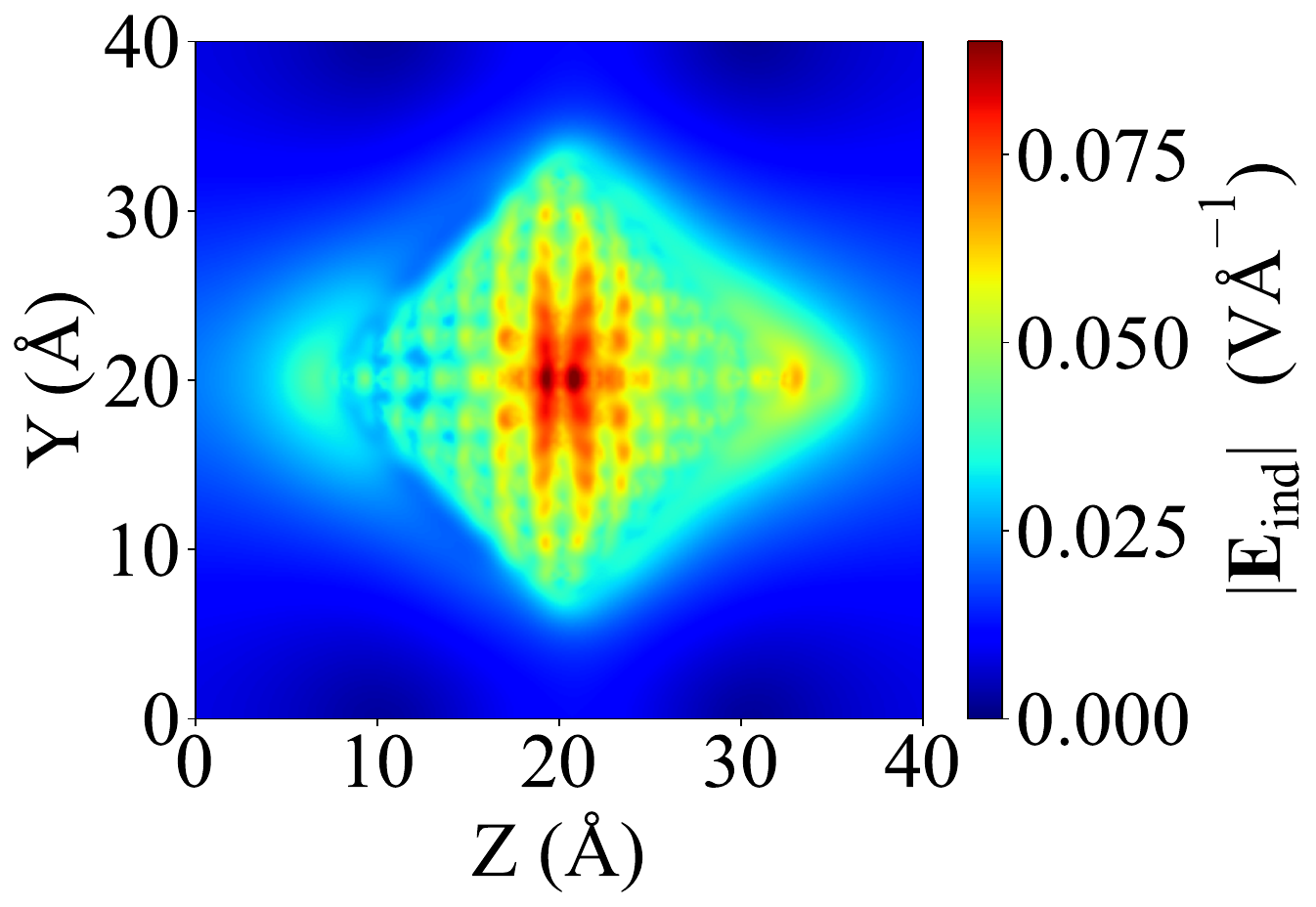}
		\caption{$(z,y)$-distribution of the $x$-averaged induced electric field $\left<\|\mathbf{E}_{\text{ind}}(y,z)\|\right>_x$ on the Ag$_{231}^{L1}$ nanoshell with H$_2$ adsorbed on the right vertex under the external pulse-1 with intensity $I_{\mathrm{max}}=2\times 10^{13}$ W cm$^{-2}$ and frequencies (left) 2.48 eV and (right) 8 eV at the time instant $t = 10$ fs. External field is polarized along the $z-$axis.
      Notice different scale.}
		\label{fig:field_enhanc}
	\end{figure}

Figure~\ref{fig:field_enhanc} shows the induced electric field on the Ag$_{231}^{L1}$ nanoshell with H$_2$ adsorbate under the external pulse-1 with intensity $I_{\mathrm{max}}=2\times 10^{13}$ W cm$^{-2}$ and frequencies (left) 2.48 eV and (right) 8 eV at the time instant $t = 10$ fs. These results are obtained from the RT-TDDFT-ED calculations for the structures surrounded by the layer of ghost atoms. We show the mean over the $x-$axis induced field in the $yz-$plane. At the plasmon resonance (2.48 eV), the nanoparticle’s collective oscillation strongly amplifies the external field, producing hot spots on the nanoshell vertices, with induced electric field in the vicinity of the H$_2$ molecule reaching 0.25 V {\AA}$^{-1}$. In contrast, at the off-resonant excitation energy (8 eV), the particle’s polarizability is much weaker and its induced field is largely out of phase with the driving field. As a result, there is a much weaker induced field at the H$_2$ location, reaching only about 0.05 V {\AA}$^{-1}$, five time weaker than in the case of the resonant frequency.

\begin{figure}[h!]
 \centering
		\includegraphics[width=0.4\linewidth]{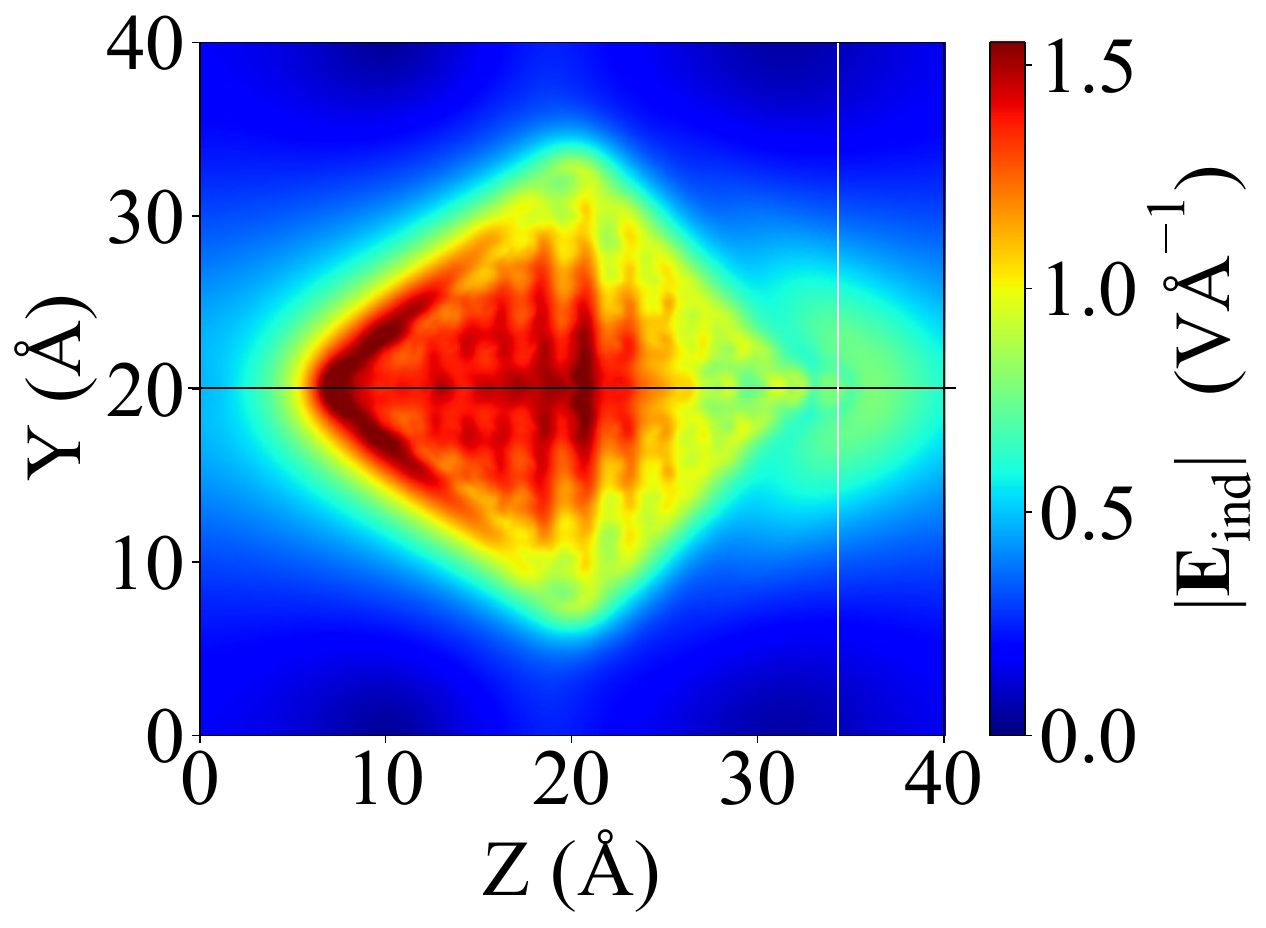}\\
        \includegraphics[width=0.45\linewidth]{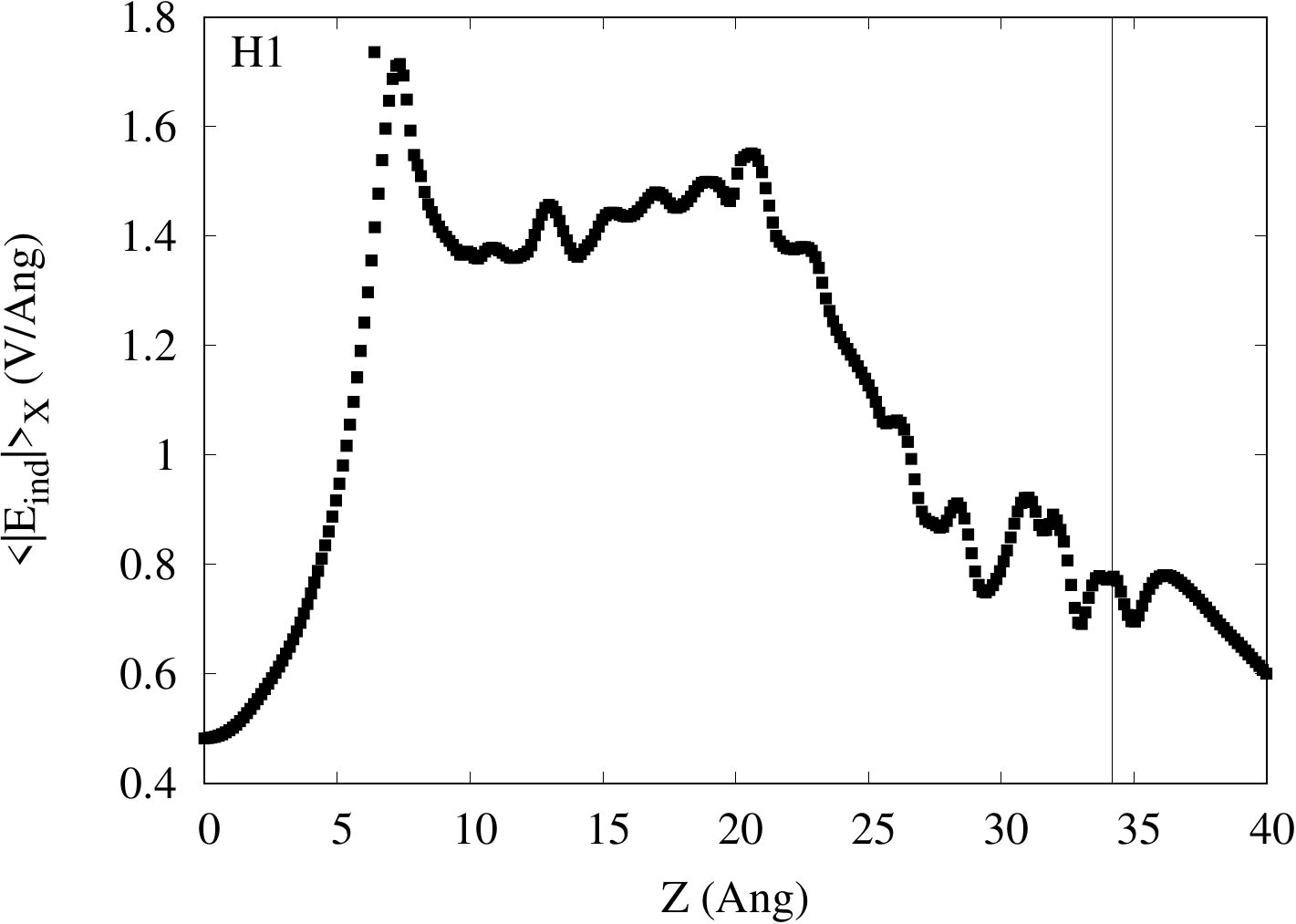} 
        \includegraphics[width=0.45\linewidth]{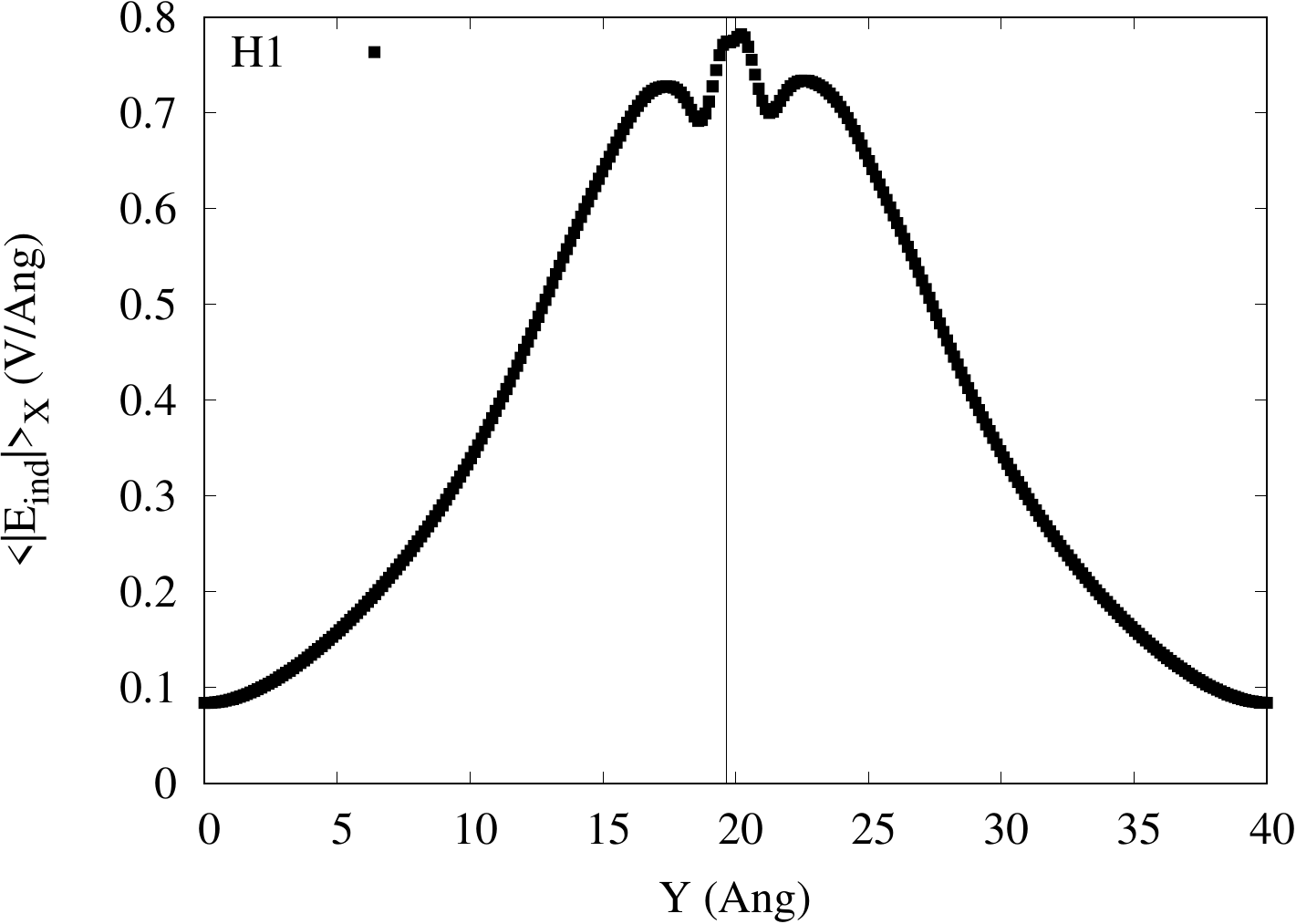}
        \includegraphics[width=0.45\linewidth]{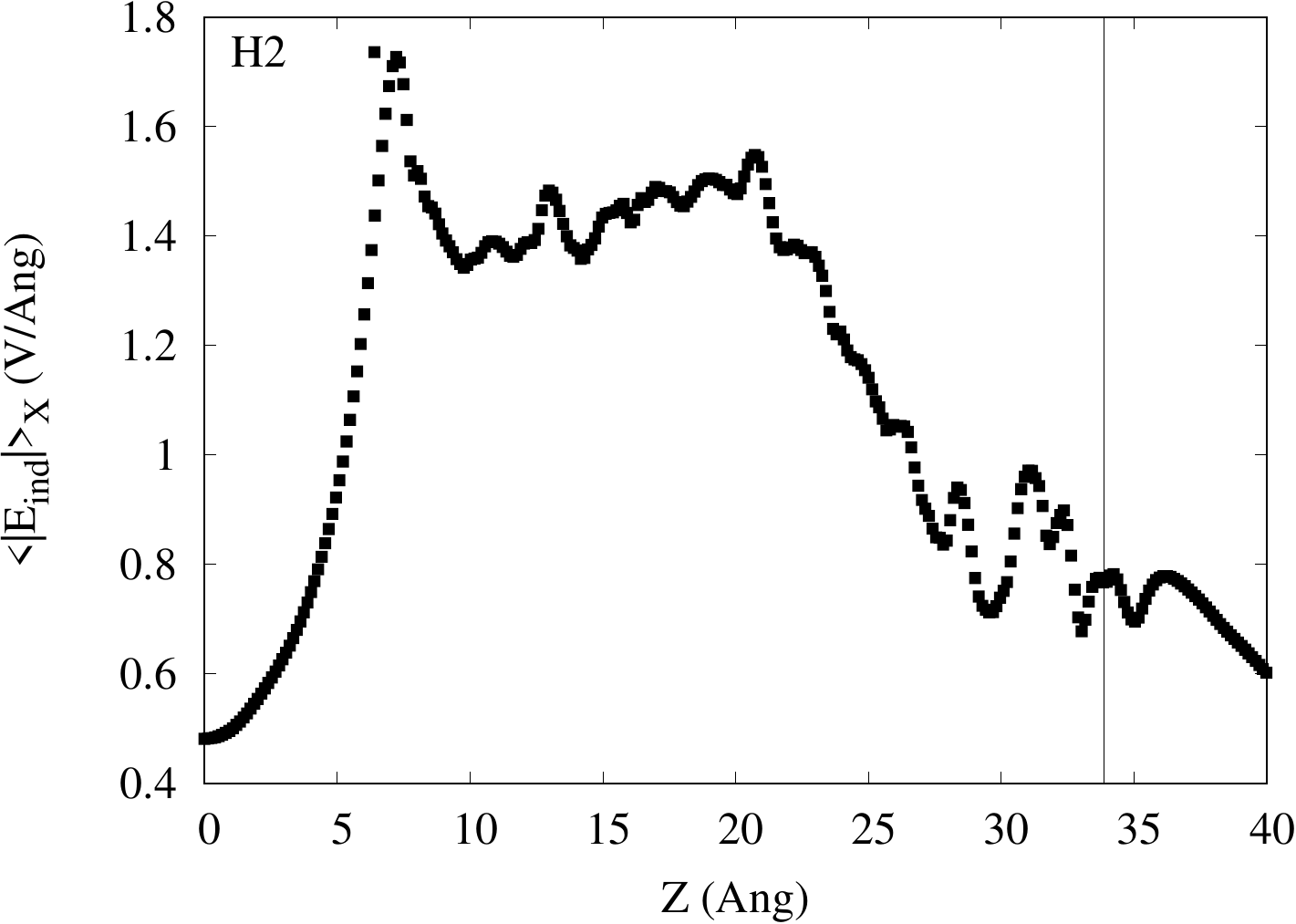} 
        \includegraphics[width=0.45\linewidth]{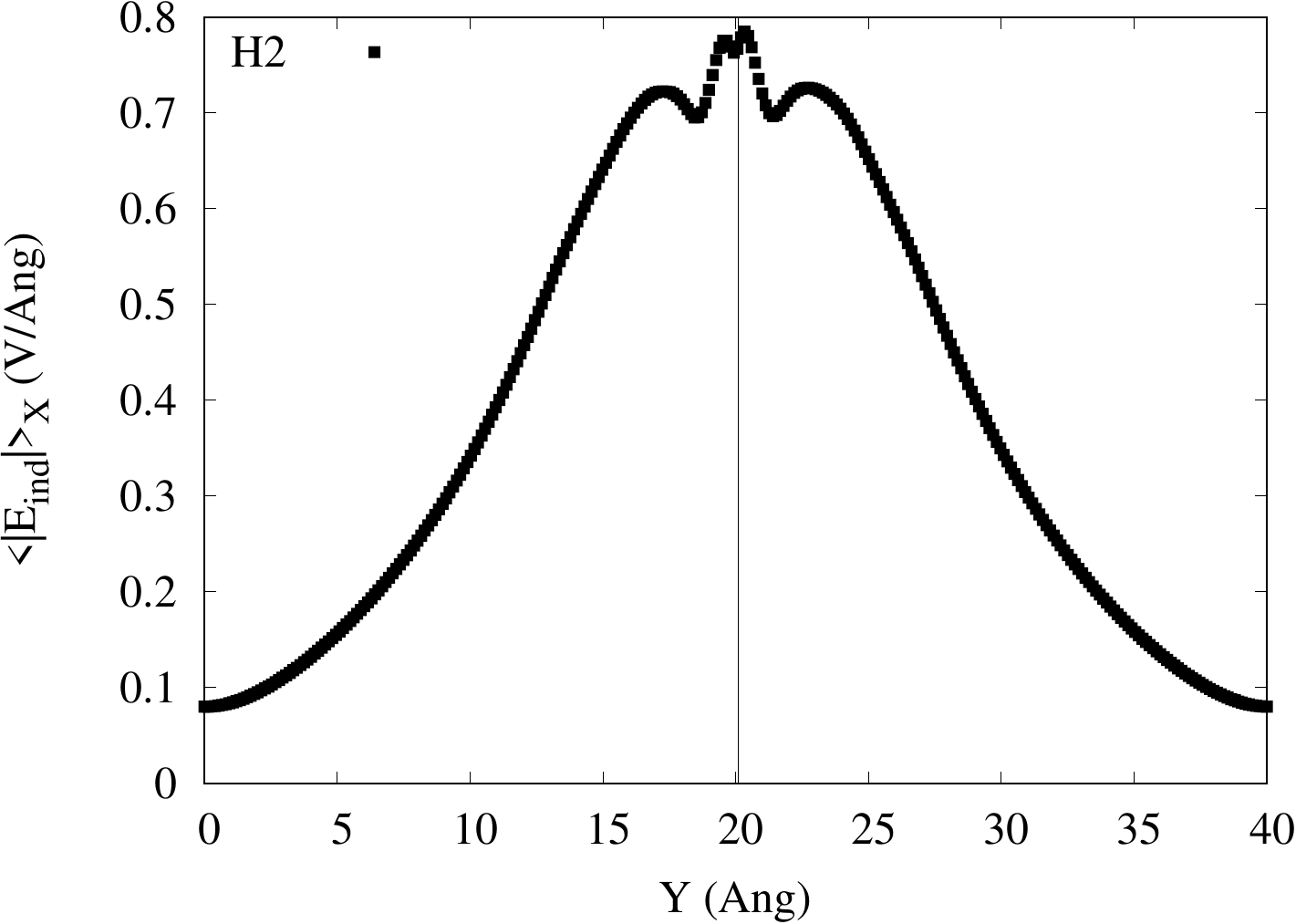}
		\caption{$x$-averaged electric field induced on Ag$^{L1}_{231}+$H$_2$ by the external pulse-2 with frequency 2.48 eV at the time instant $t = 34.25$ fs. The molecule is adsorbed on the right vertex. The Cartesian coordinates of the atoms forming the molecule, denoted as H1 and H2 in the figure, are (20.17091, 19.65791, 34.16472) and  (19.41151, 20.09346, 33.86330), respectively. 
        (Upper panel) $(z,y)$-distribution of the $x$-averaged induced  electric field. Horizontal (vertical) black (white) line approximately shows the cut through H1 and H2 at their $y$ ($z$) positions, for which lower left (lower right) panels show the induced electric field as a function of $z ~(y)$.} 
		\label{fig:field_2res}
	\end{figure}

 The electron oscillations due to plasmon translate into a symmetrical induced field distribution around both vertices along the $z-$axis in Fig.~\ref{fig:field_enhanc}(left) due to the fact that we show the modulus $\left<\|\mathbf{E}_{\text{ind}}\|\right>_x$. Since strong field of the pulses leads to the superposition of both plasmonic and nonlinear effects such as ionization, the observed symmetry is lost at the time instances when the nanoshell has lost electrons (Figs.~\ref{fig:field_2res}, \ref{fig:field_2res_1}, \ref{fig:field_4res}, \ref{fig:field_4res_1}, and \ref{fig:field_1res} below). The emitted electrons occupy ghost atoms in the time-dependent density, while all the electrons are on the Ag$_{231}^{L1}+$ H$_2$ in the ground-state density (for the current snapshot atomic structure) which we subtract to find the induced density. This means that the positive electron change on one vertex will be smaller than the negative electron change on the other vertex by the number of emitted electrons.

\clearpage

\section{External electric field and induced dipole moment for pulses 2 and 4}
\begin{figure}[h!]
 \centering
		\includegraphics[width=0.8\linewidth]{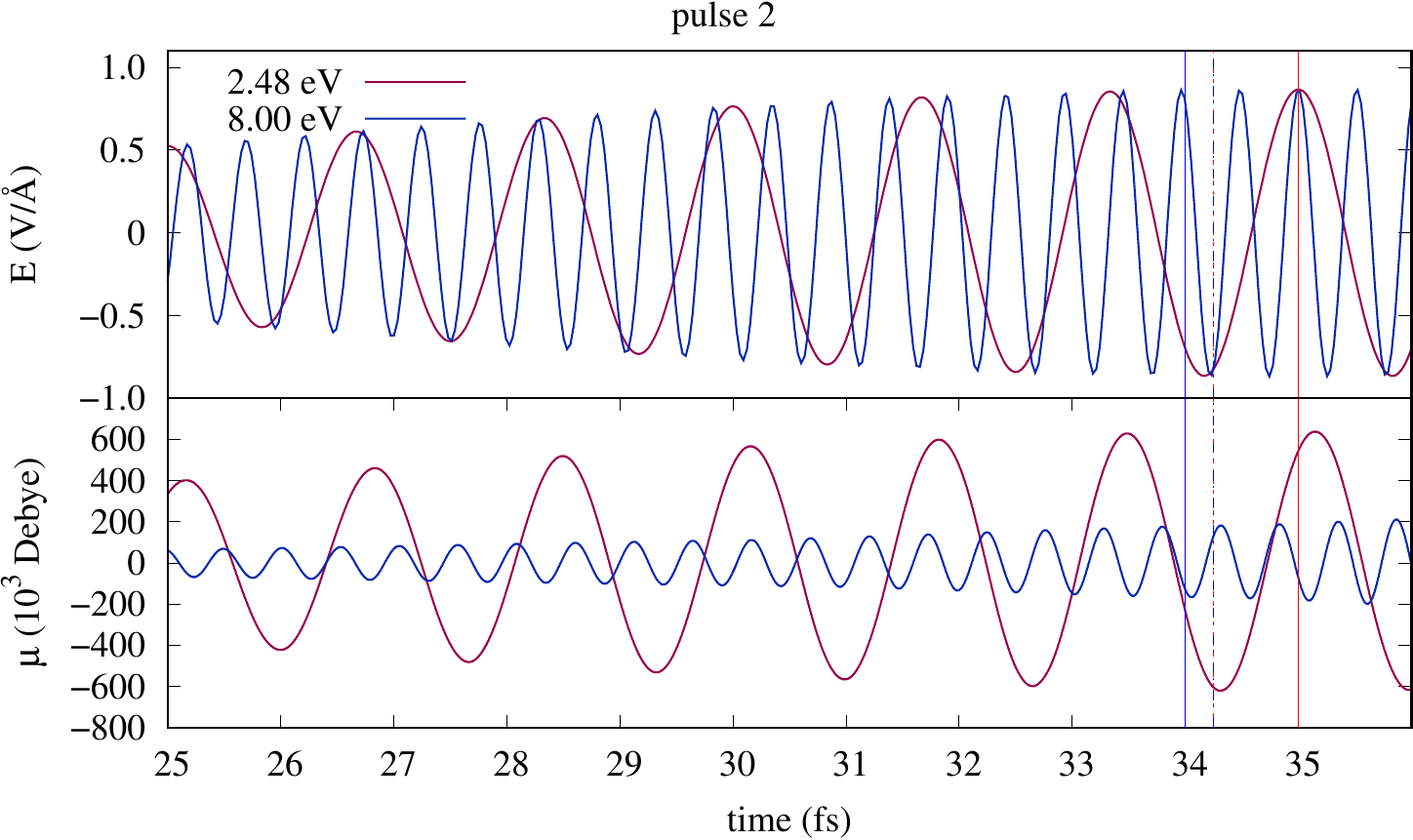}
      \caption{Upper panel: Electric field strength for pulse-2 with frequencies 2.48 eV (resonant) and 8 eV (off-resonant). Lower panel: Corresponding induced dipole moment as a function of time. Vertical red (blue) lines show the time instances used for plotting the induced electric field at 2.48 eV (8 eV) in Fig. 6 of the main article and in Figs~\ref{fig:field_2res} to \ref{fig:field_2nonres_1} below.}
\label{fig:efield_dip_2}
	\end{figure}

\begin{figure}[h!]
 \centering
		\includegraphics[width=0.8\linewidth]{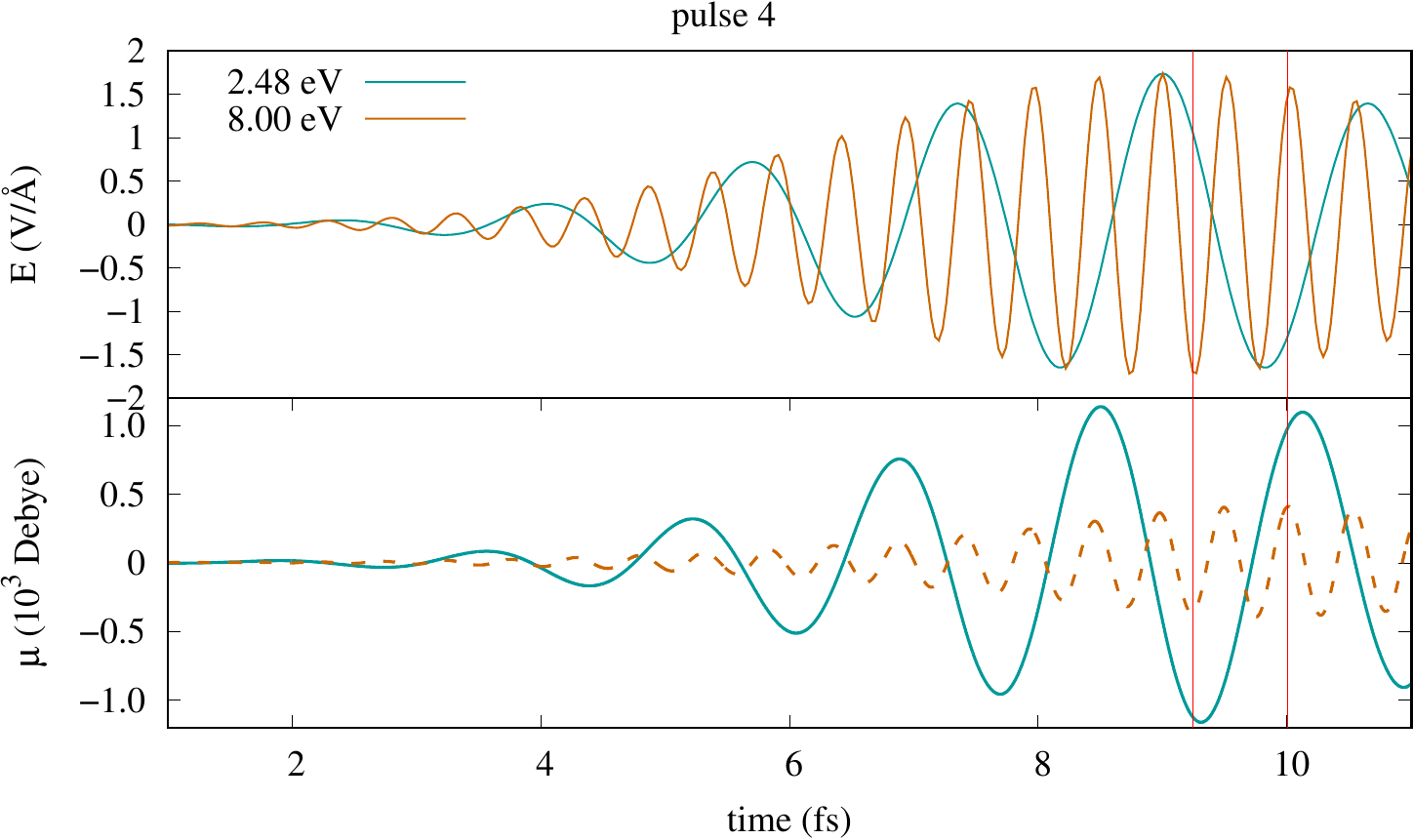}
      \caption{Same as Fig.~\ref{fig:efield_dip_2} for pulse-4 corresponding to Fig. 7 of the main article and to Figs~\ref{fig:field_4res} to \ref{fig:field_4nonres_1} below.}
\label{fig:efield_dip_4}
	\end{figure}

\clearpage

\section{Induced electric field analysis at the position of H$_2$}

\begin{figure}[h!]
 \centering
		\includegraphics[width=0.4\linewidth]{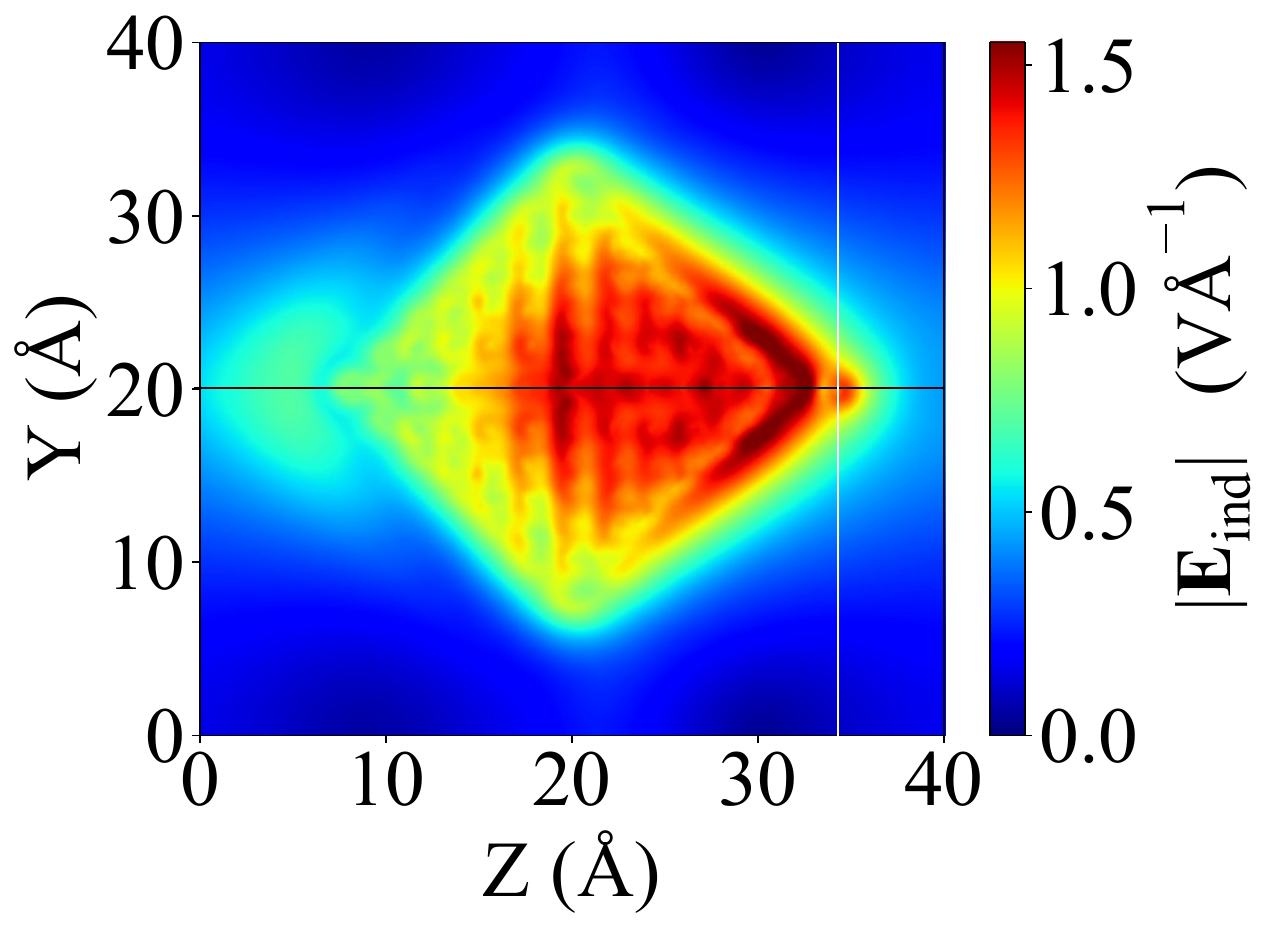}\\
        \includegraphics[width=0.45\linewidth]{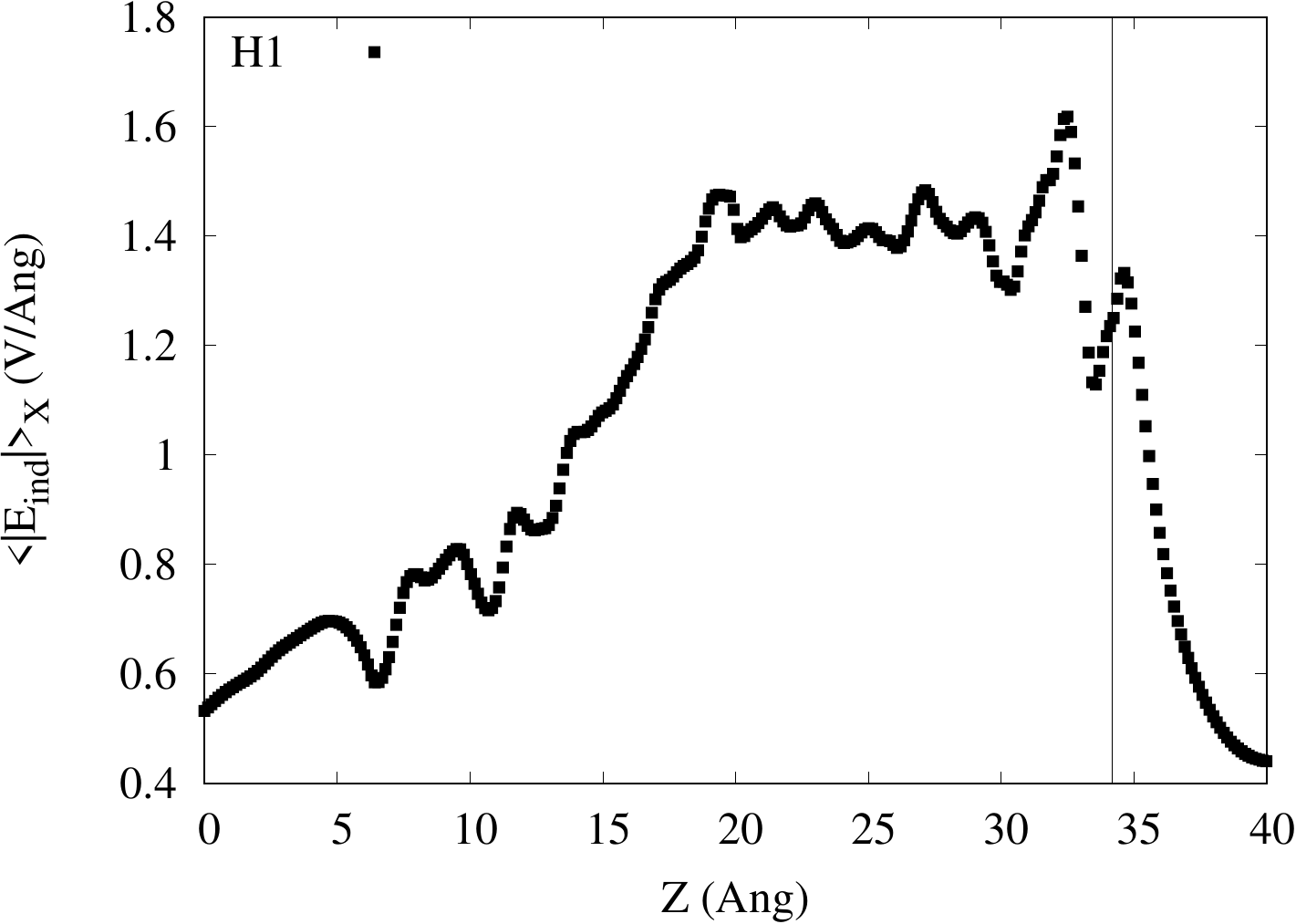} 
        \includegraphics[width=0.45\linewidth]{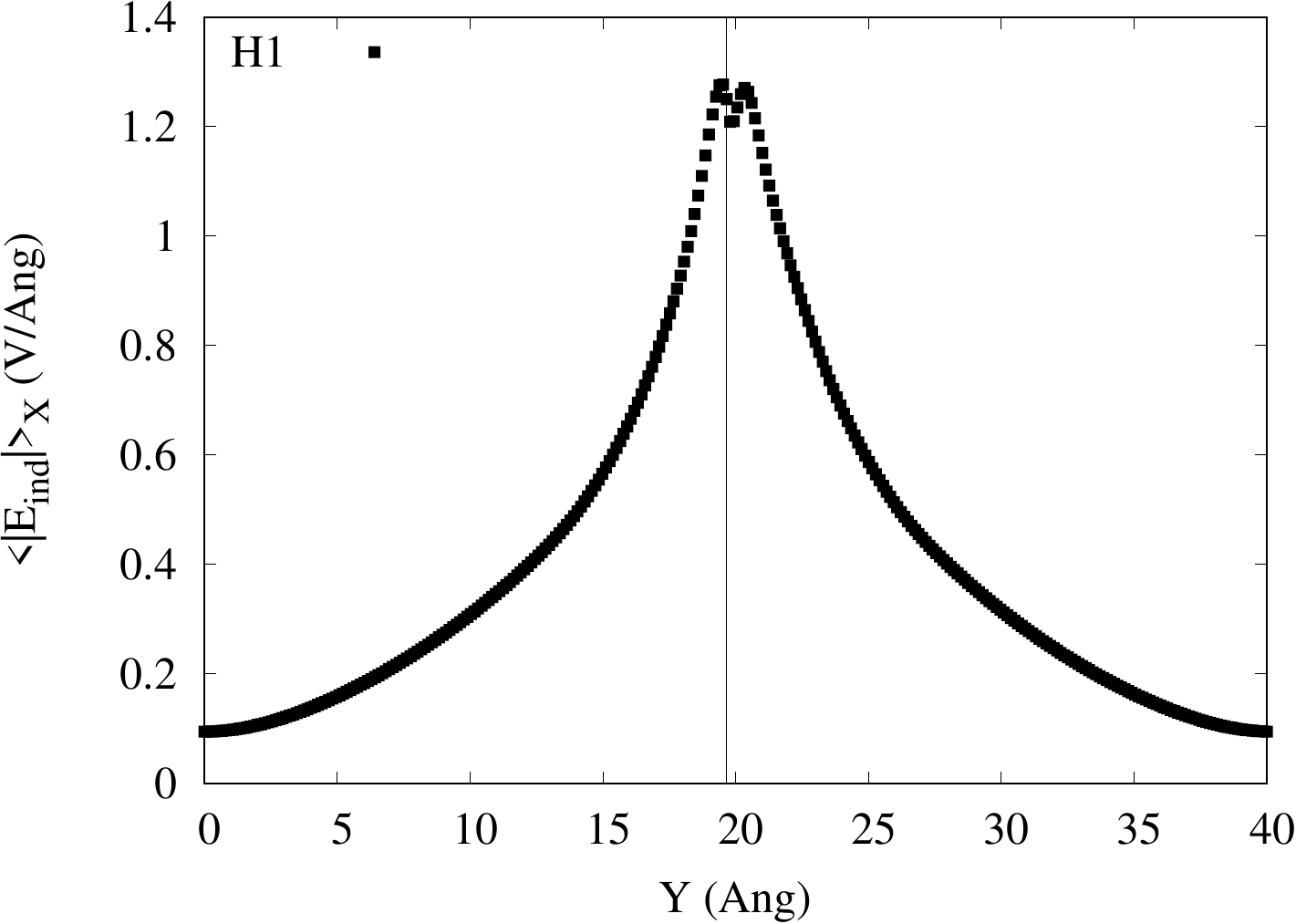}
        \includegraphics[width=0.45\linewidth]{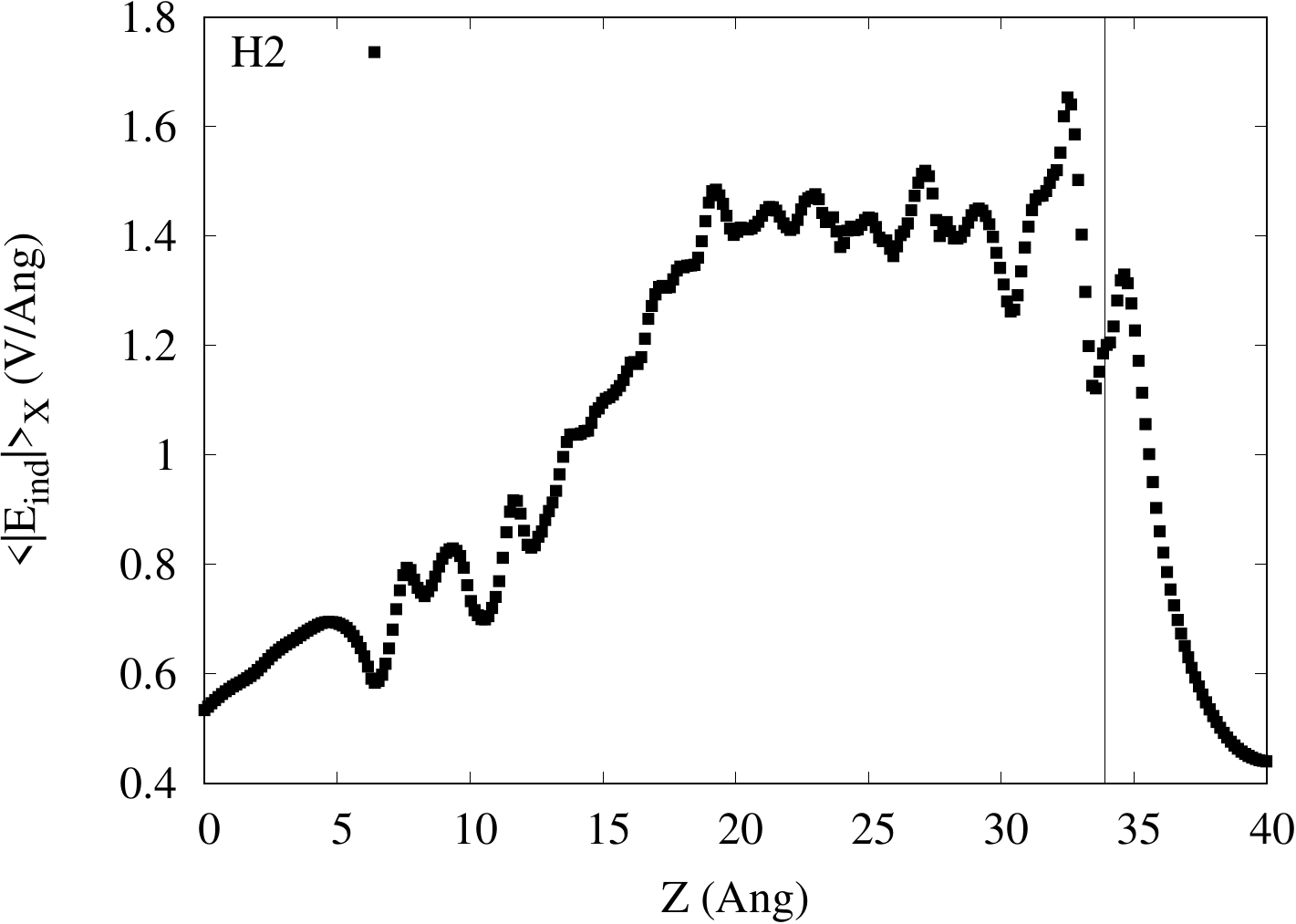} 
        \includegraphics[width=0.45\linewidth]{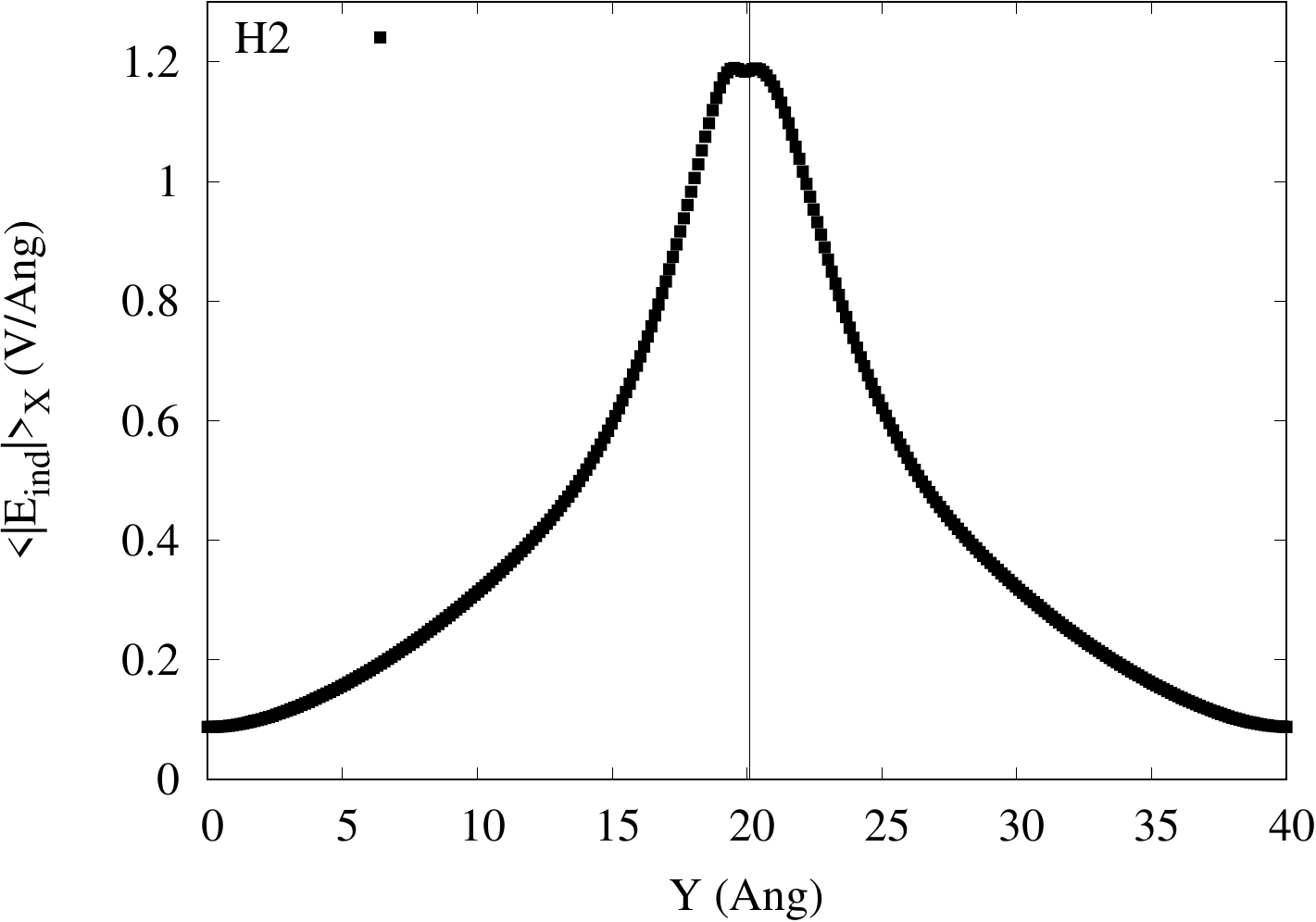}
		\caption{Same as Fig.~\ref{fig:field_2res} but for pulse-2 with frequency 2.48 eV at the time instant $t = 35$ fs. The molecule Cartesian coordinates are H1: (20.16494 19.66330 34.18505) and H2: (19.39228, 20.08282, 33.88200).}
		\label{fig:field_2res_1}
	\end{figure}

   \begin{figure}[h!]
 \centering
		\includegraphics[width=0.4\linewidth]{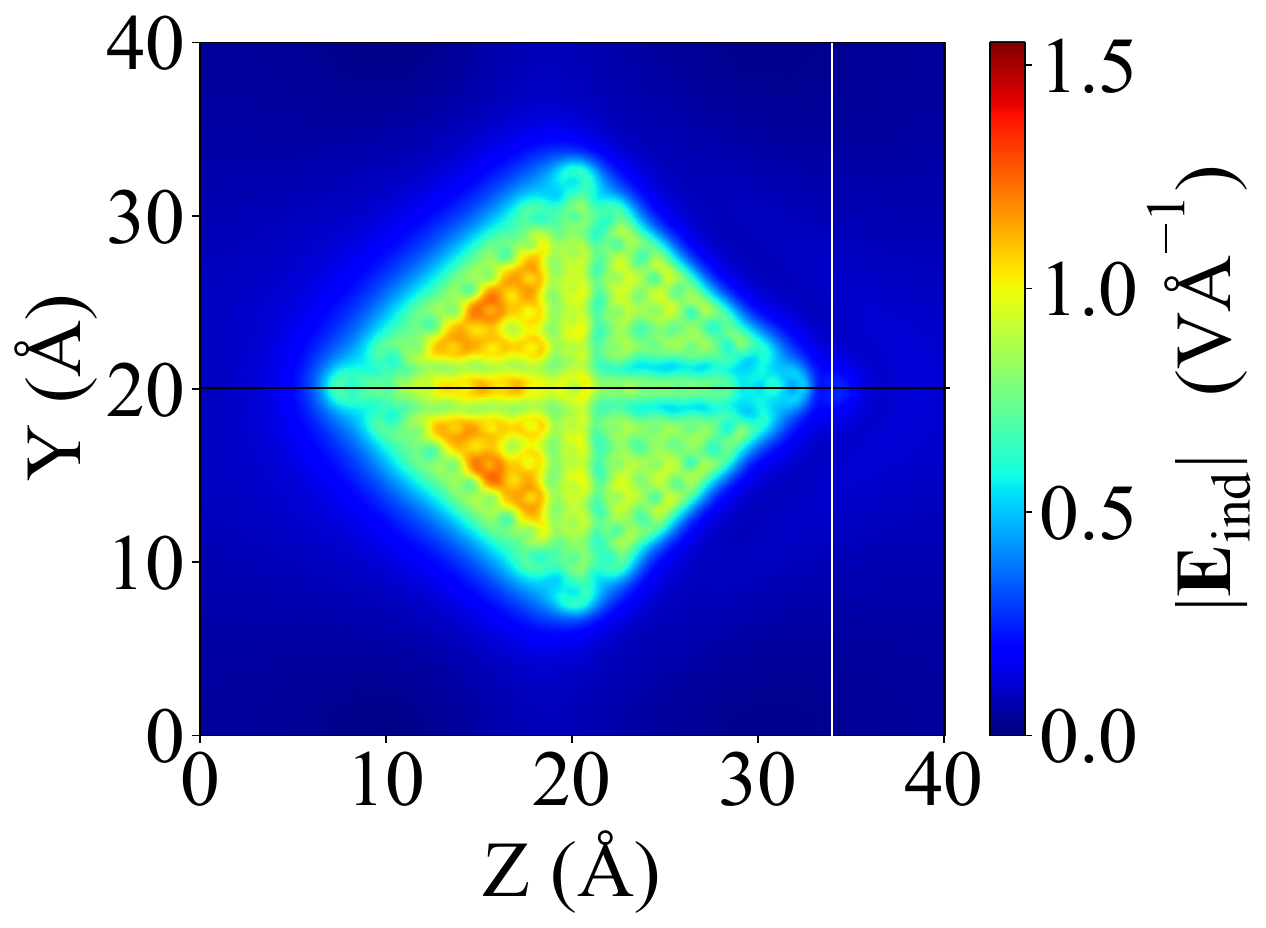}\\
        \includegraphics[width=0.45\linewidth]{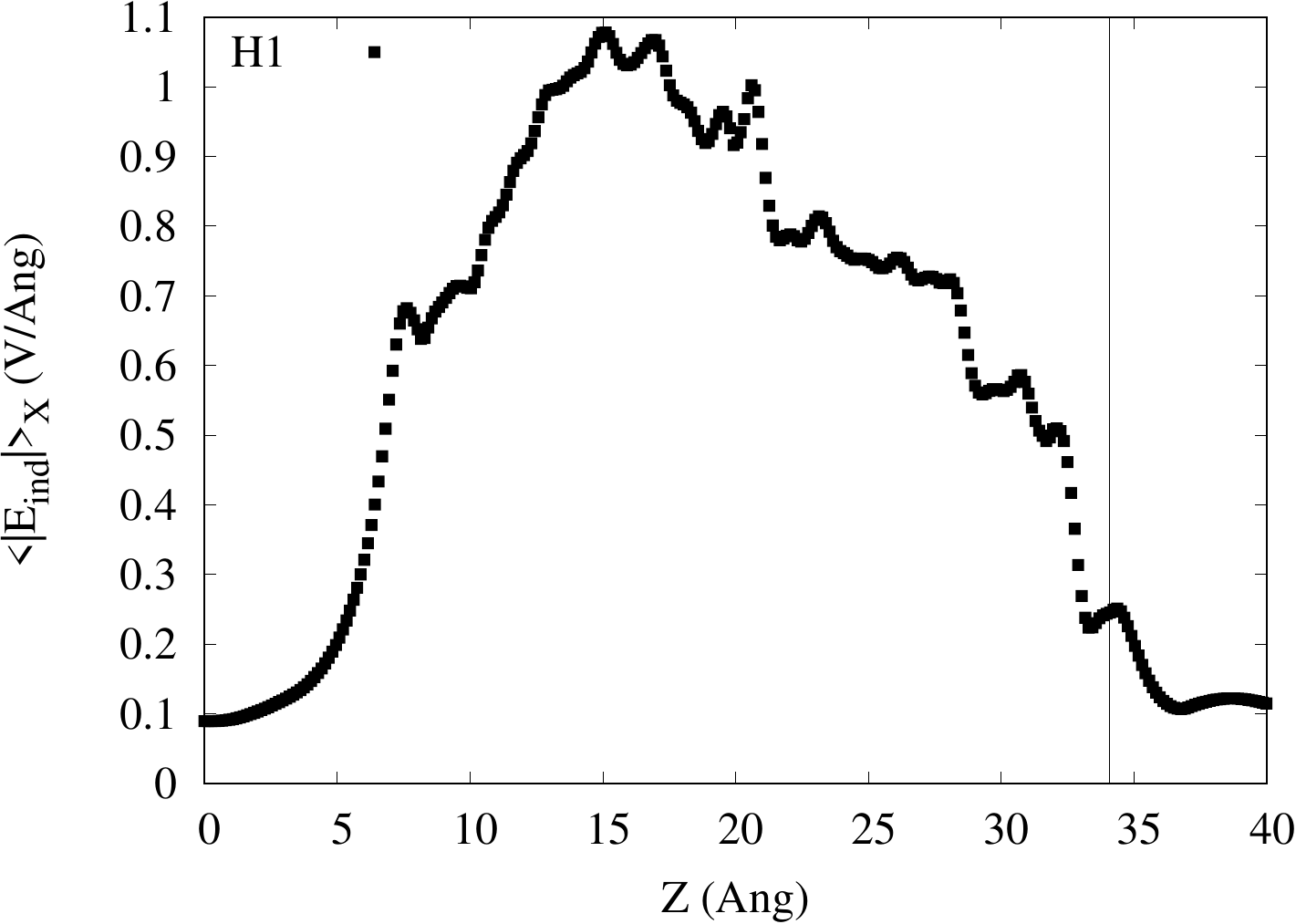} 
        \includegraphics[width=0.45\linewidth]{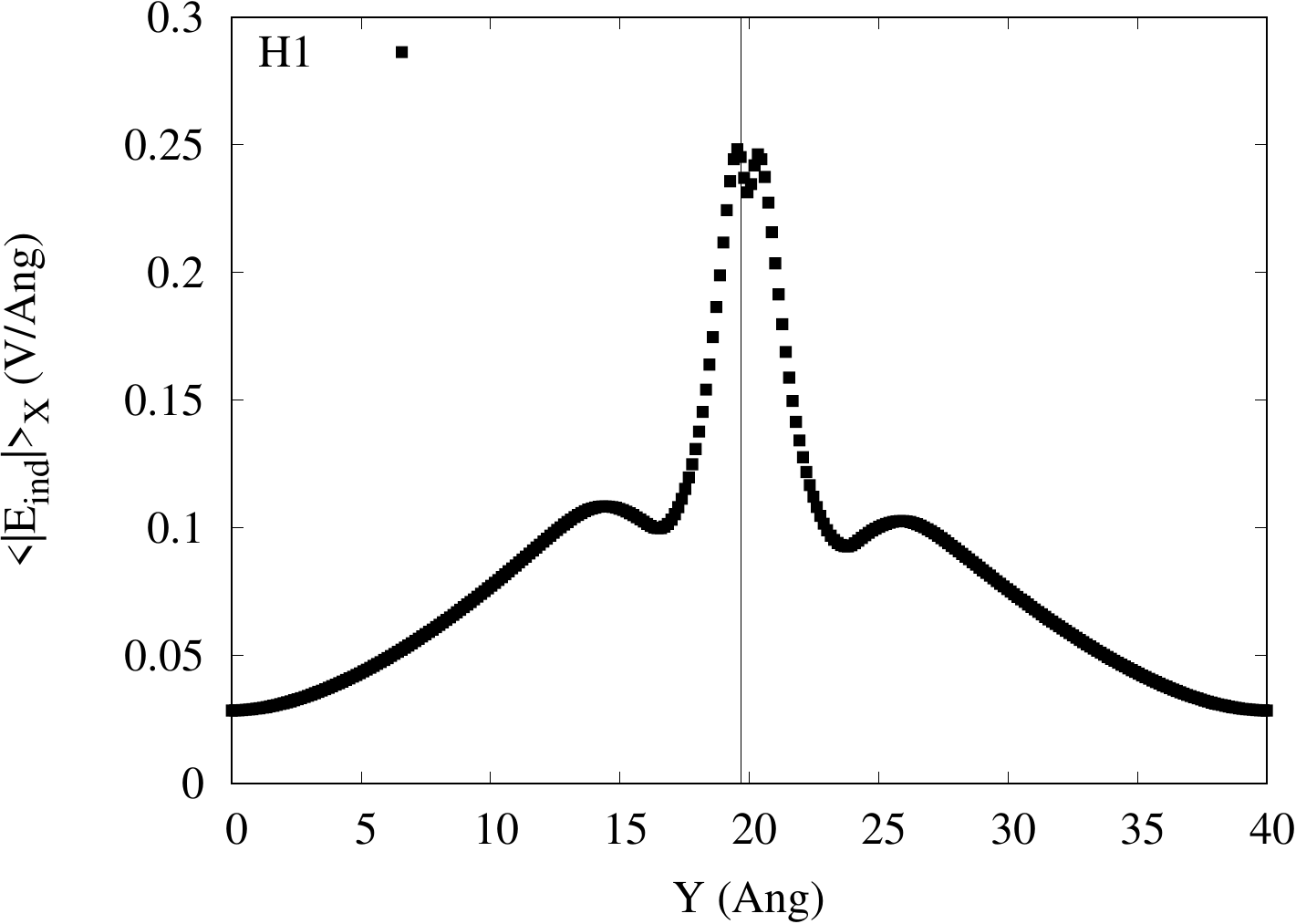}
        \includegraphics[width=0.45\linewidth]{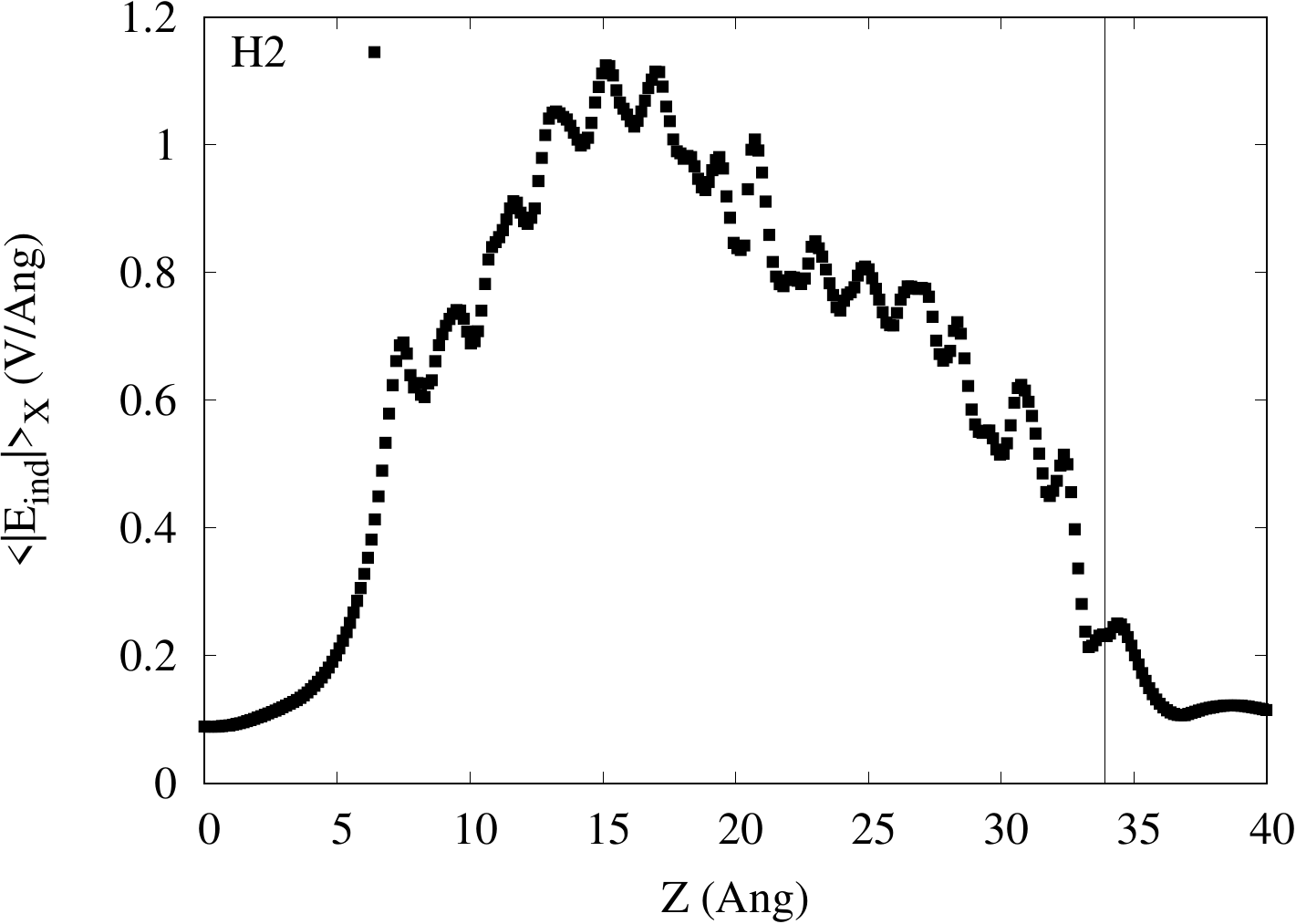} 
        \includegraphics[width=0.45\linewidth]{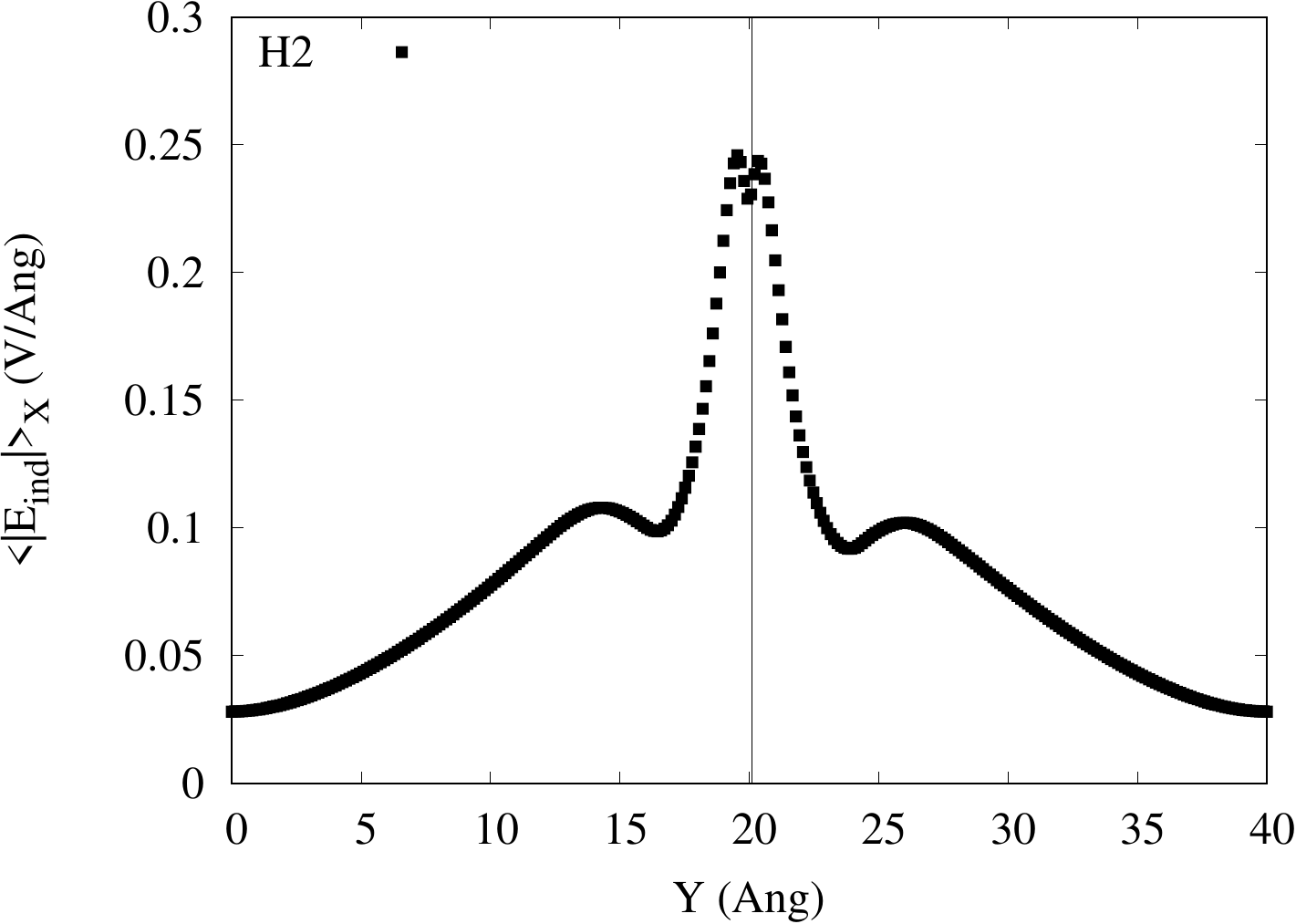}
		\caption{Same as Fig.~\ref{fig:field_2res} but for pulse-2 with frequency 8 eV at the time instant $t = 34$ fs. The molecule Cartesian coordinates are H1: (20.13996, 19.67609, 34.05631) and H2: (19.45773, 20.08741, 33.91462).}
		\label{fig:field_2nonres}
	\end{figure}

    \begin{figure}[h!]
 \centering
		\includegraphics[width=0.4\linewidth]{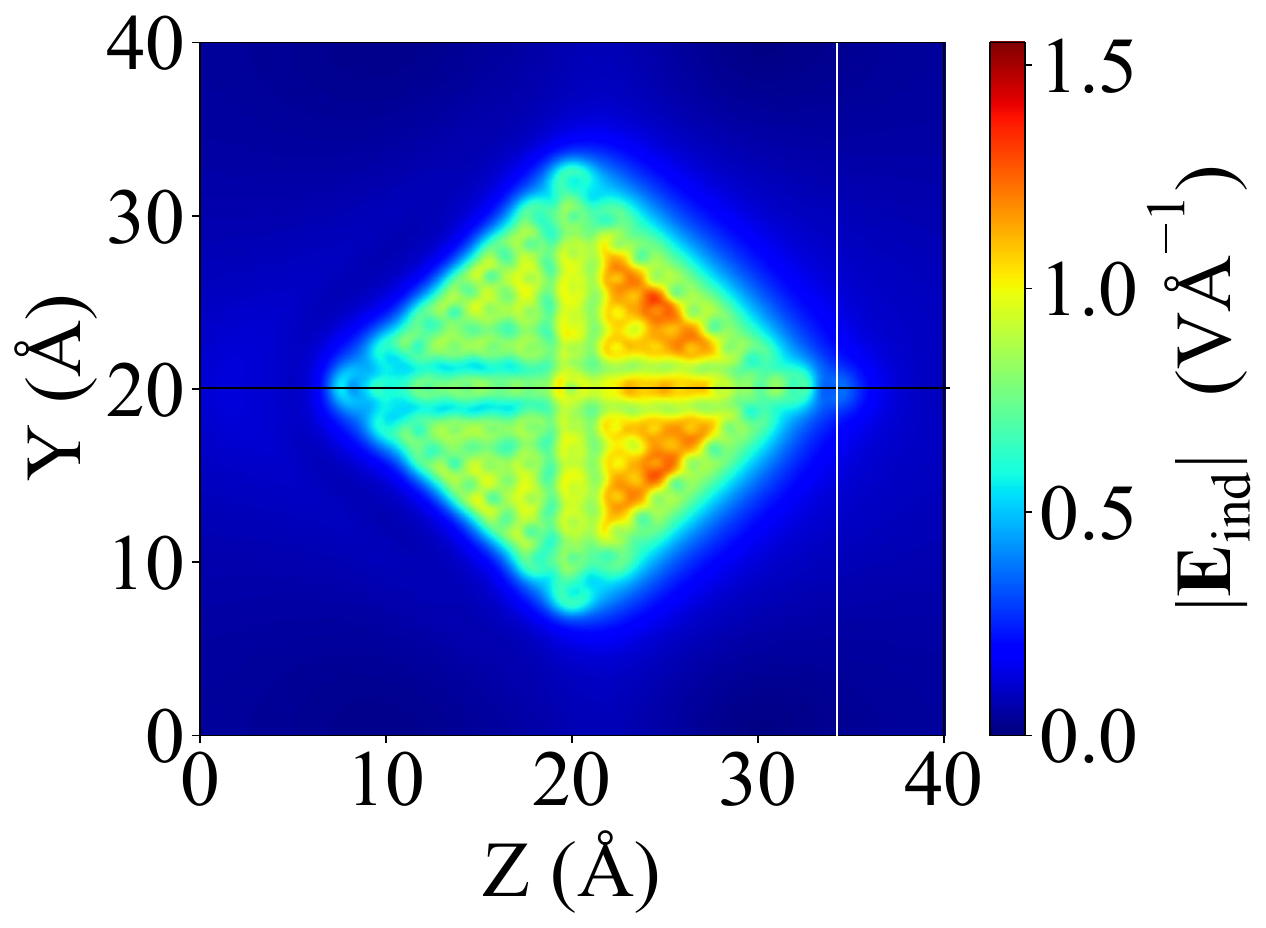}\\
        \includegraphics[width=0.45\linewidth]{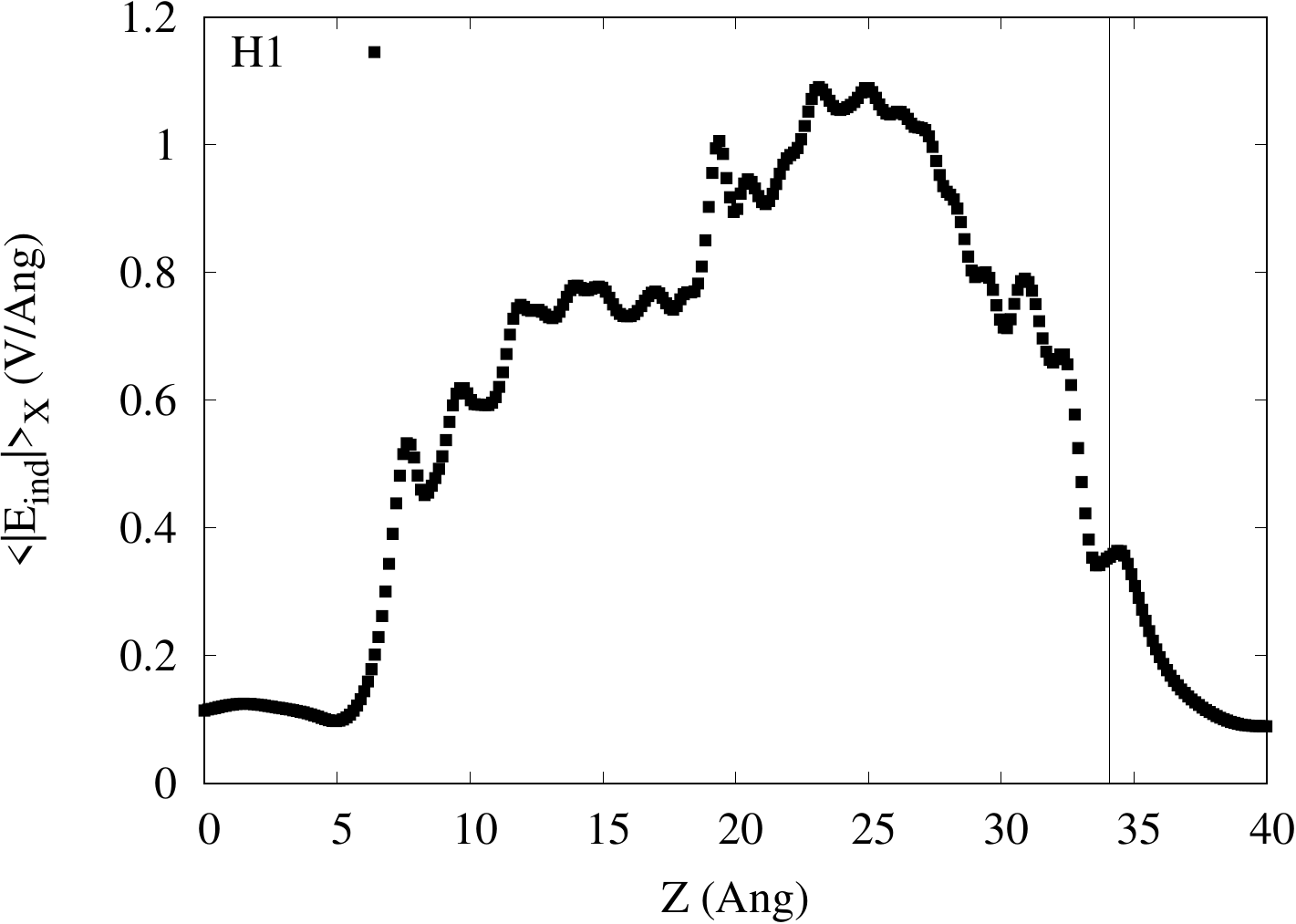} 
        \includegraphics[width=0.45\linewidth]{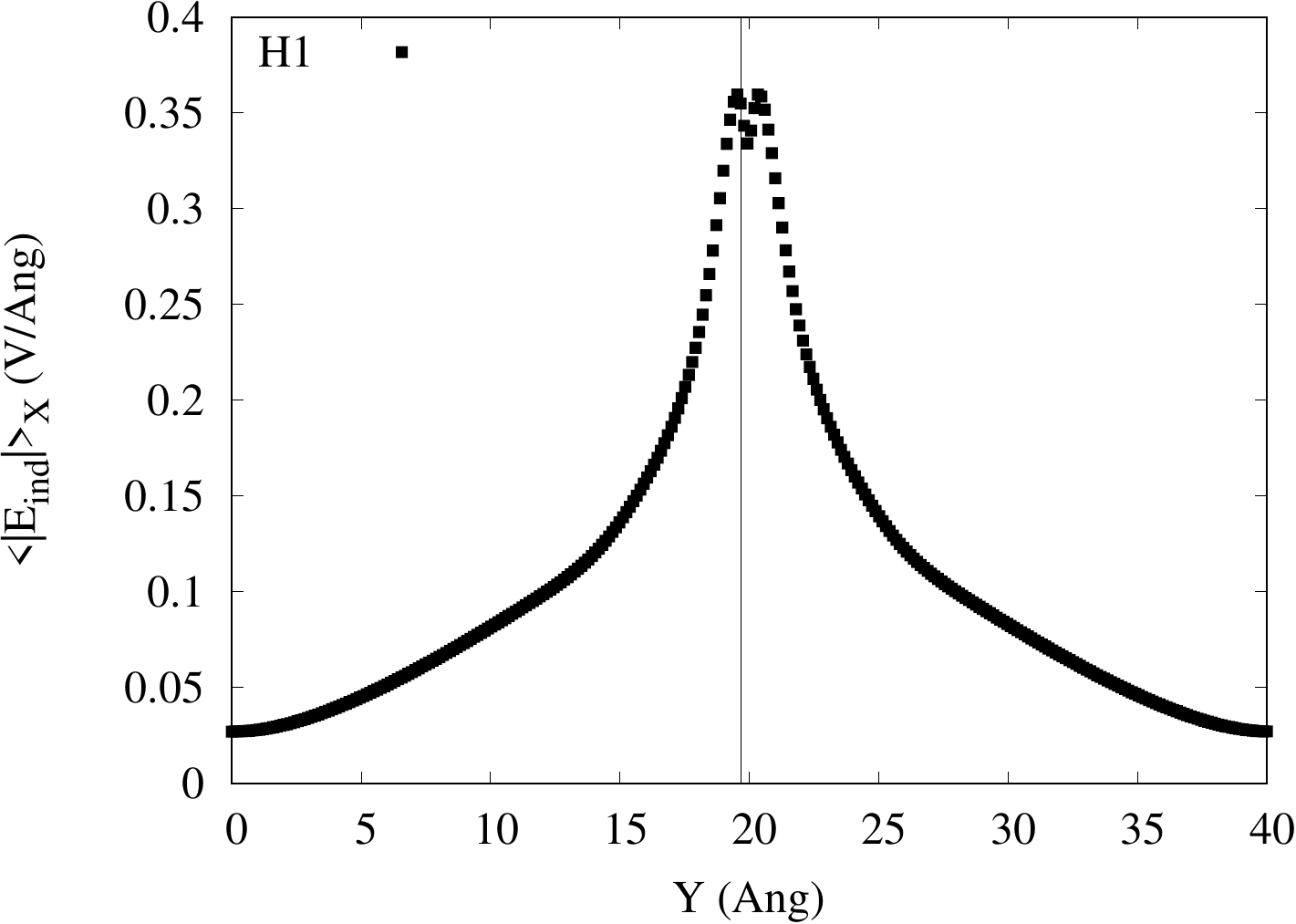}
        \includegraphics[width=0.45\linewidth]{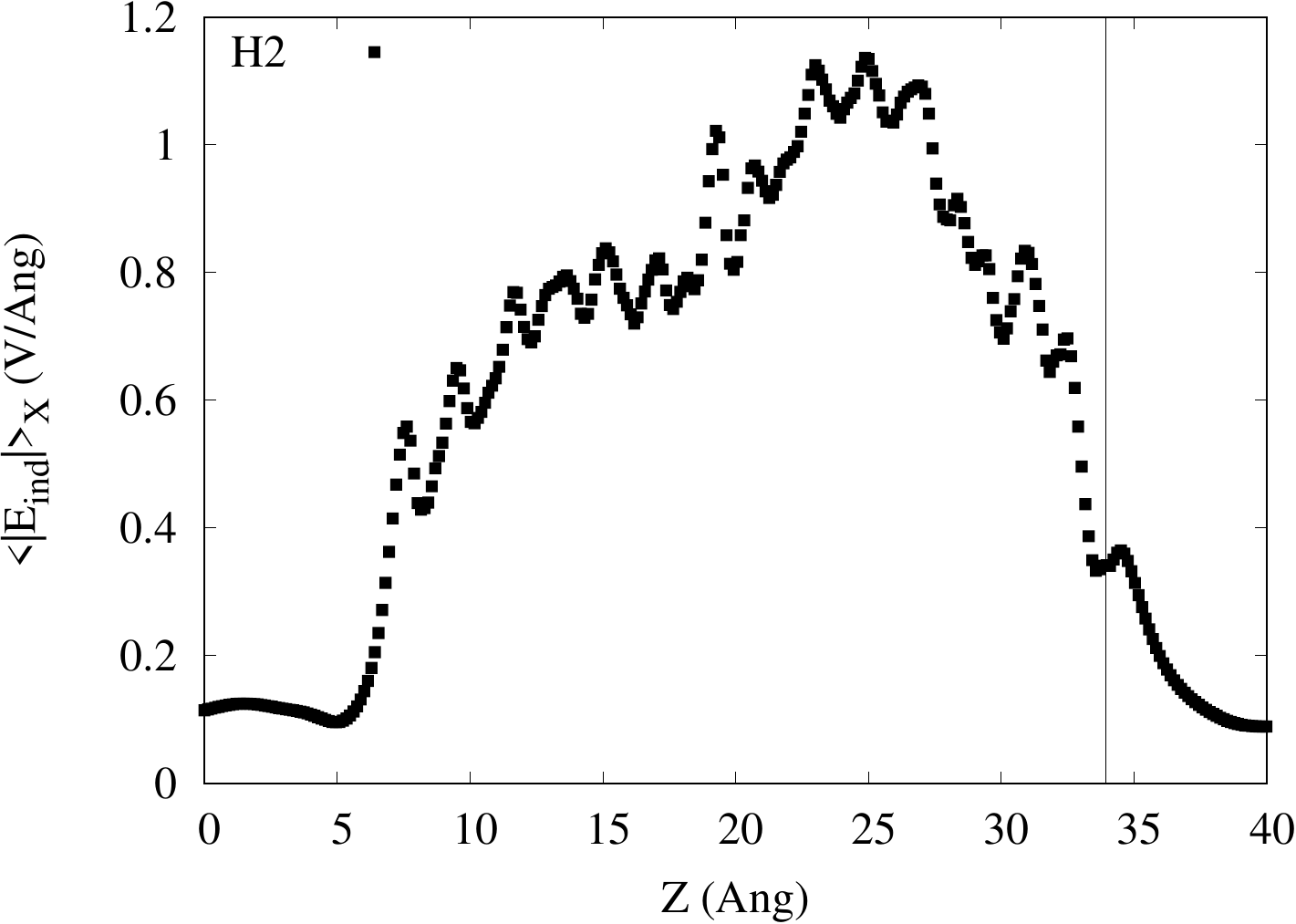} 
        \includegraphics[width=0.45\linewidth]{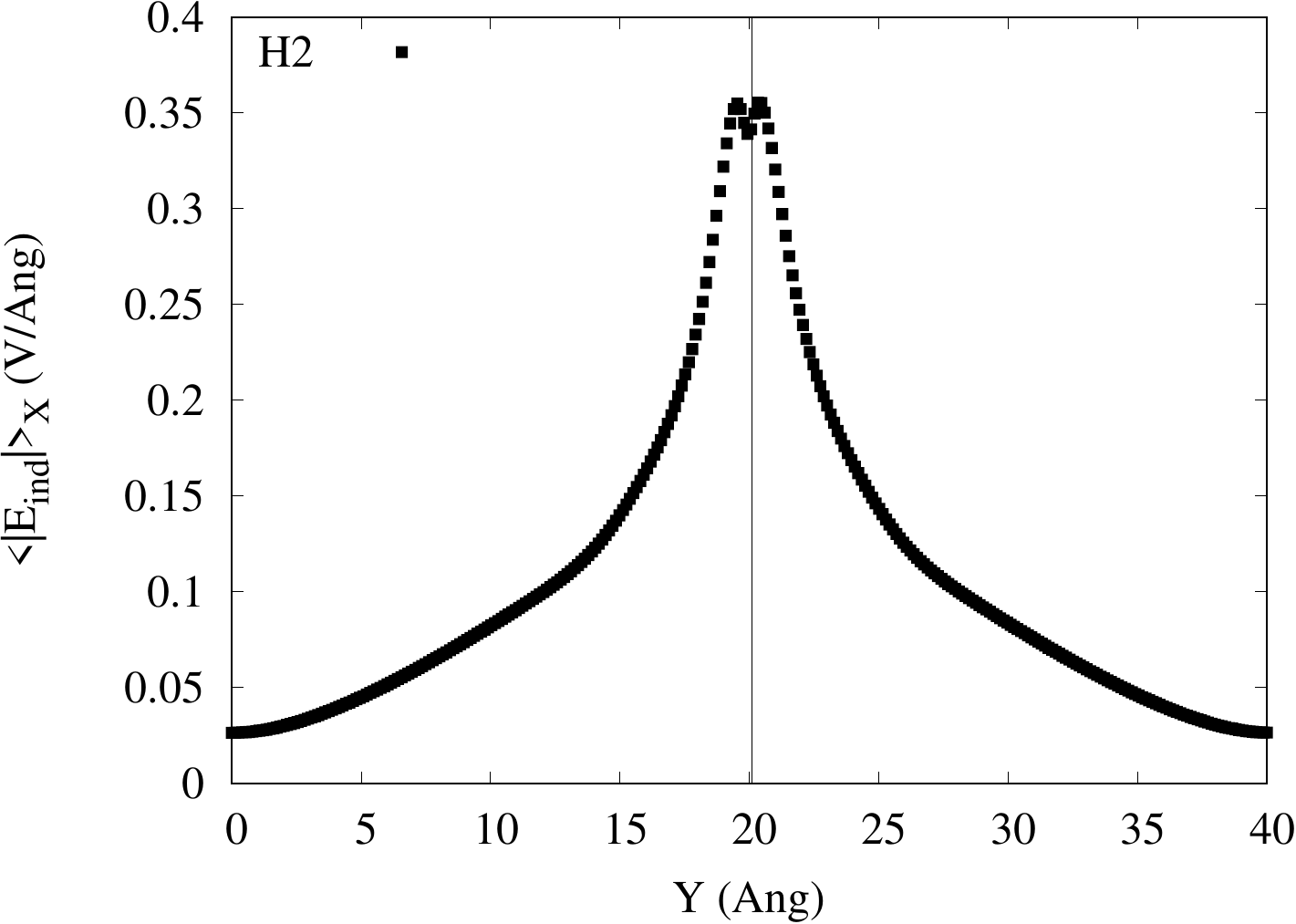}
		\caption{Same as Fig.~\ref{fig:field_2res} but for pulse-2 with frequency 8 eV at the time instant $t = 34.25$ fs. The molecule Cartesian coordinates are H1: (20.13695, 19.67874, 34.05676) and H2: (19.45385, 20.08326, 33.91865).}
		\label{fig:field_2nonres_1}
	\end{figure}

    \begin{figure}[h!]
 \centering
		\includegraphics[width=0.4\linewidth]{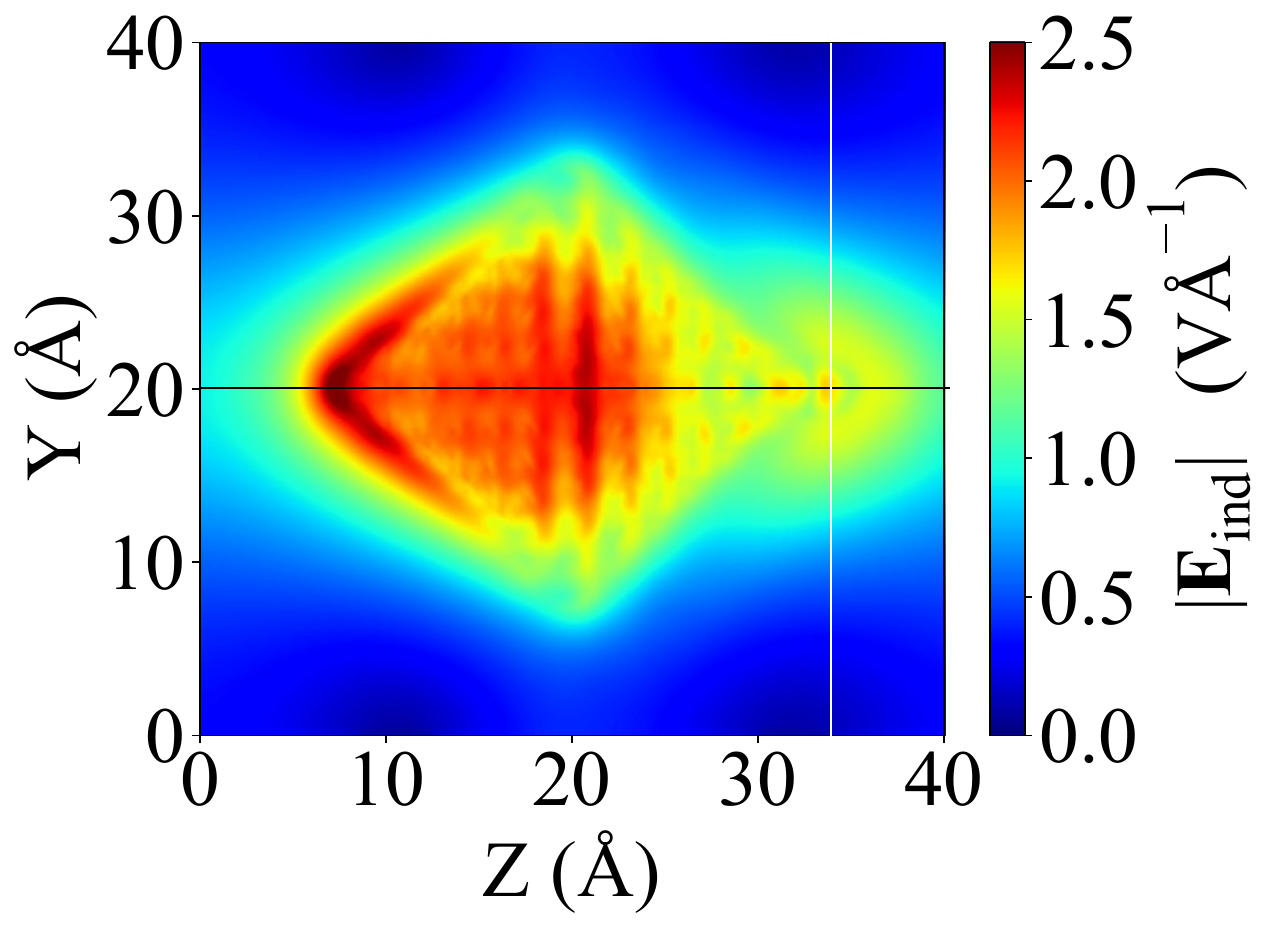}\\
        \includegraphics[width=0.45\linewidth]{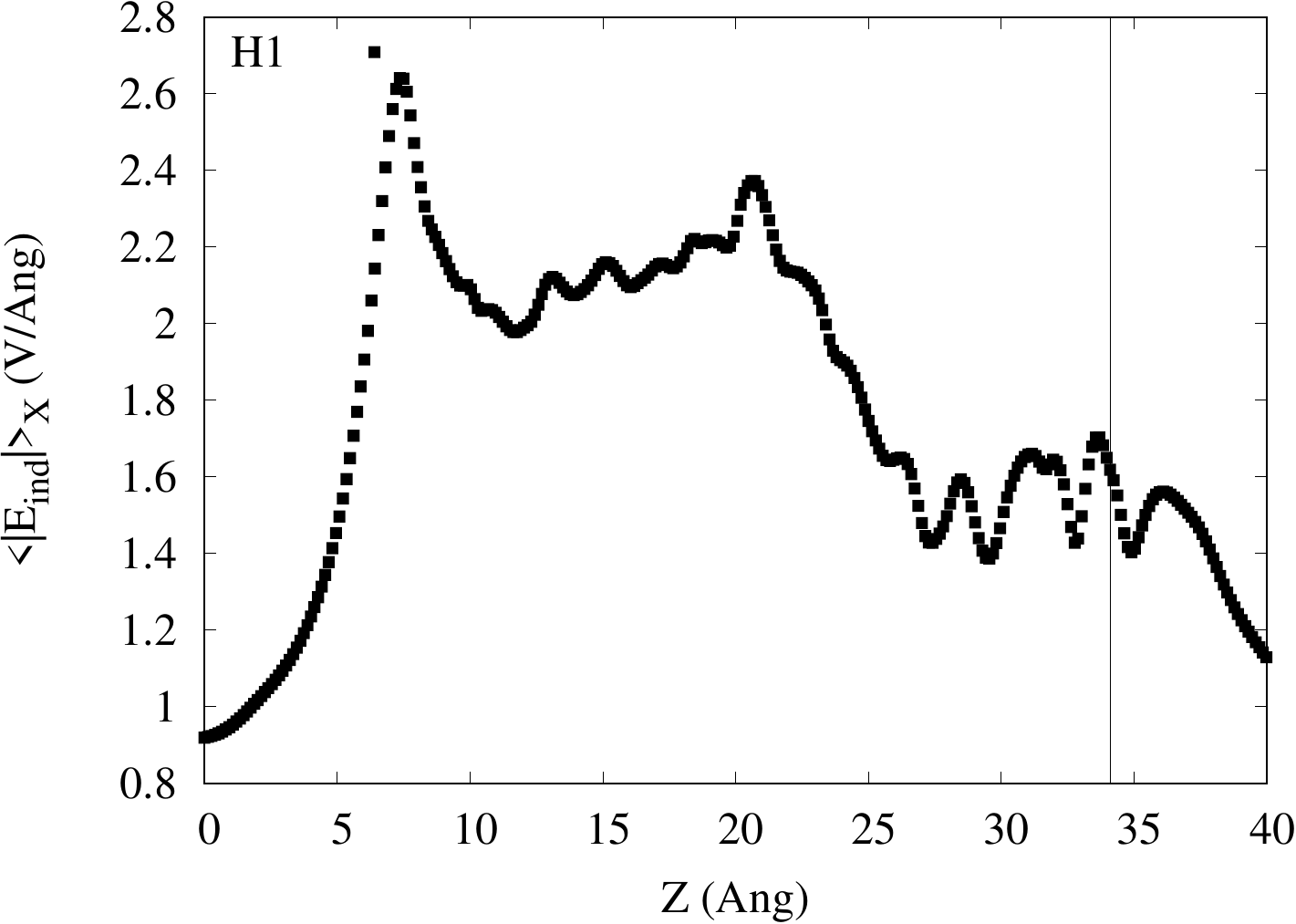} 
        \includegraphics[width=0.45\linewidth]{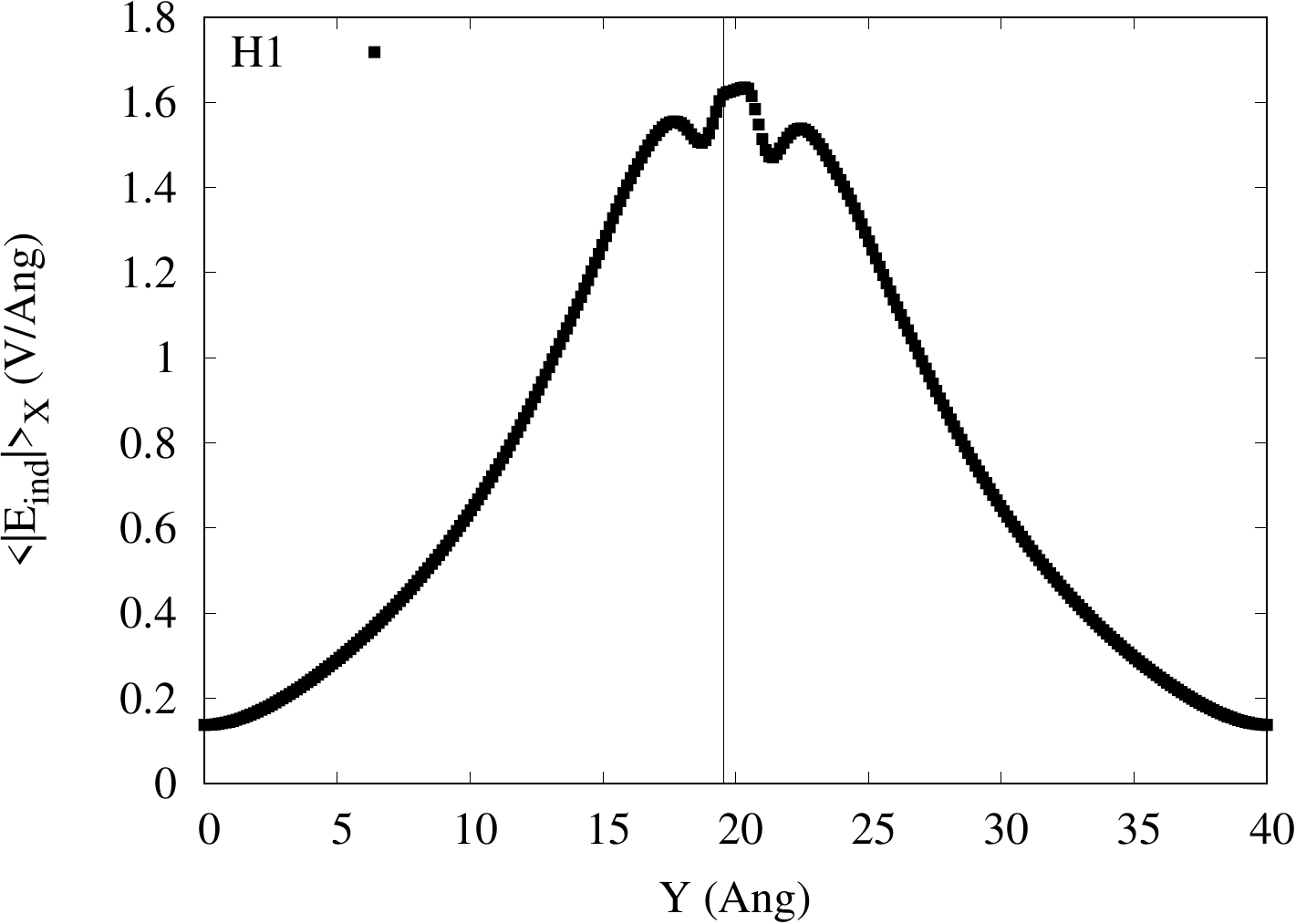}
        \includegraphics[width=0.45\linewidth]{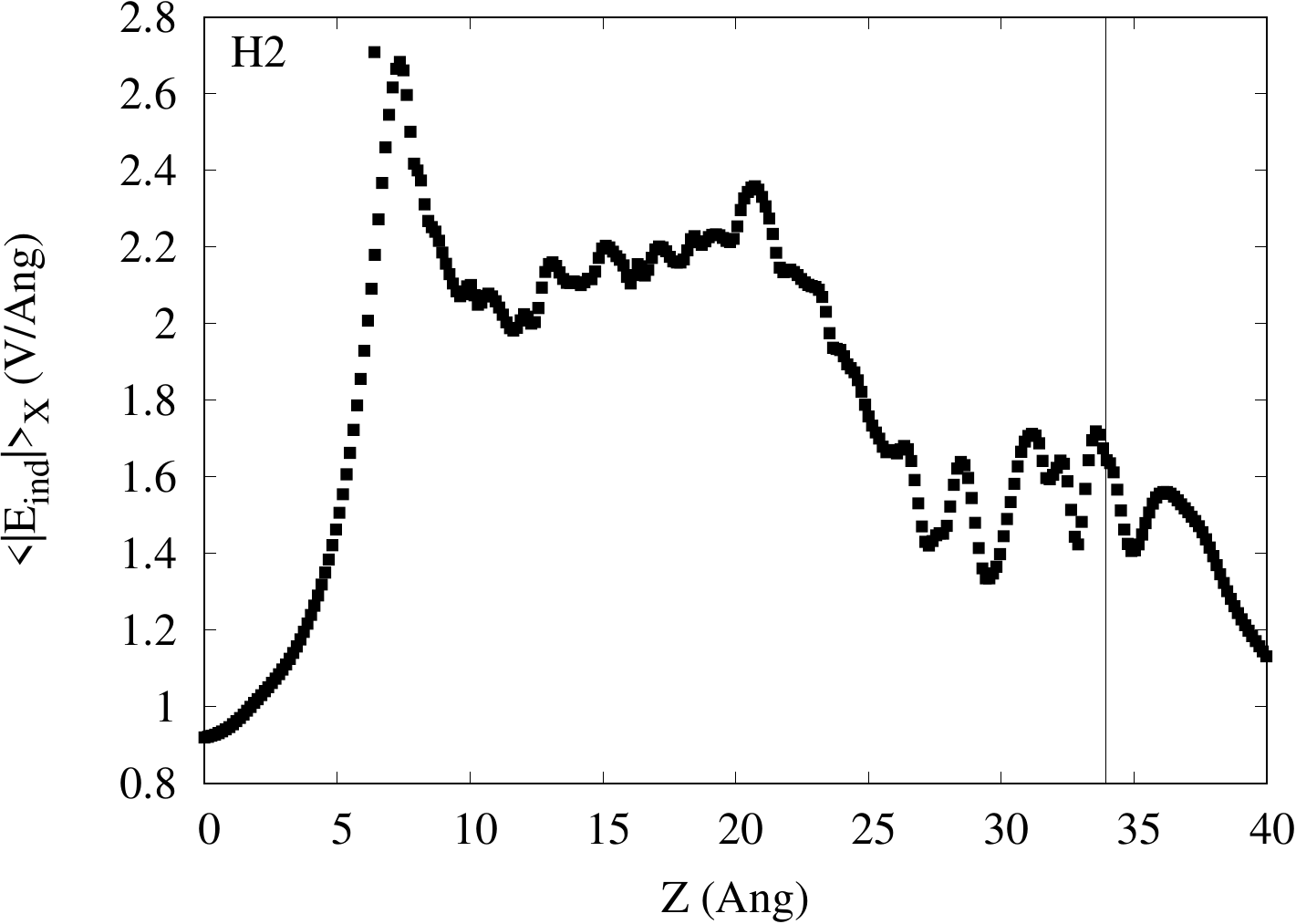} 
        \includegraphics[width=0.45\linewidth]{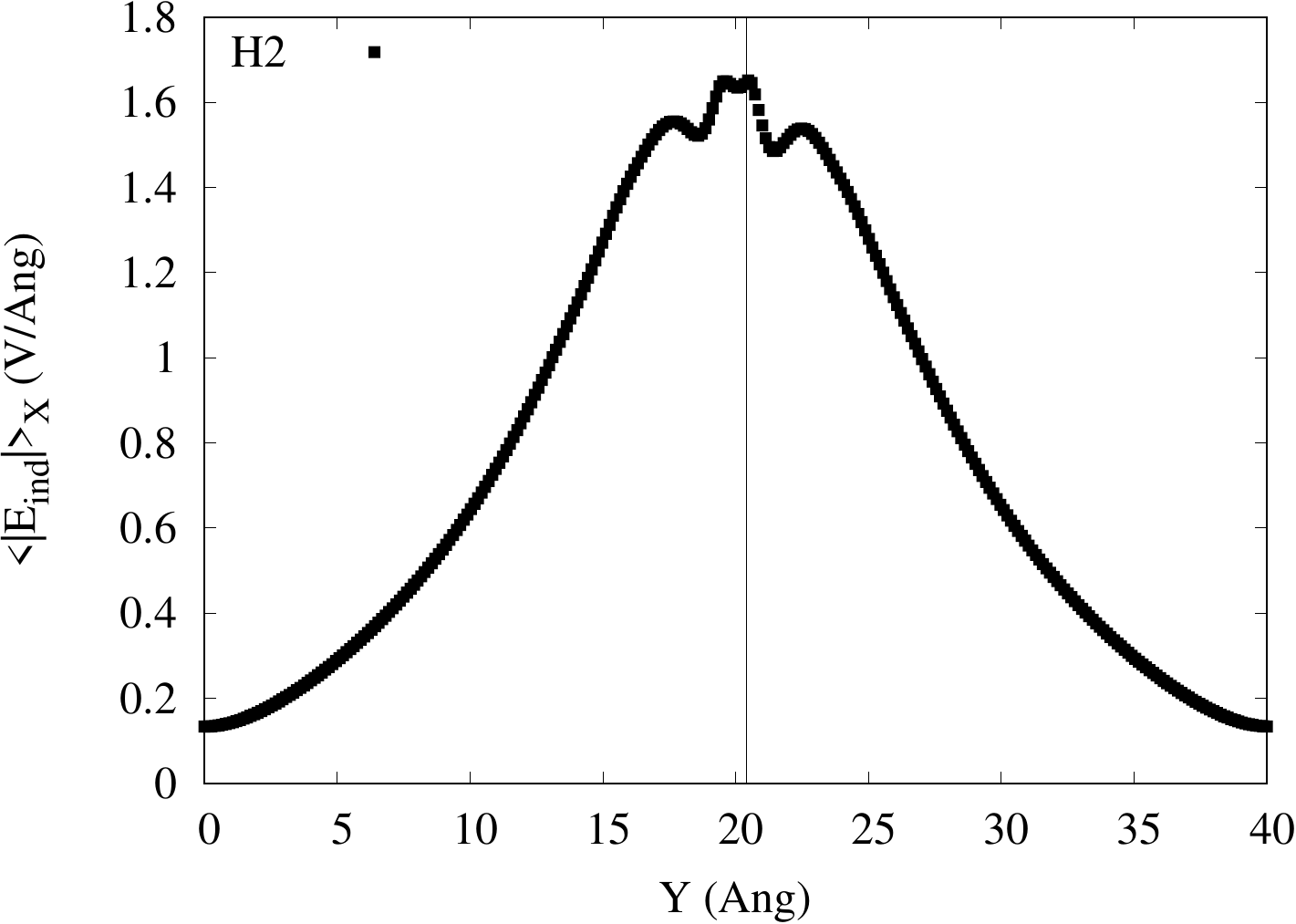}
		\caption{Same as Fig.~\ref{fig:field_2res} but for pulse-4 with frequency 2.48 eV at the time instant $t = 9.24$ fs. The molecule Cartesian coordinates are H1: (20.17667, 19.53189, 34.10876) and H2: (20.04871, 20.39224, 33.92203).}
		\label{fig:field_4res}
	\end{figure}

    \begin{figure}[h!]
 \centering
		\includegraphics[width=0.4\linewidth]{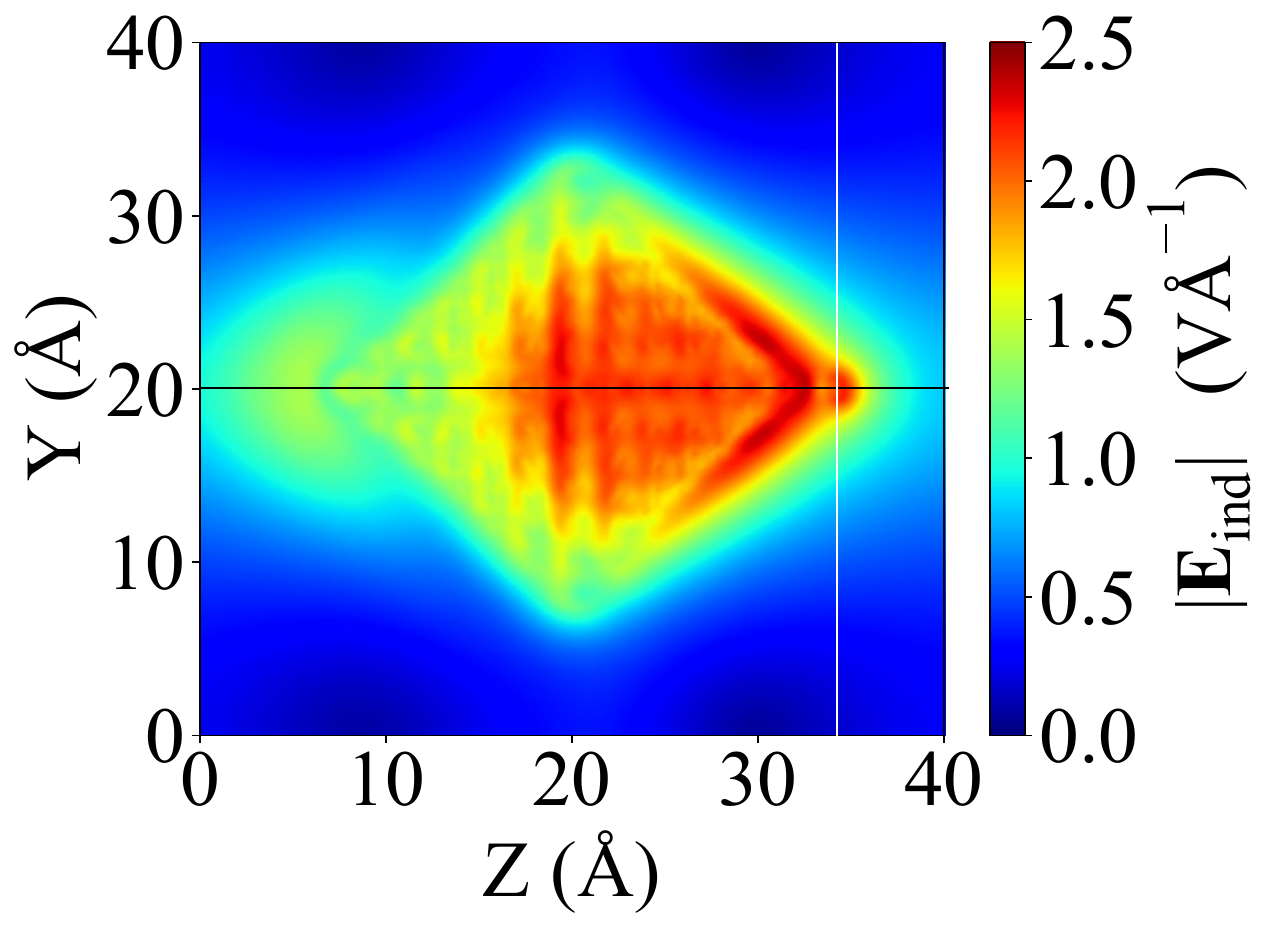}\\
        \includegraphics[width=0.45\linewidth]{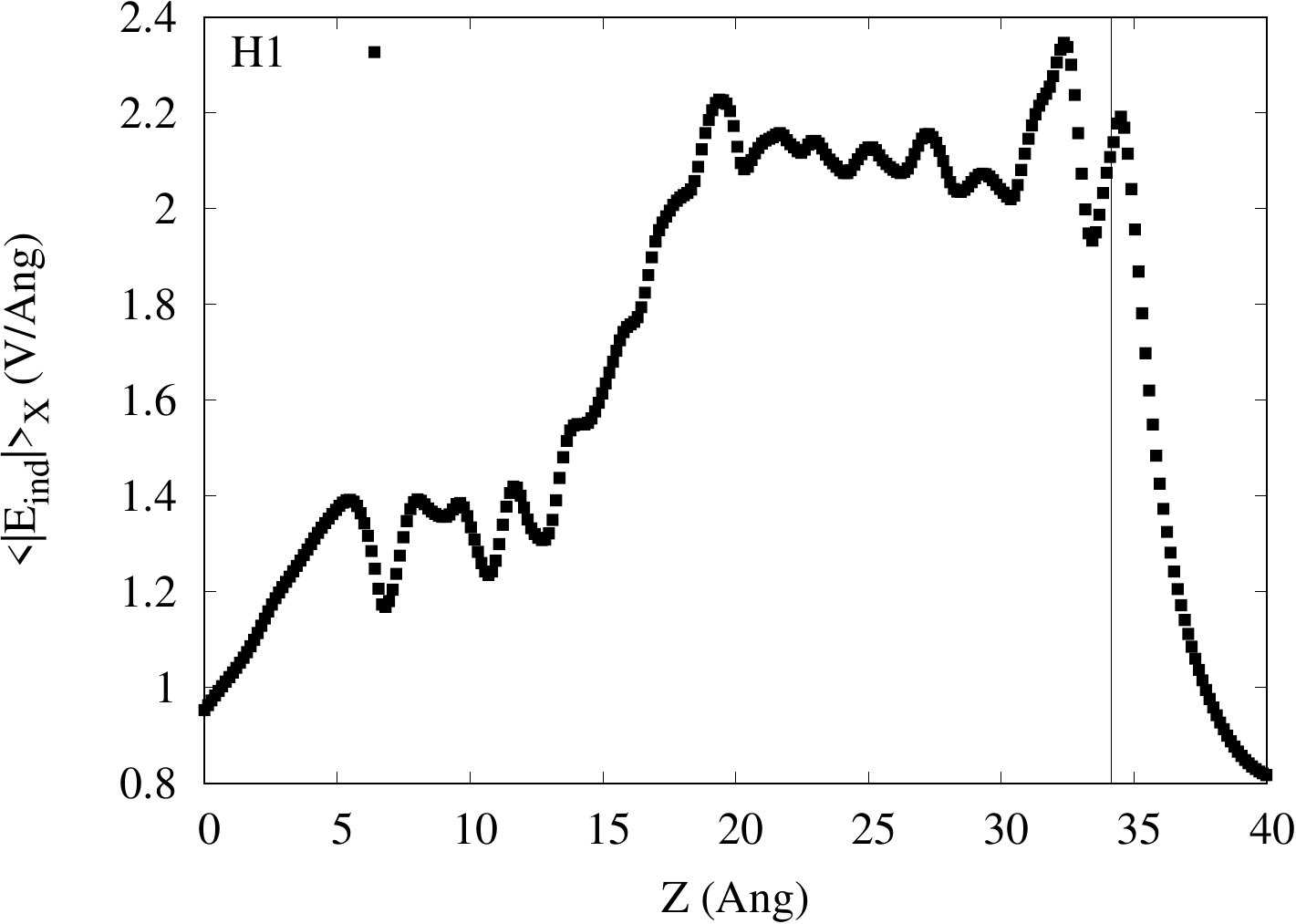} 
        \includegraphics[width=0.45\linewidth]{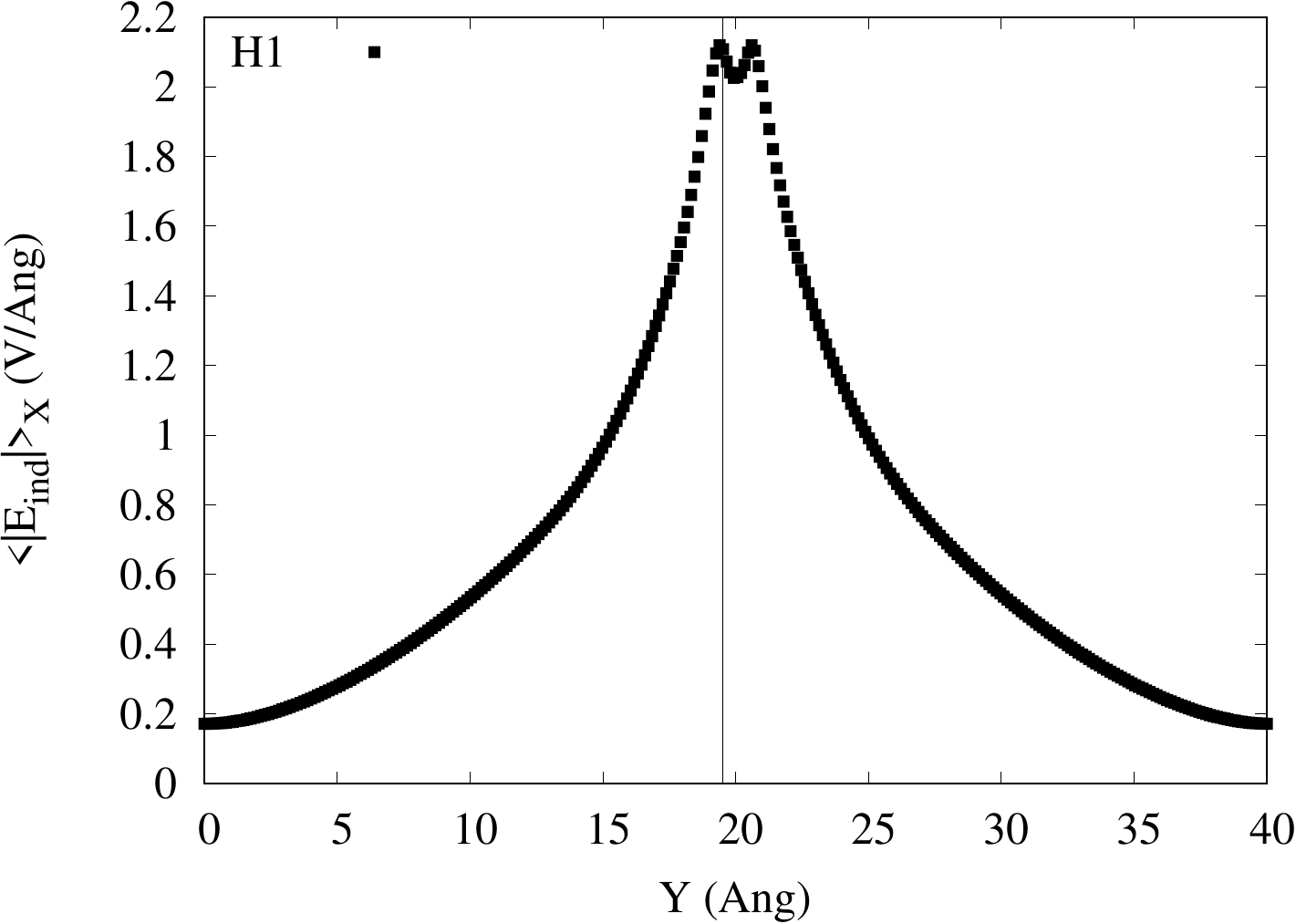}
        \includegraphics[width=0.45\linewidth]{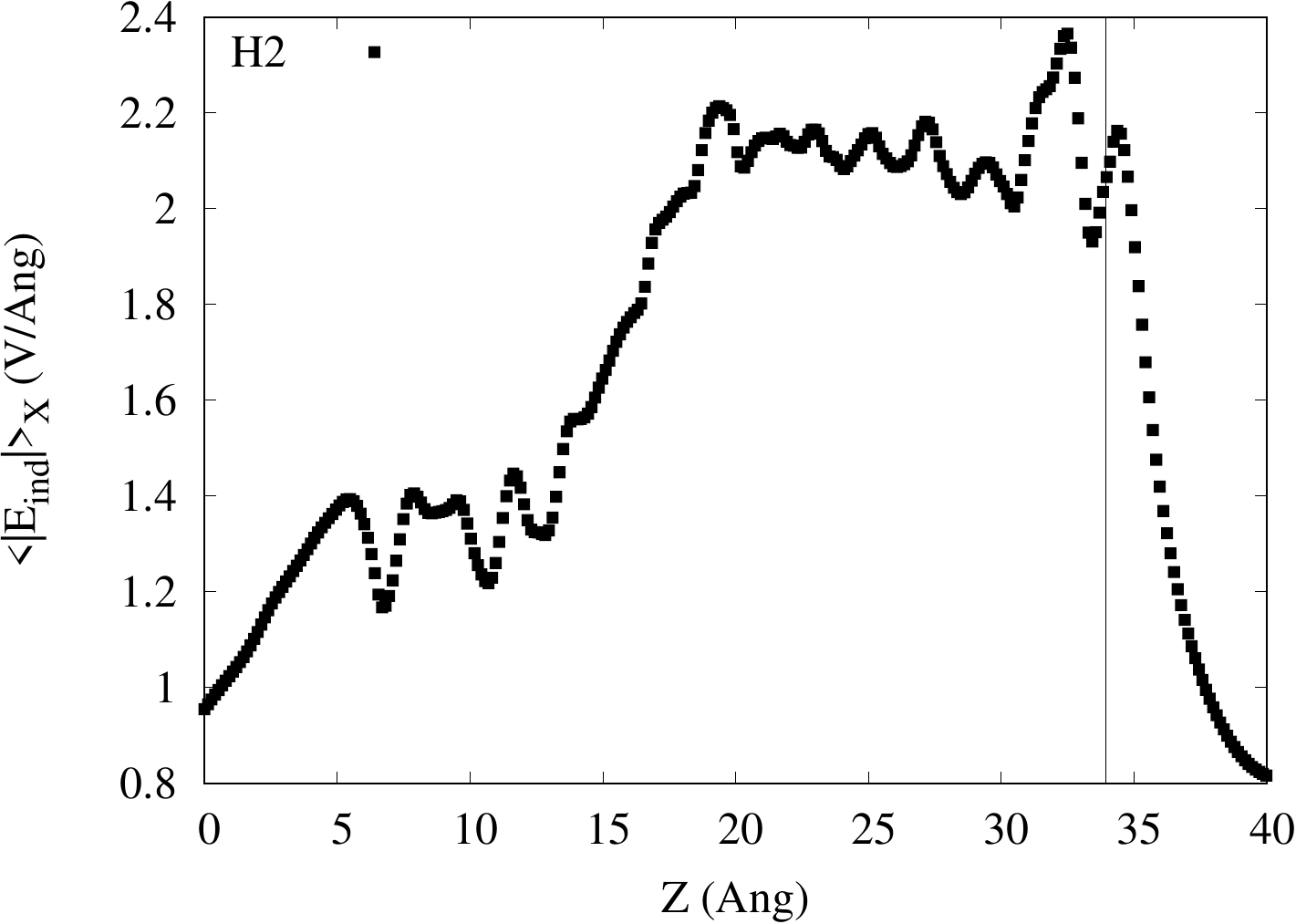} 
        \includegraphics[width=0.45\linewidth]{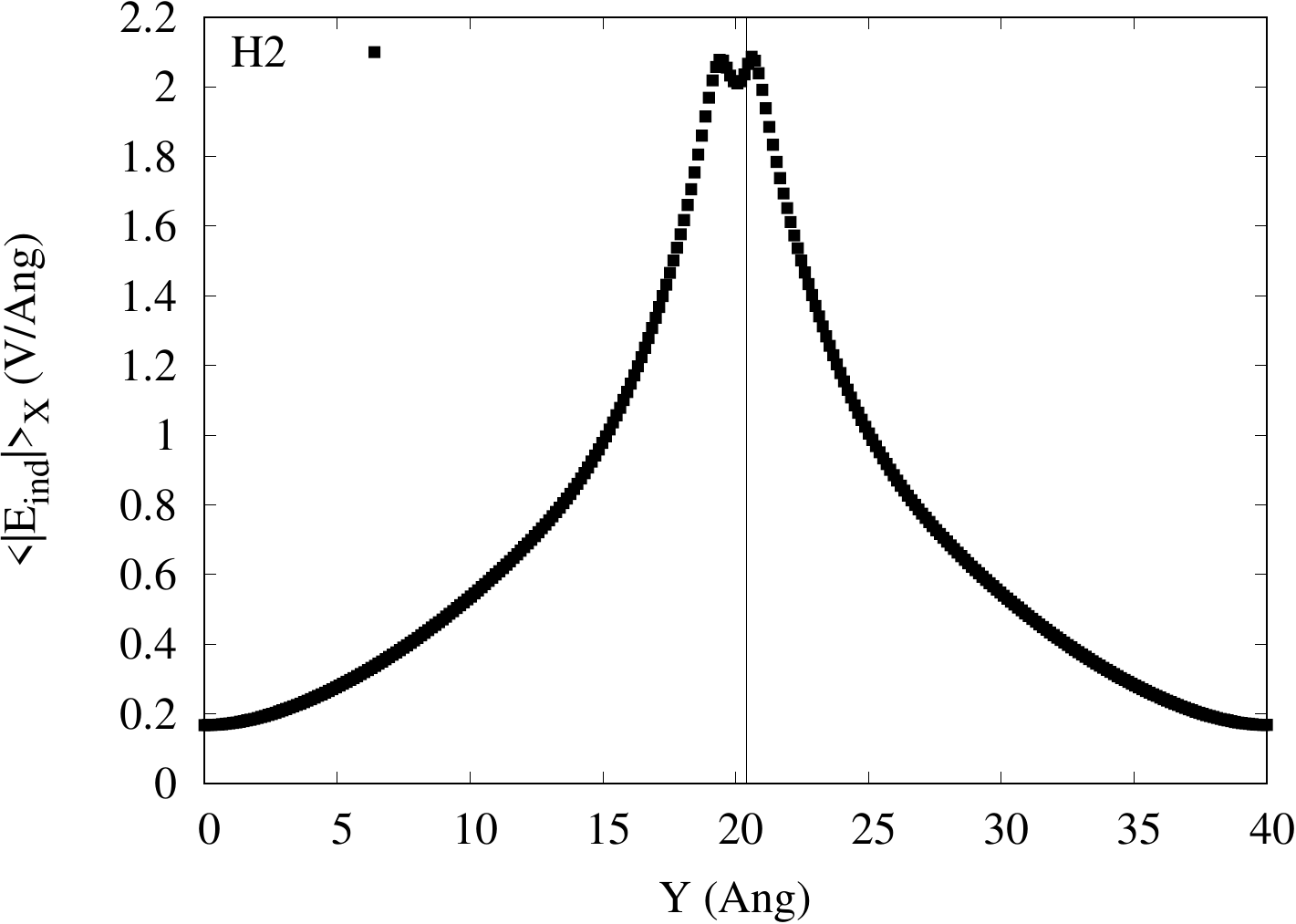}
		\caption{Same as Fig.~\ref{fig:field_2res} but for pulse-4 with frequency 2.48 eV at the time instant $t = 10$ fs. The molecule Cartesian coordinates are H1: (20.18015, 19.49985, 34.13747) and H2: (20.02655, 20.42064, 33.92358).}
		\label{fig:field_4res_1}
	\end{figure}

    \begin{figure}[h!]
 \centering
		\includegraphics[width=0.4\linewidth]{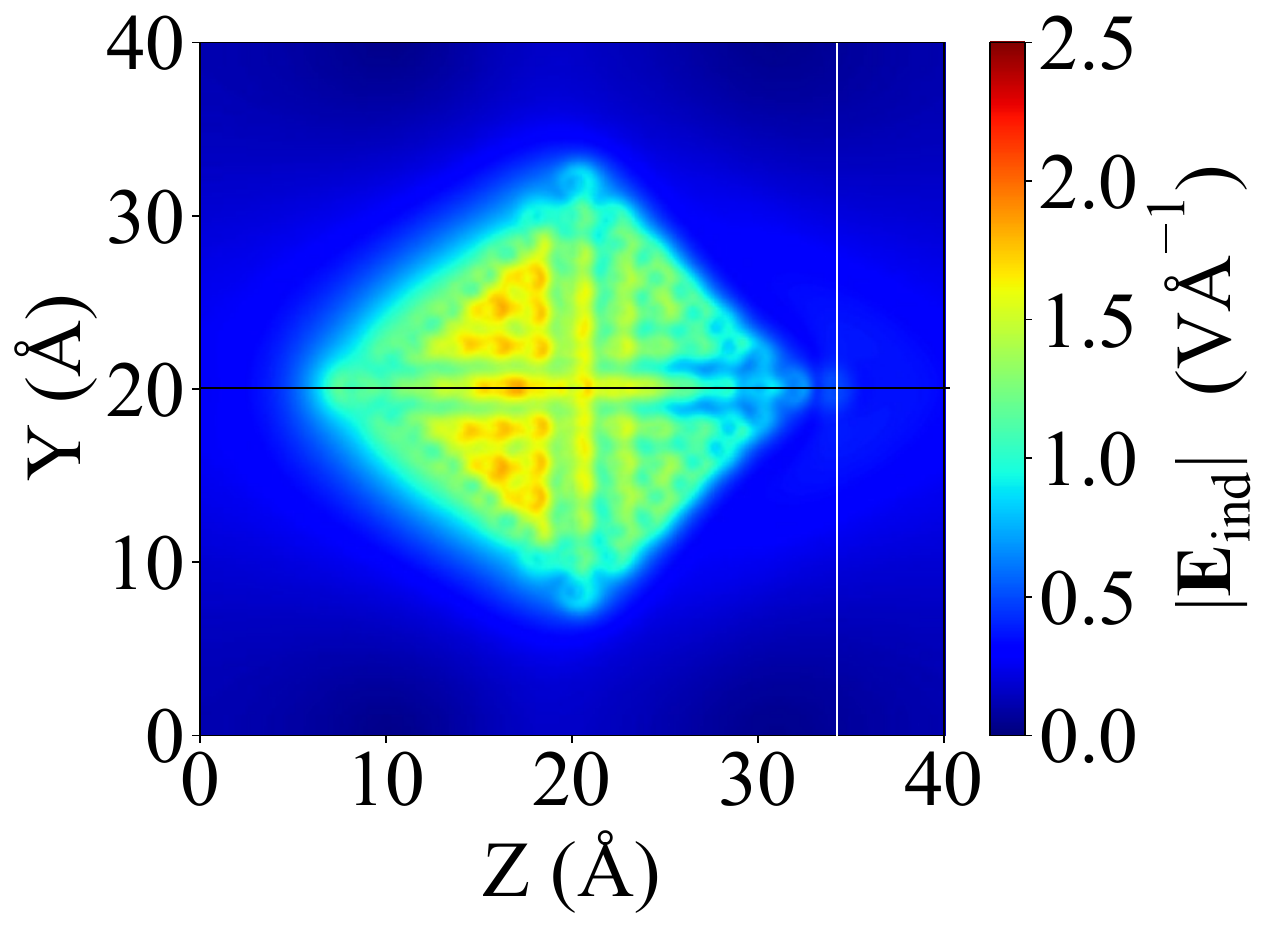}\\
        \includegraphics[width=0.45\linewidth]{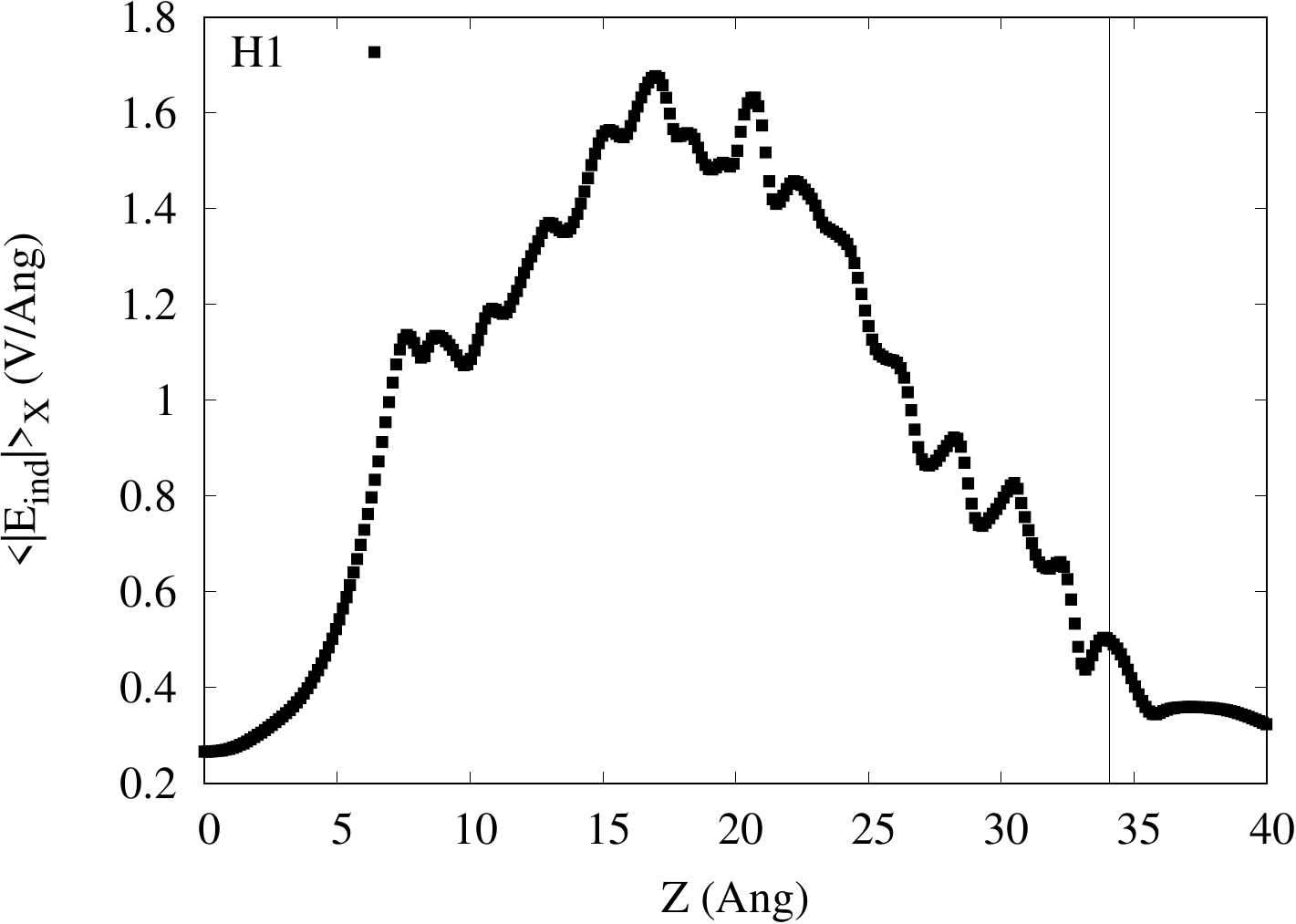} 
        \includegraphics[width=0.45\linewidth]{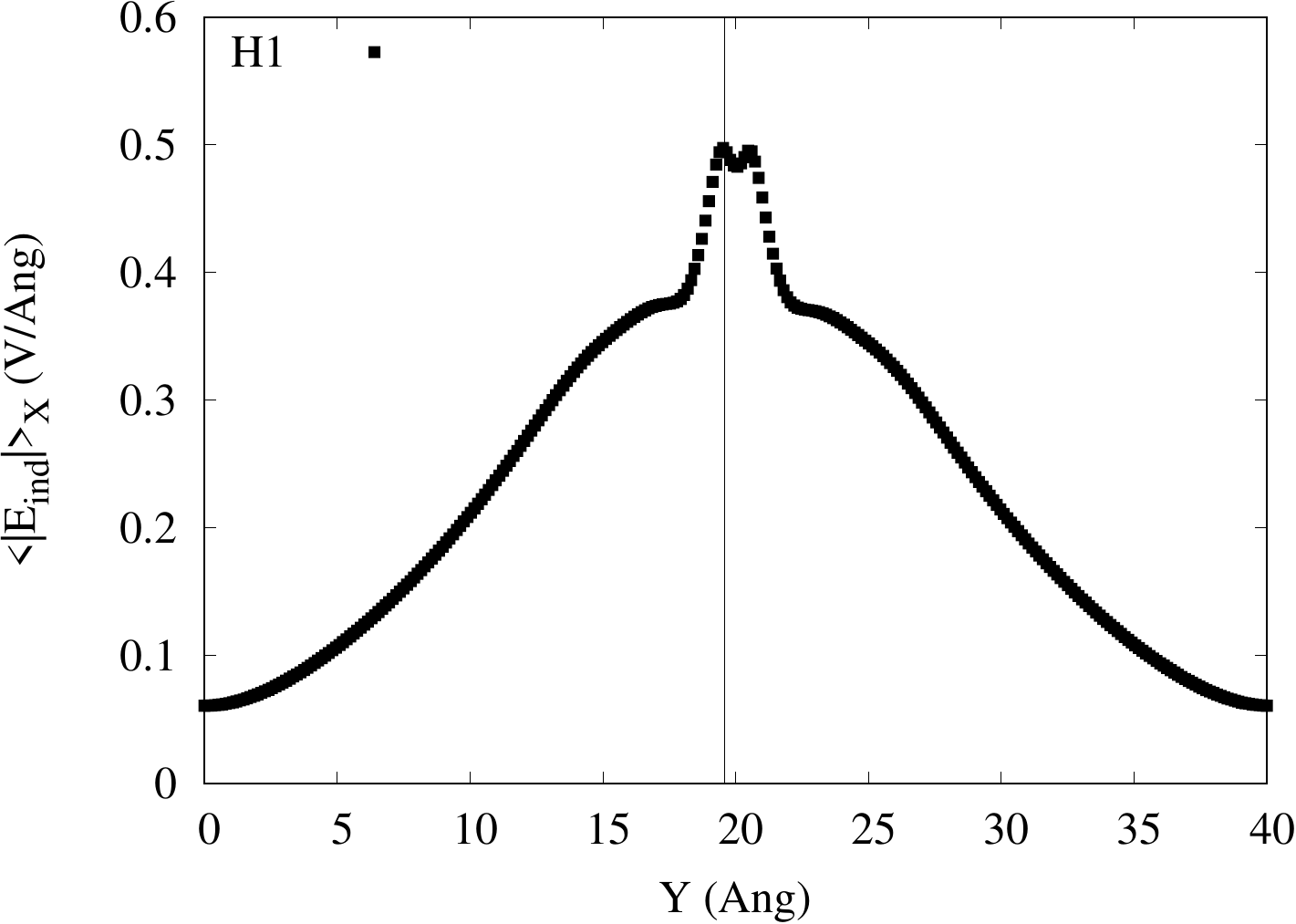}
        \includegraphics[width=0.45\linewidth]{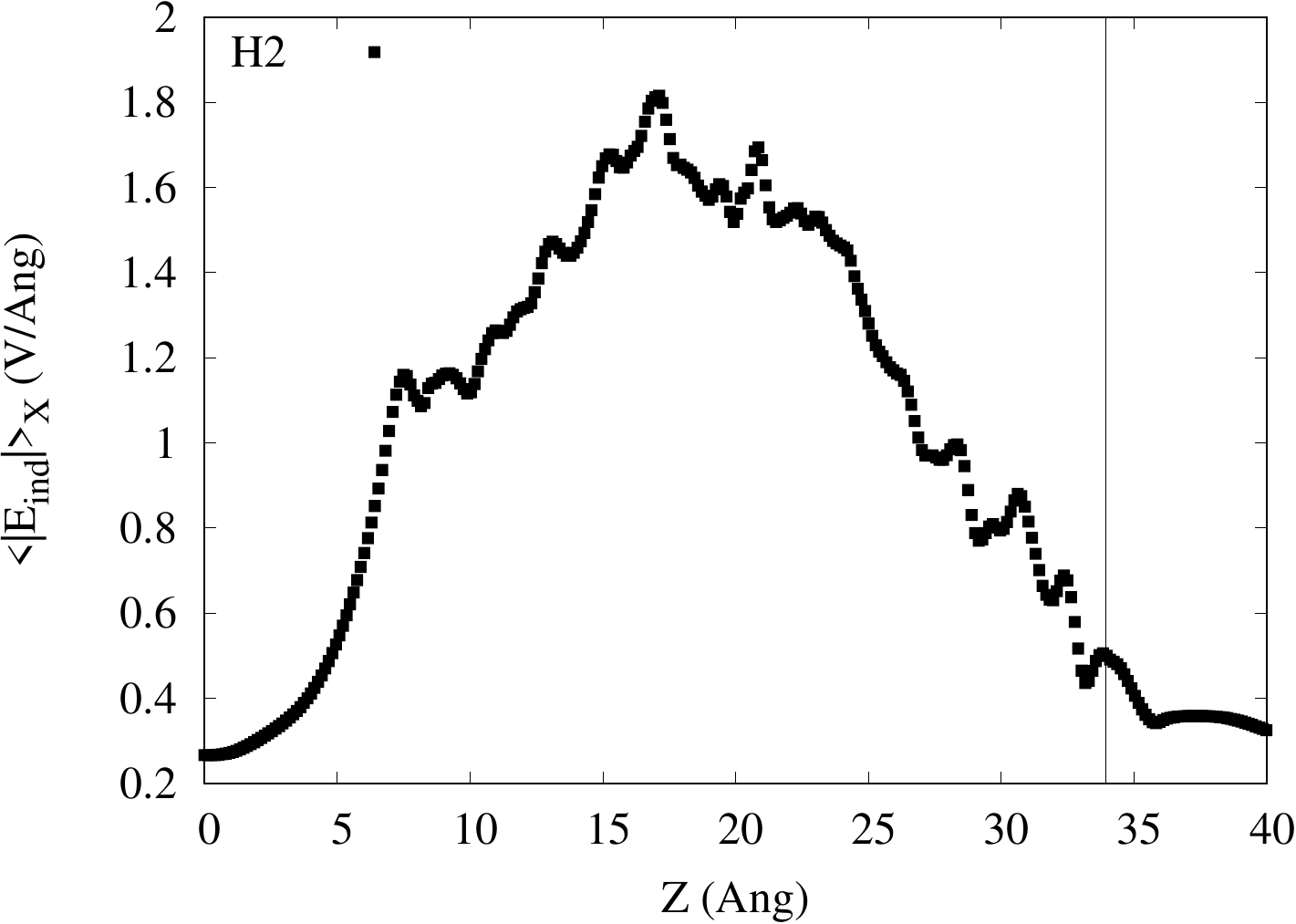} 
        \includegraphics[width=0.45\linewidth]{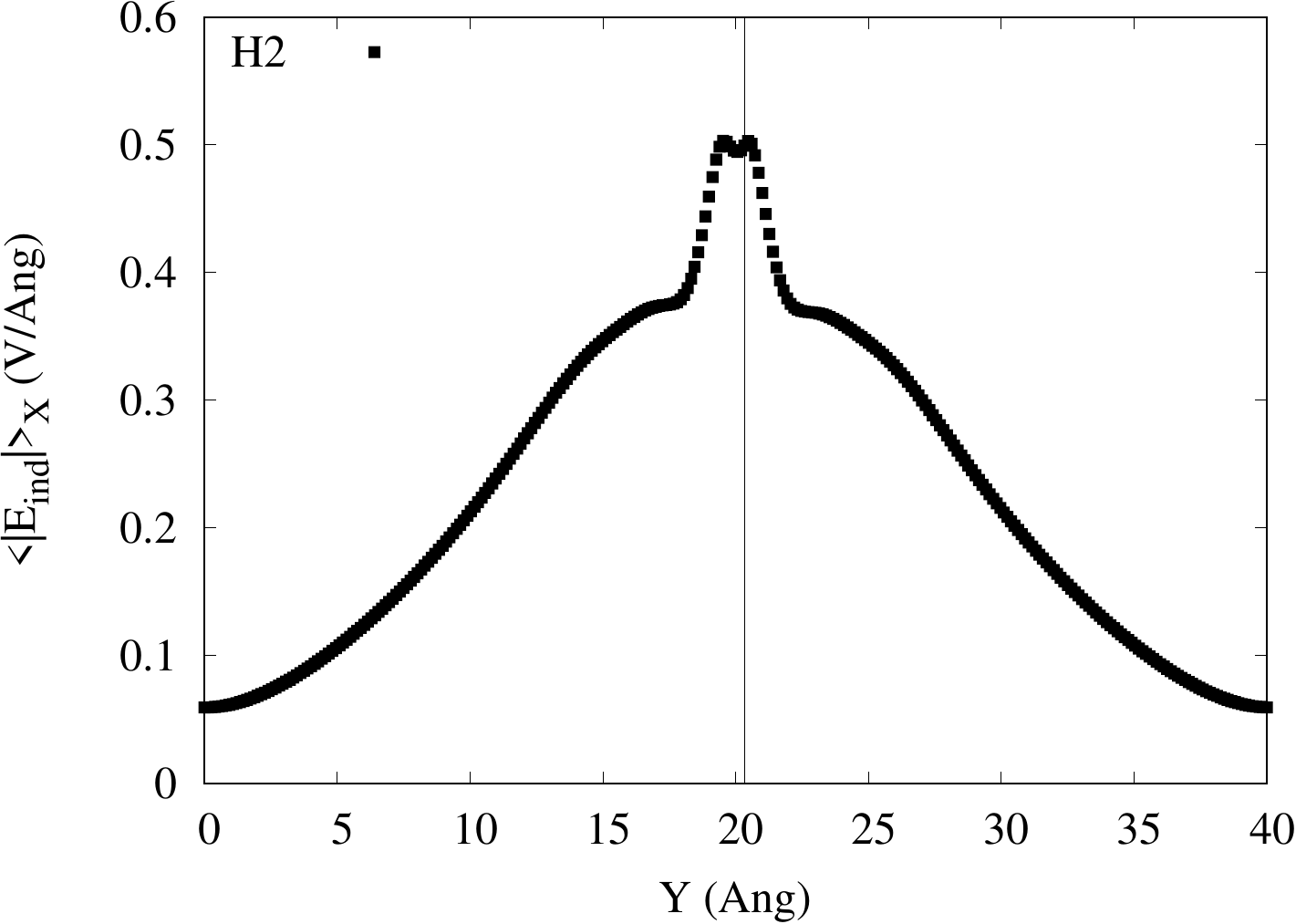}
		\caption{Same as Fig.~\ref{fig:field_2res} but for pulse-4 with frequency 8 eV at the time instant $t = 9.24$ fs. The molecule Cartesian coordinates are H1: (20.17310, 19.56901, 34.08232) and H2: (20.05217, 20.35282, 33.92159).}
		\label{fig:field_4nonres}
	\end{figure}

    \begin{figure}[h!]
 \centering
		\includegraphics[width=0.4\linewidth]{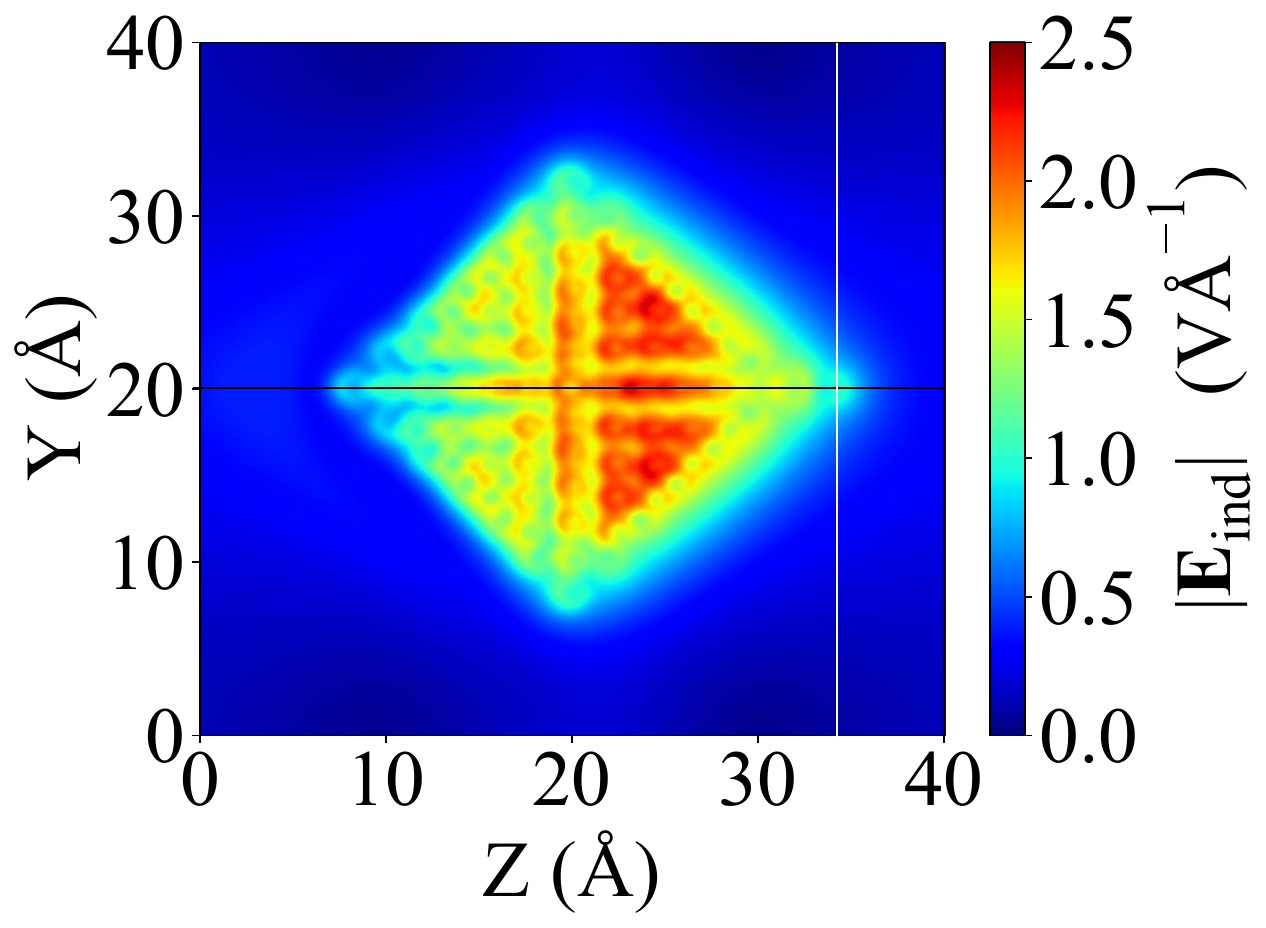}\\
        \includegraphics[width=0.45\linewidth]{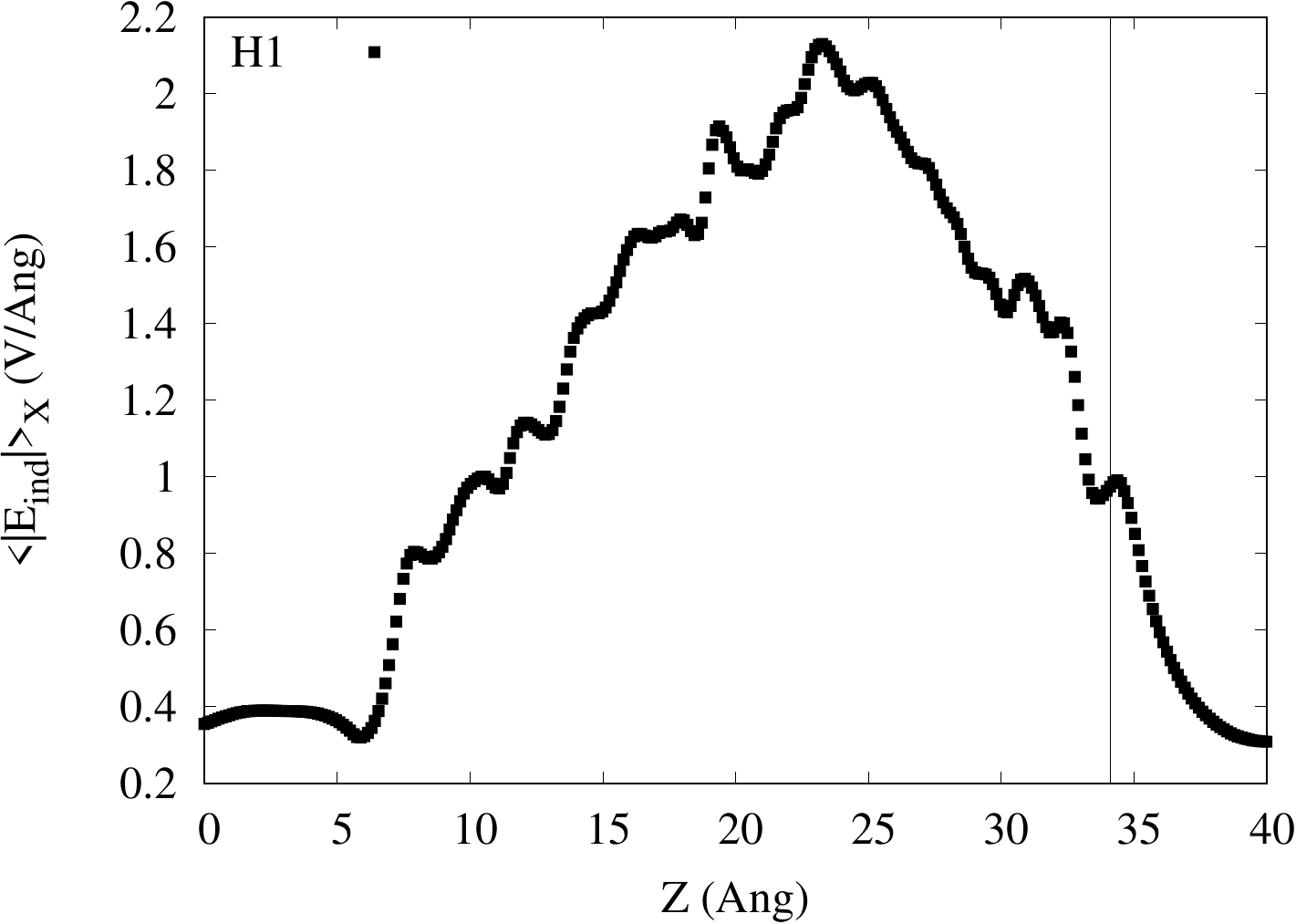} 
        \includegraphics[width=0.45\linewidth]{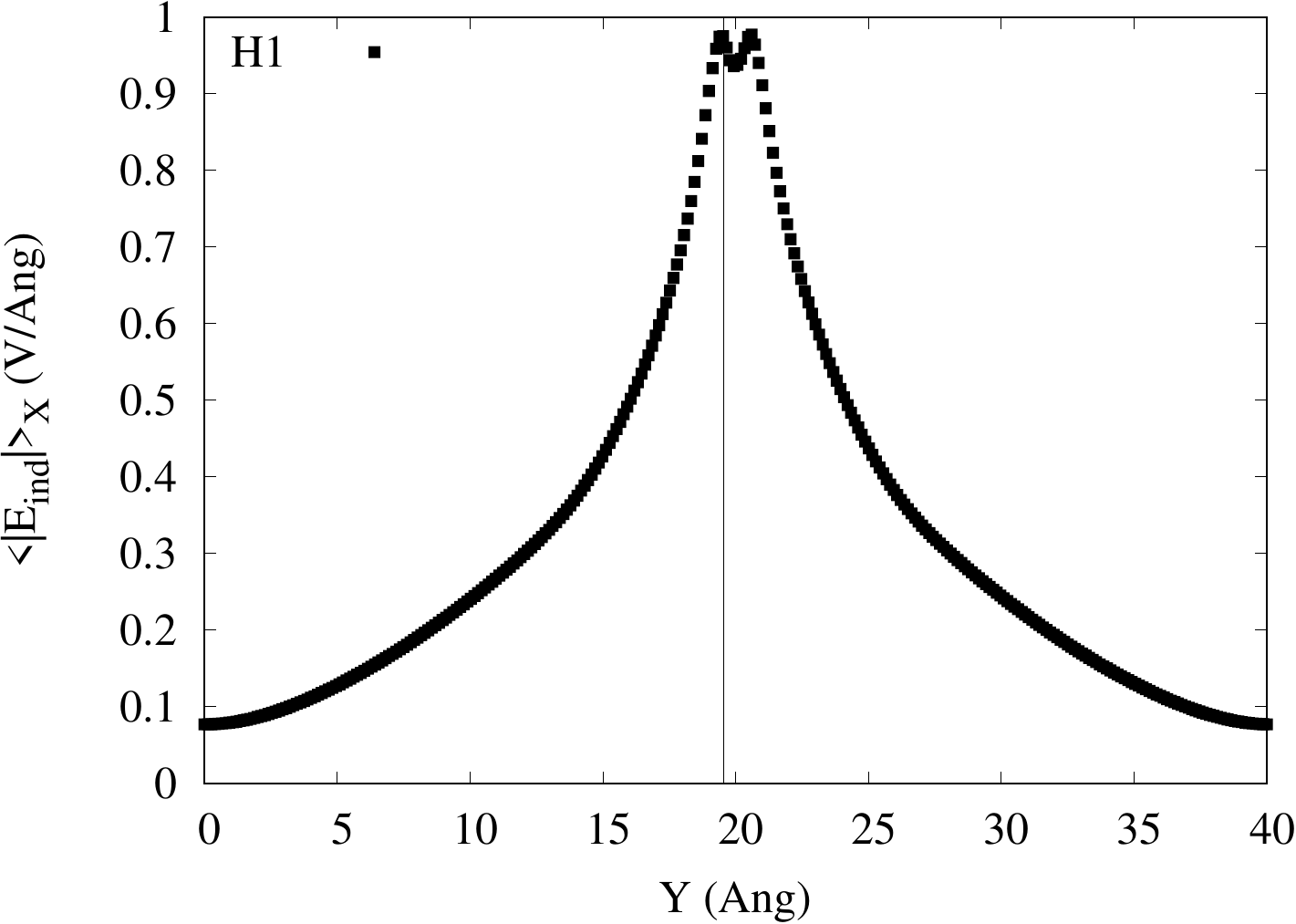}
        \includegraphics[width=0.45\linewidth]{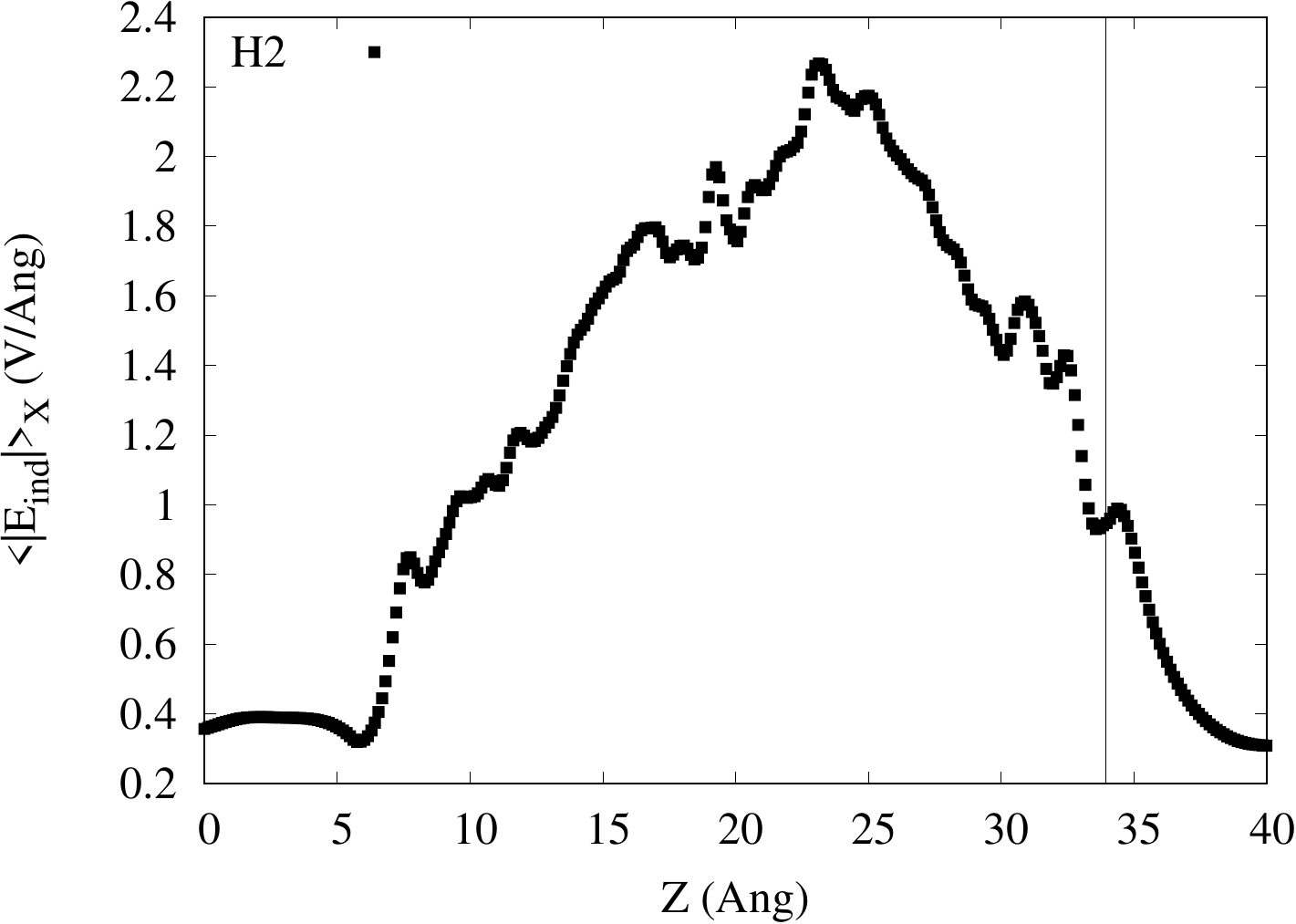} 
        \includegraphics[width=0.45\linewidth]{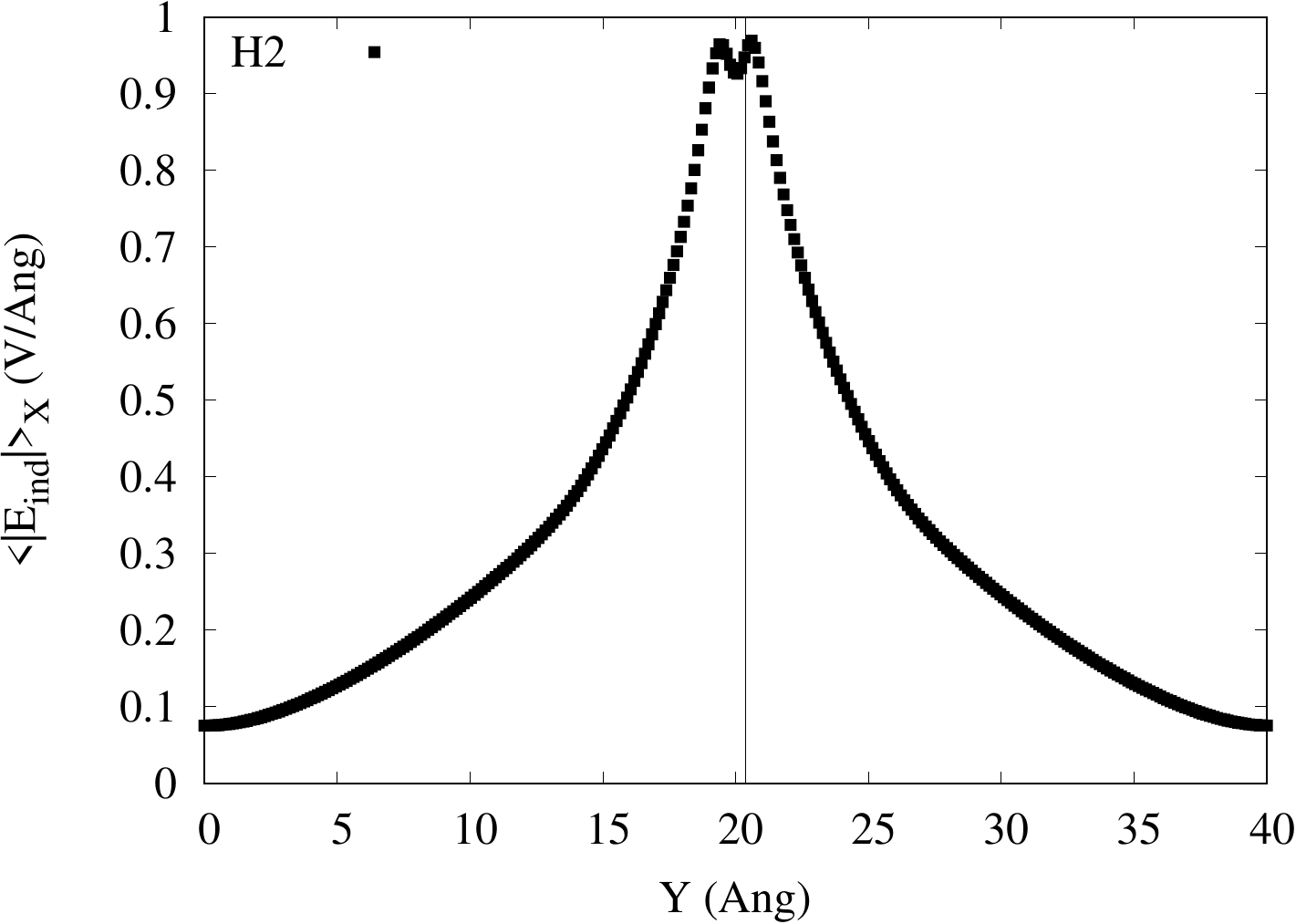}
		\caption{Same as Fig.~\ref{fig:field_2res} but for pulse-4 with frequency 8 eV at the time instant $t = 10$ fs. The molecule Cartesian coordinates are H1: (20.17441, 19.55348, 34.09393) and H2: (20.03204, 20.36180, 33.91831).}
		\label{fig:field_4nonres_1}
	\end{figure}

    \begin{figure}[h!]
 \centering
		\includegraphics[width=0.4\linewidth]{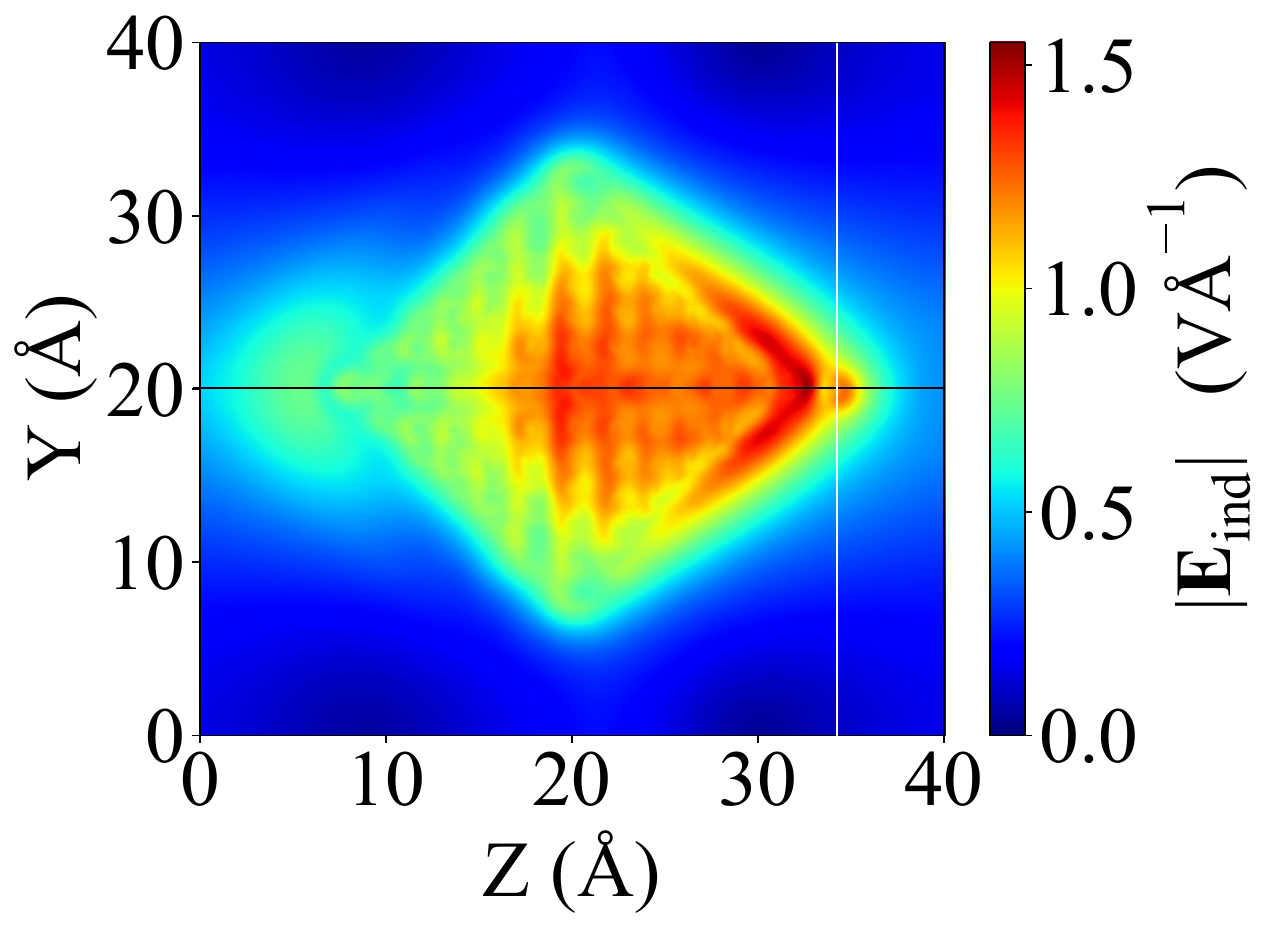}\\
        \includegraphics[width=0.45\linewidth]{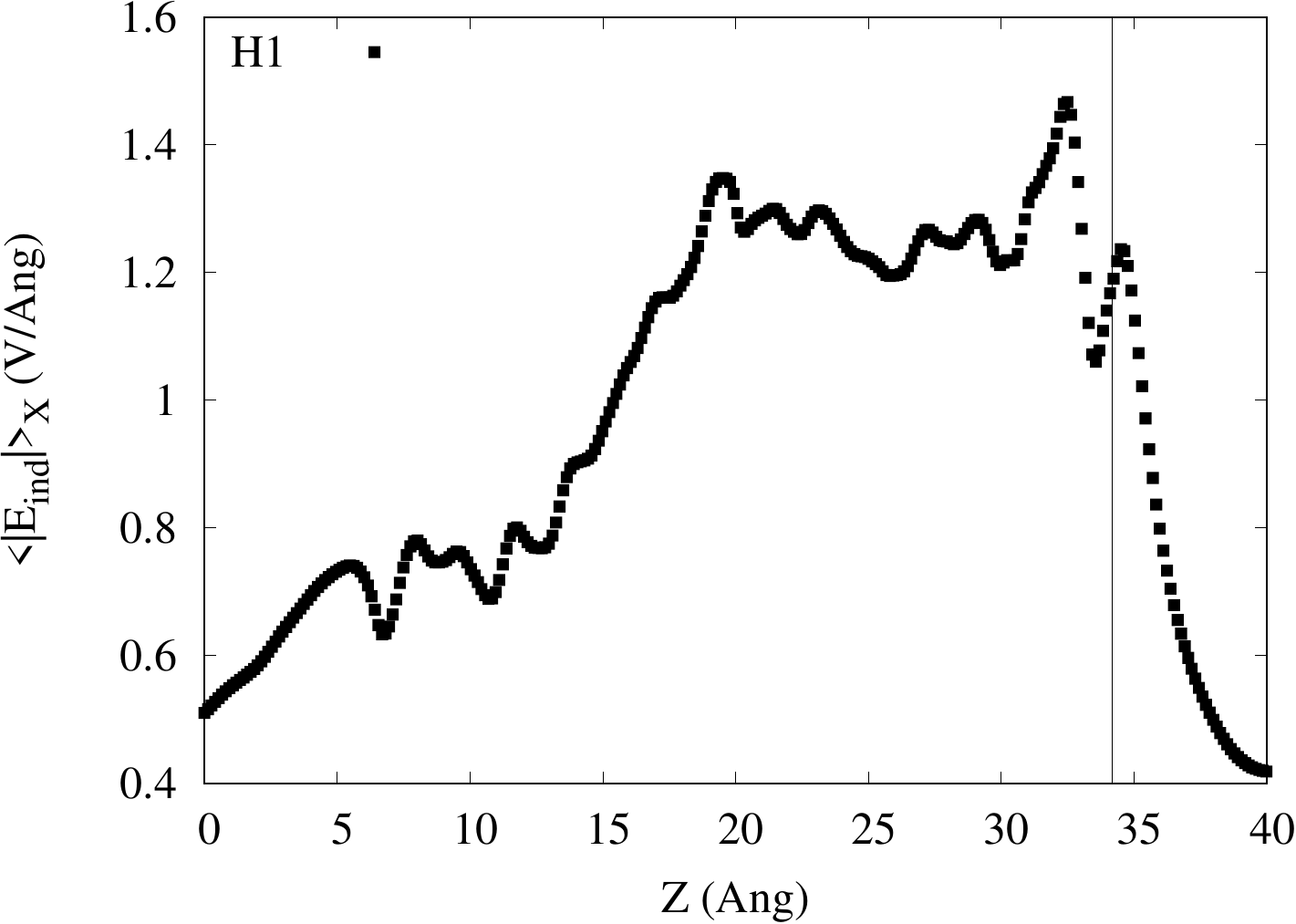} 
        \includegraphics[width=0.45\linewidth]{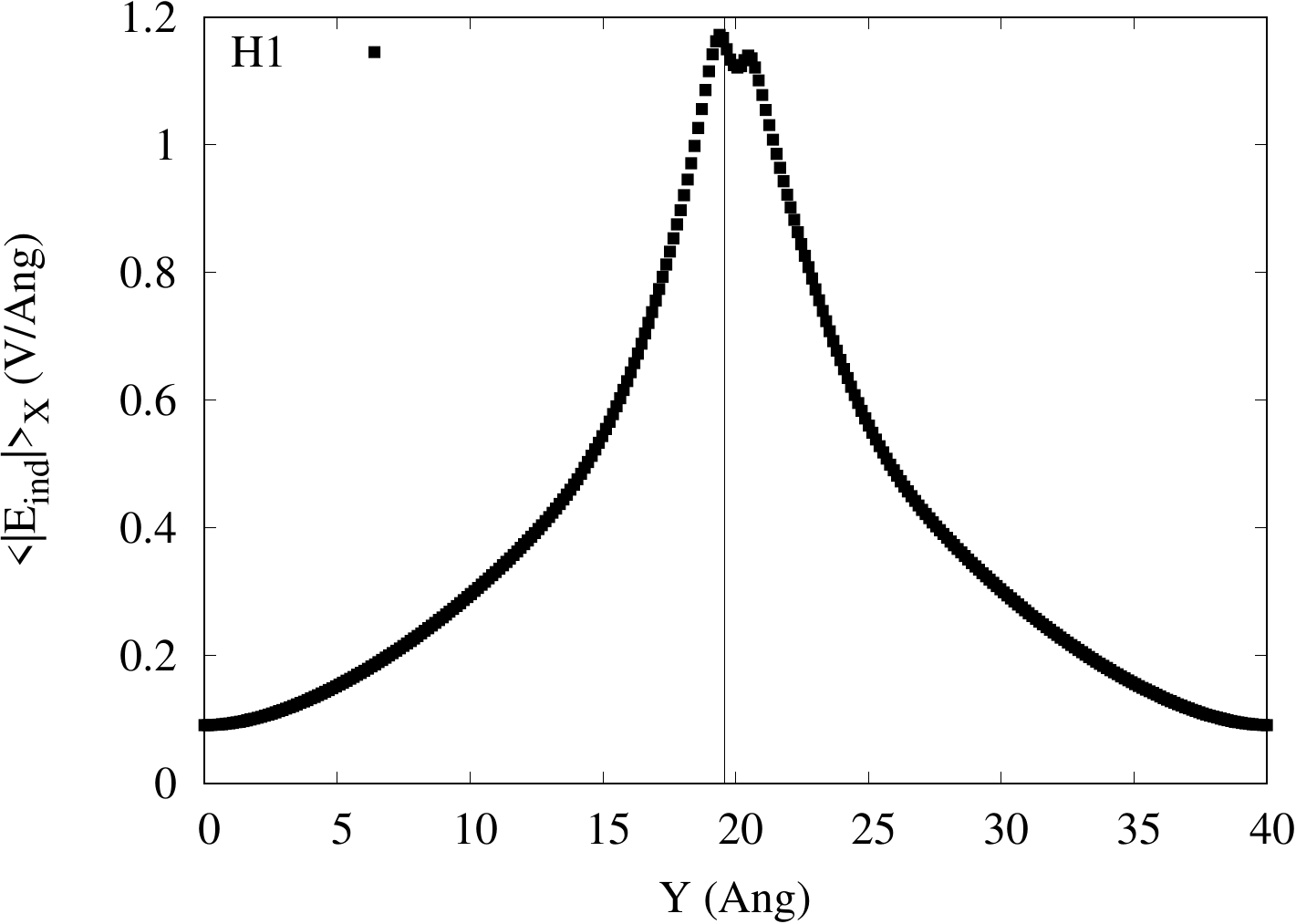}
        \includegraphics[width=0.45\linewidth]{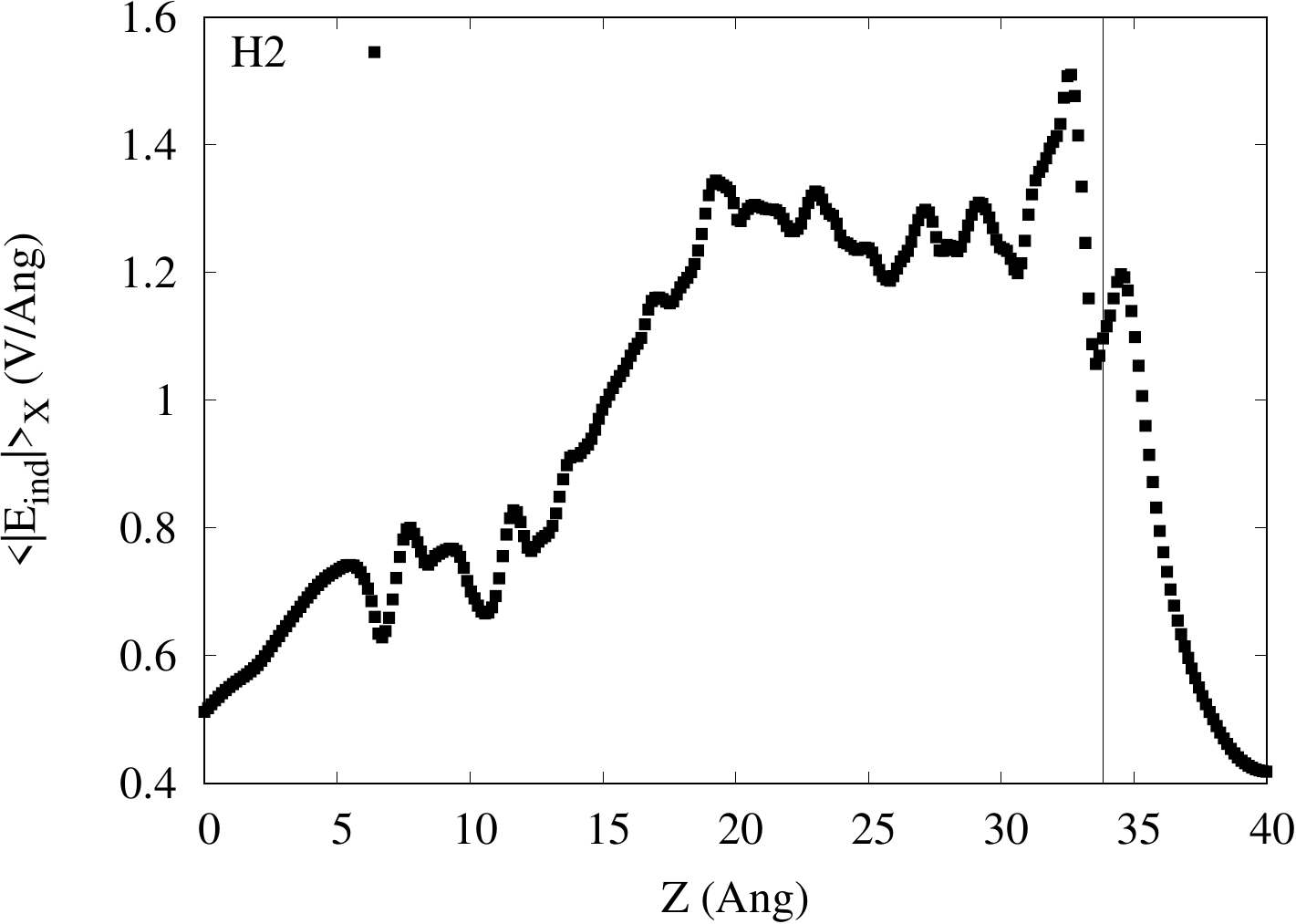} 
        \includegraphics[width=0.45\linewidth]{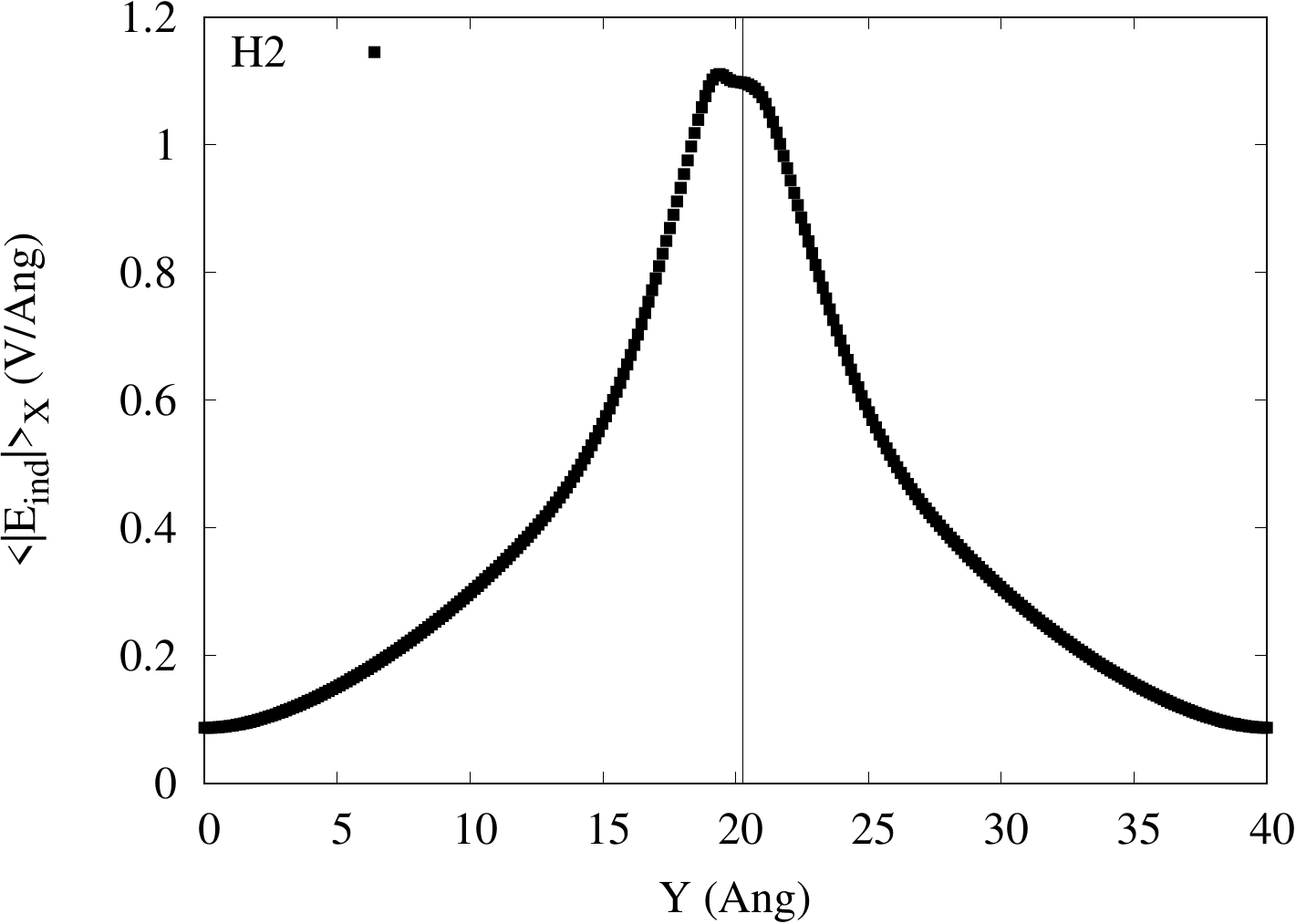}
		\caption{Same as Fig.~\ref{fig:field_2res} but for pulse-1 with frequency 2.48 eV at the time instant $t = 20$ fs. The molecule Cartesian coordinates are H1: (20.17377, 19.56827, 34.15810) and H2: (19.77894, 20.27348, 33.84442).}
		\label{fig:field_1res}
	\end{figure}

\begin{figure}[h!]
 \centering
		\includegraphics[width=0.4\linewidth]{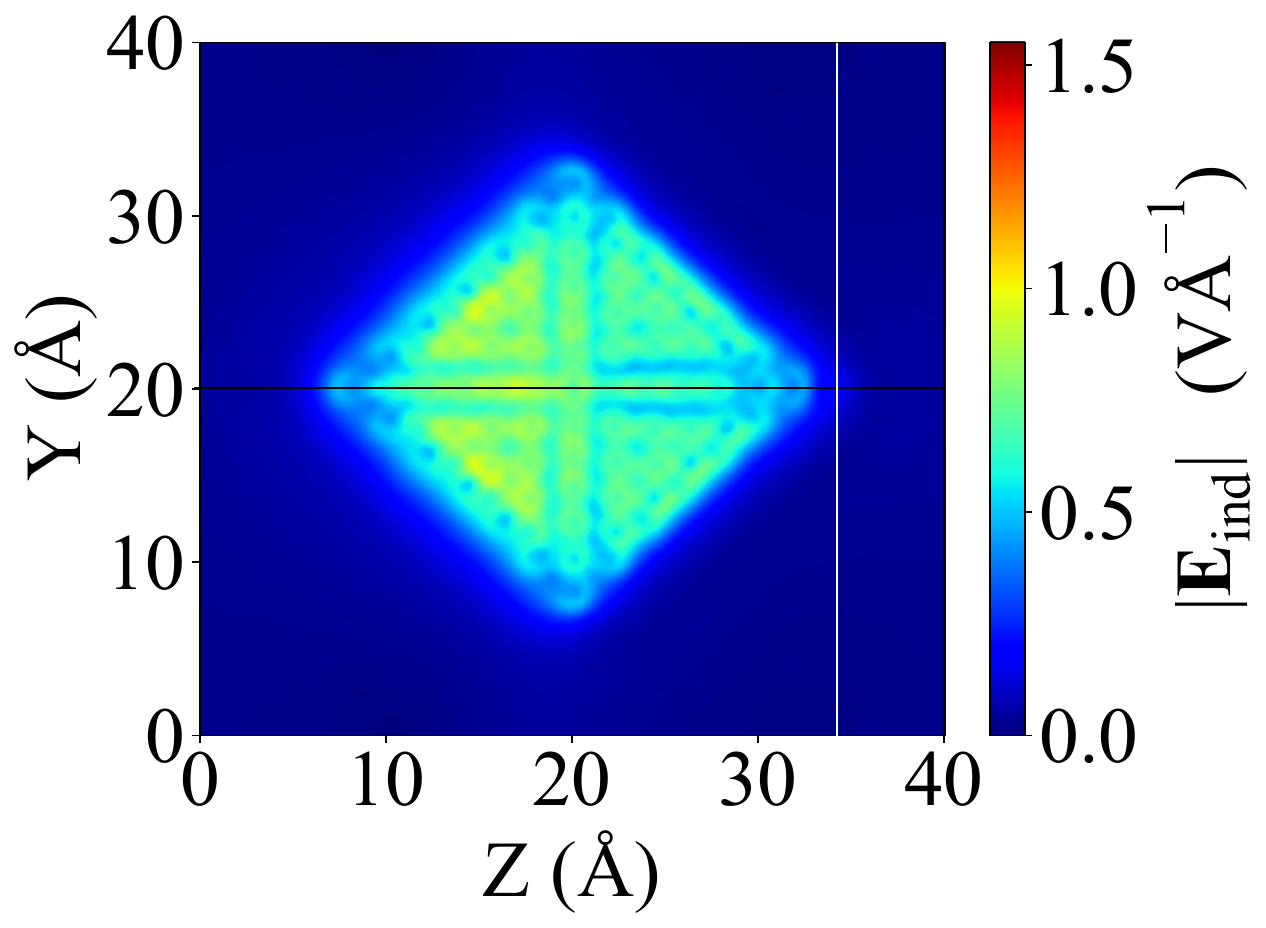}\\
        \includegraphics[width=0.45\linewidth]{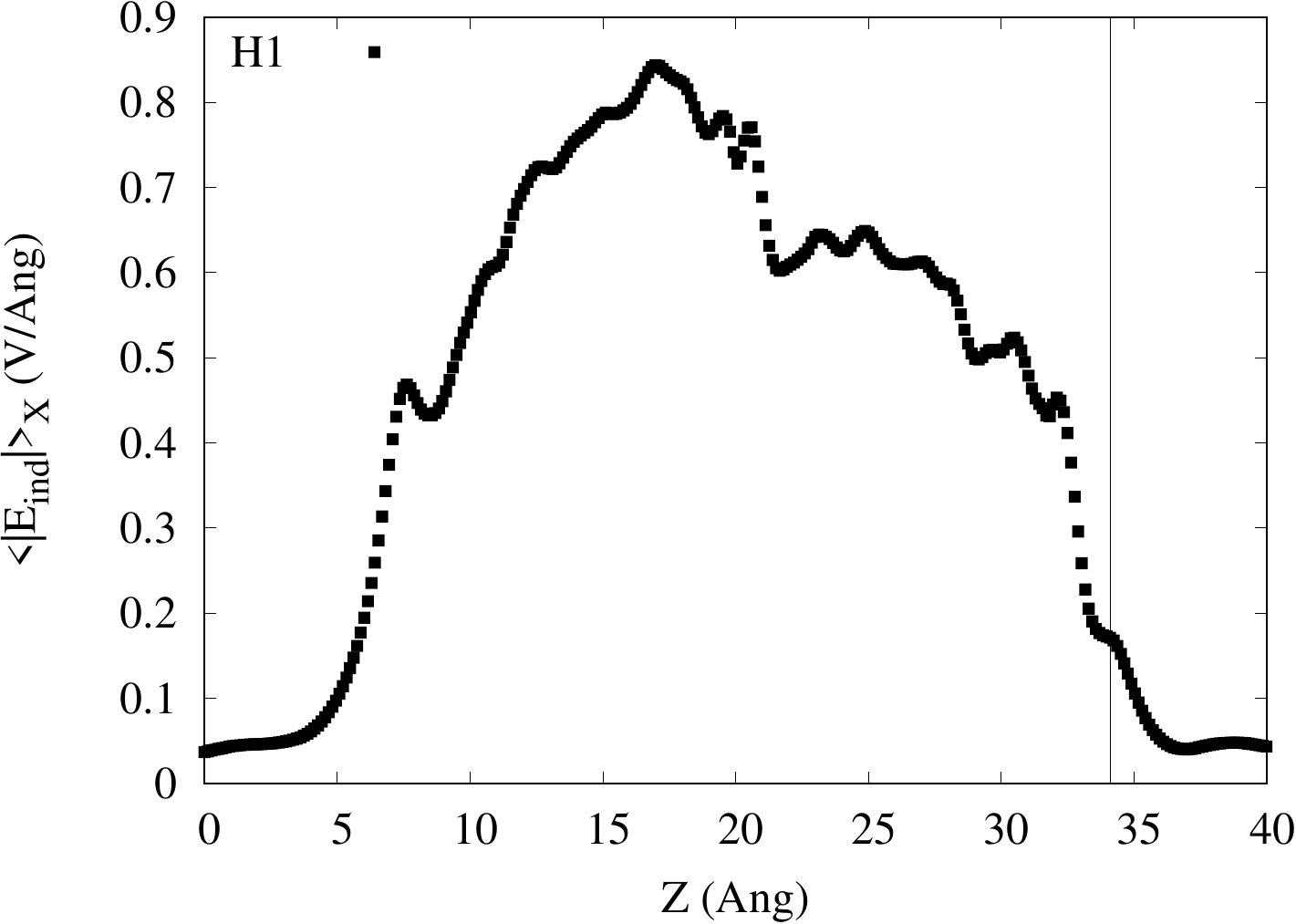} 
        \includegraphics[width=0.45\linewidth]{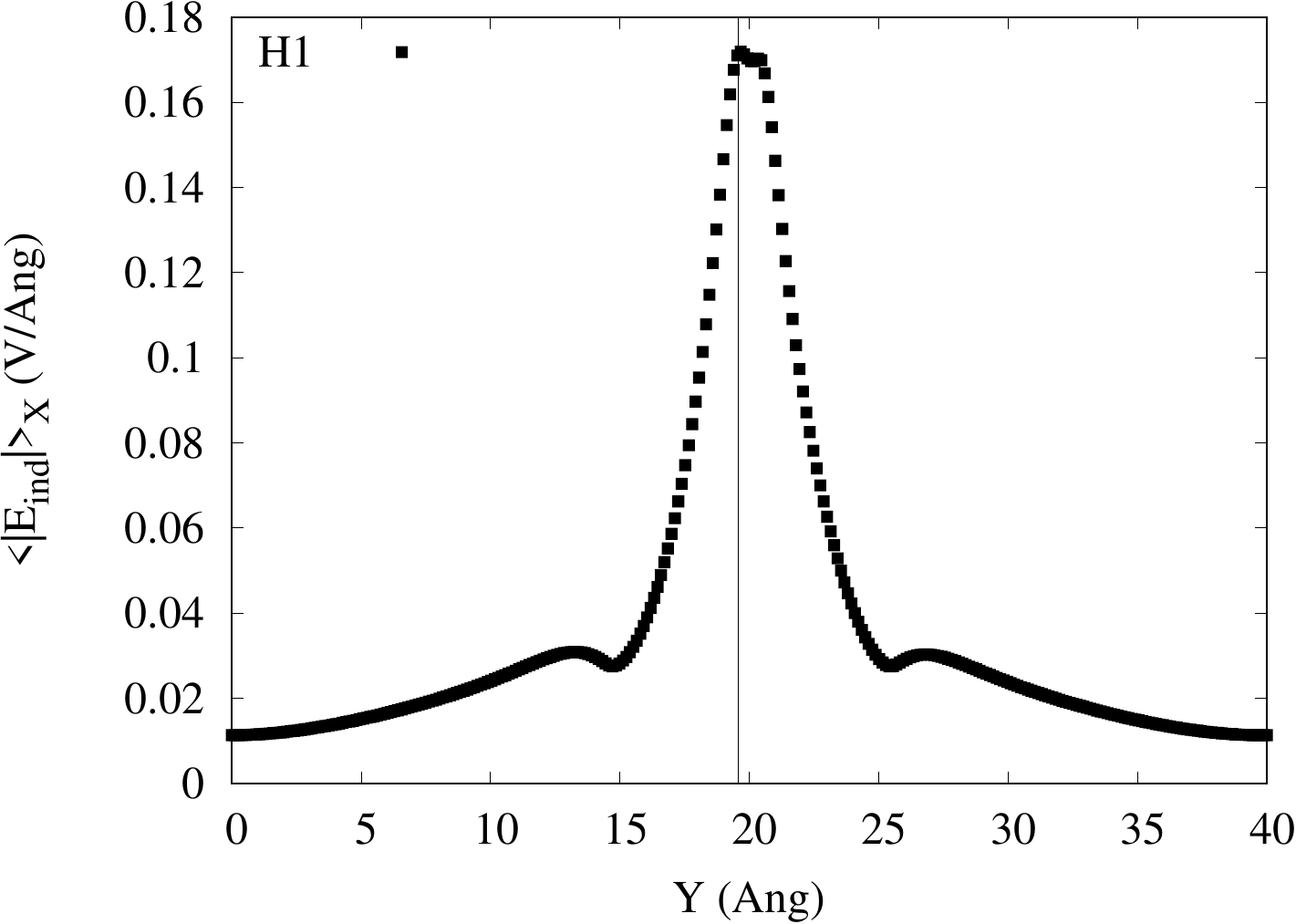}
        \includegraphics[width=0.45\linewidth]{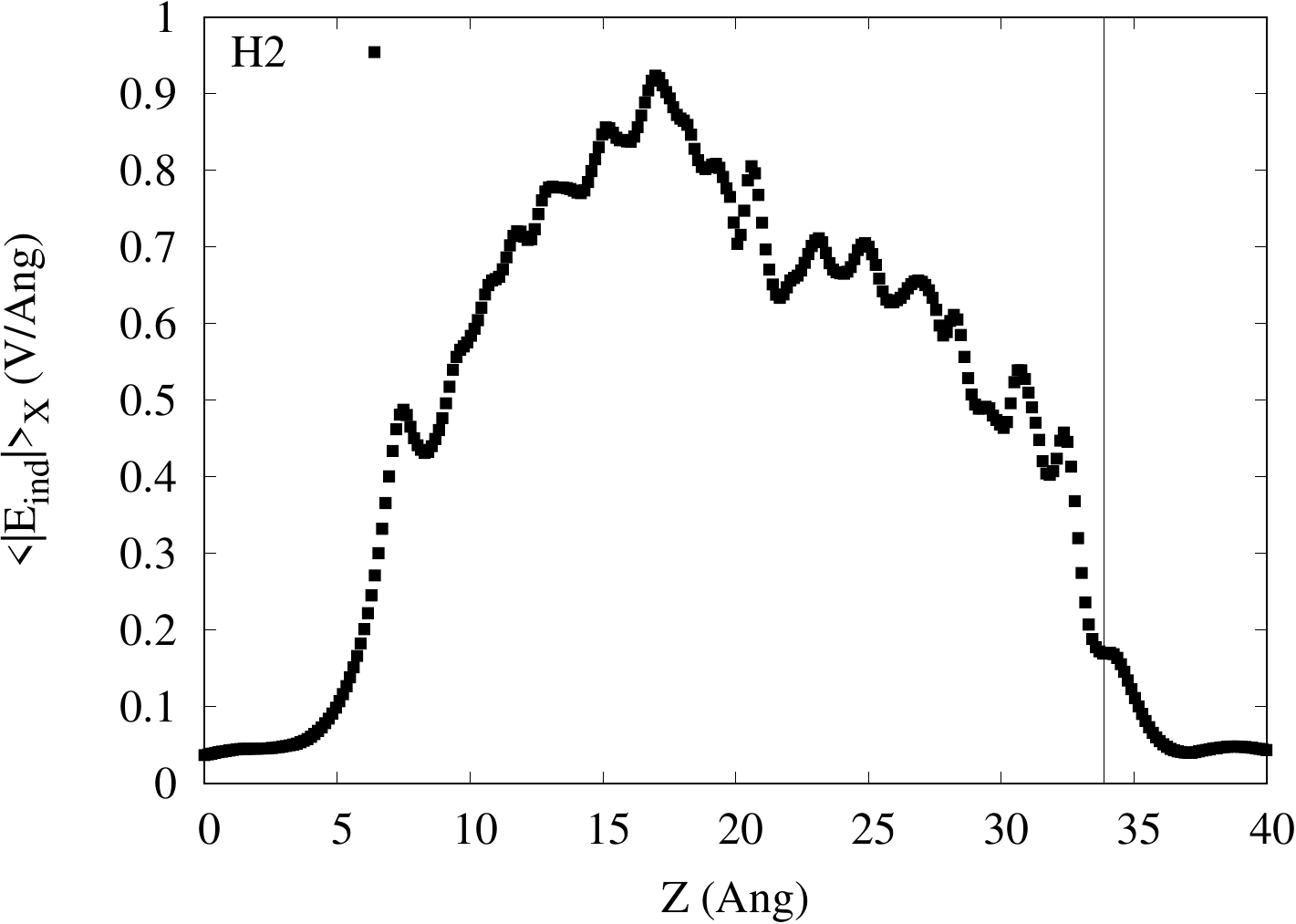} 
        \includegraphics[width=0.45\linewidth]{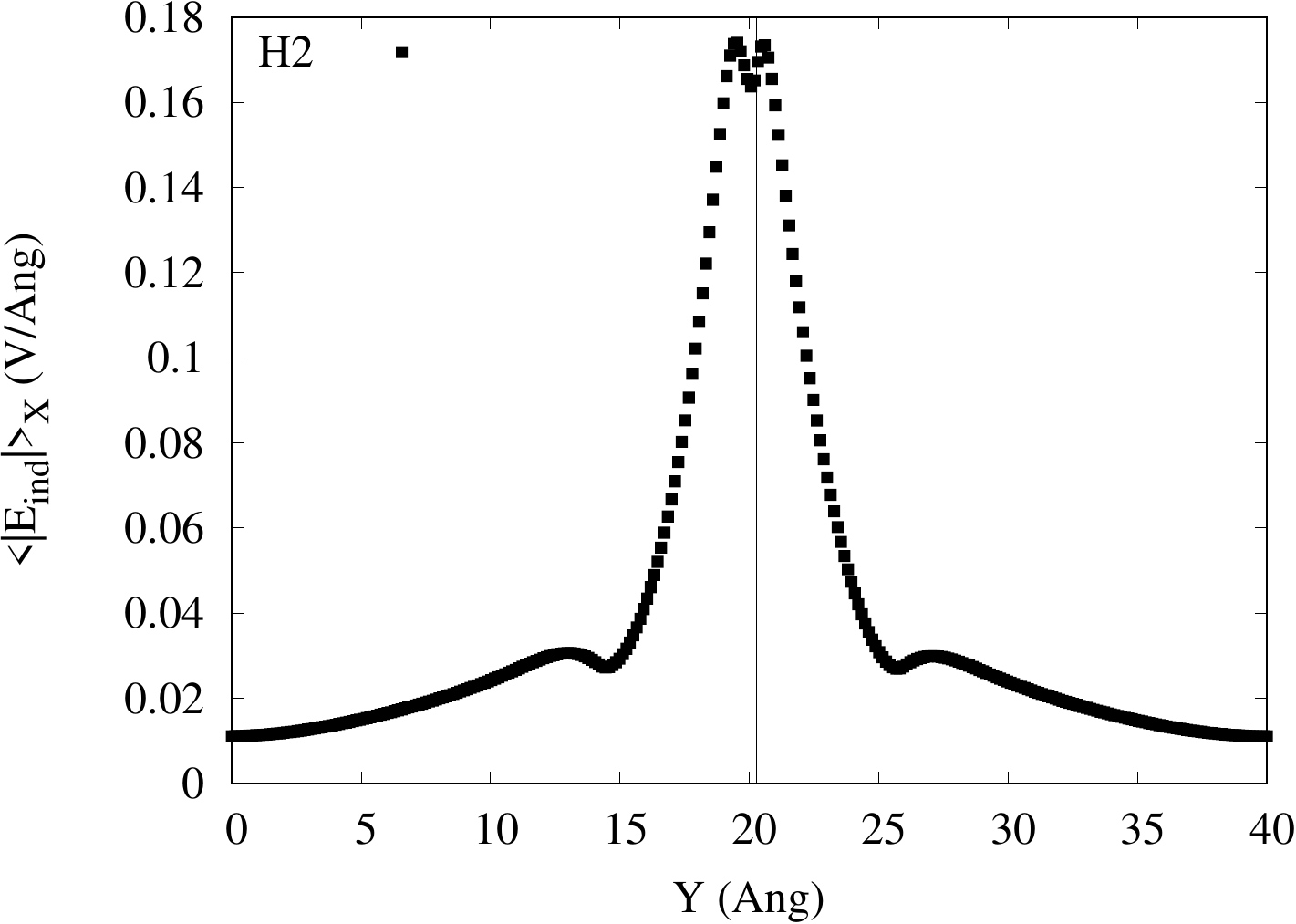}
		\caption{Same as Fig.~\ref{fig:field_2res} but for pulse-1 with frequency 8 eV at the time instant $t = 20$ fs. The molecule Cartesian coordinates are H1: (20.17020, 19.57946, 34.10292) and H2: (19.78535, 20.26223, 33.86742).}
		\label{fig:field_1nonres}
	\end{figure}

    \begin{figure}[h!]
 \centering
		\includegraphics[width=0.4\linewidth]{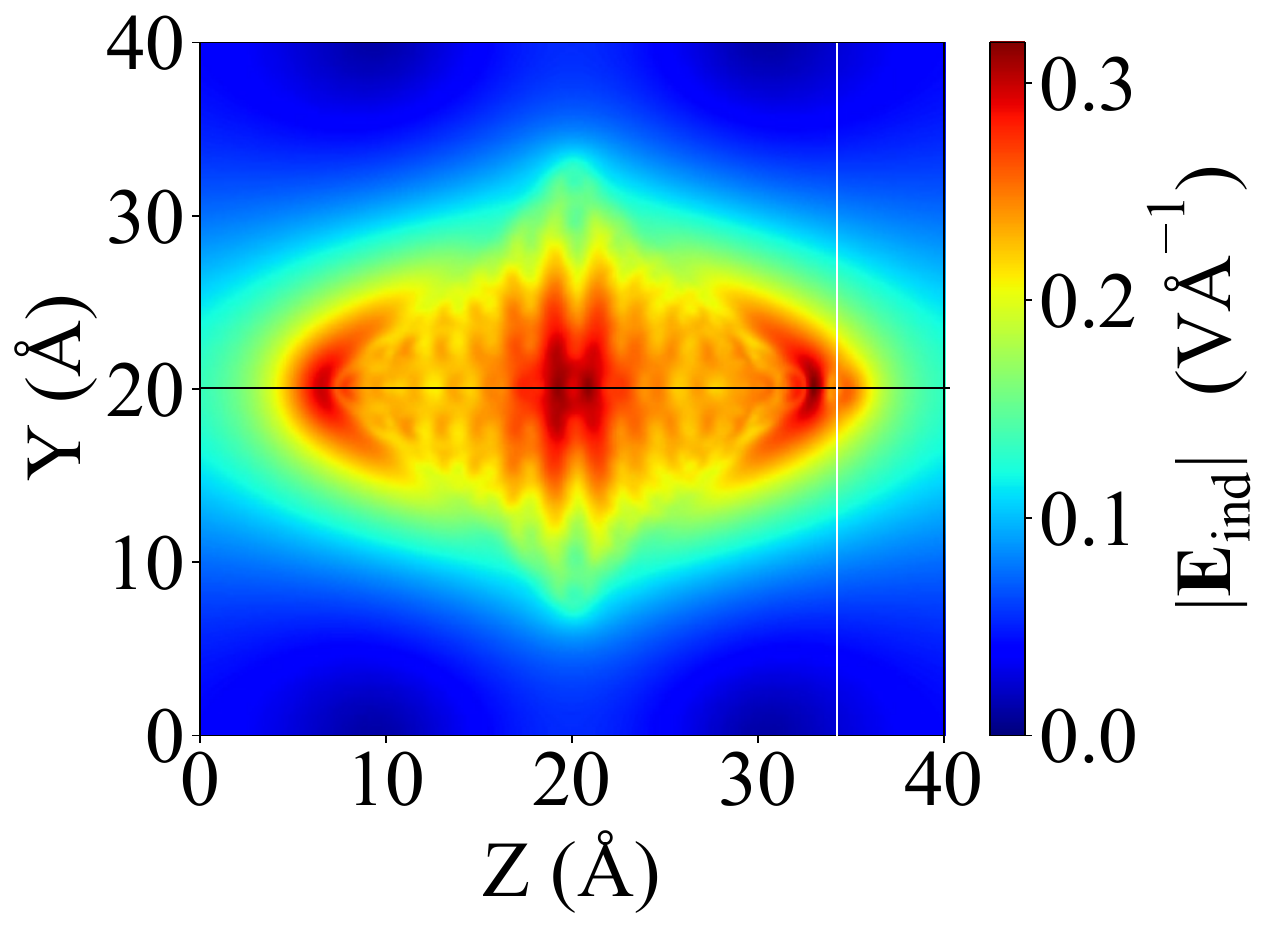}\\
        \includegraphics[width=0.45\linewidth]{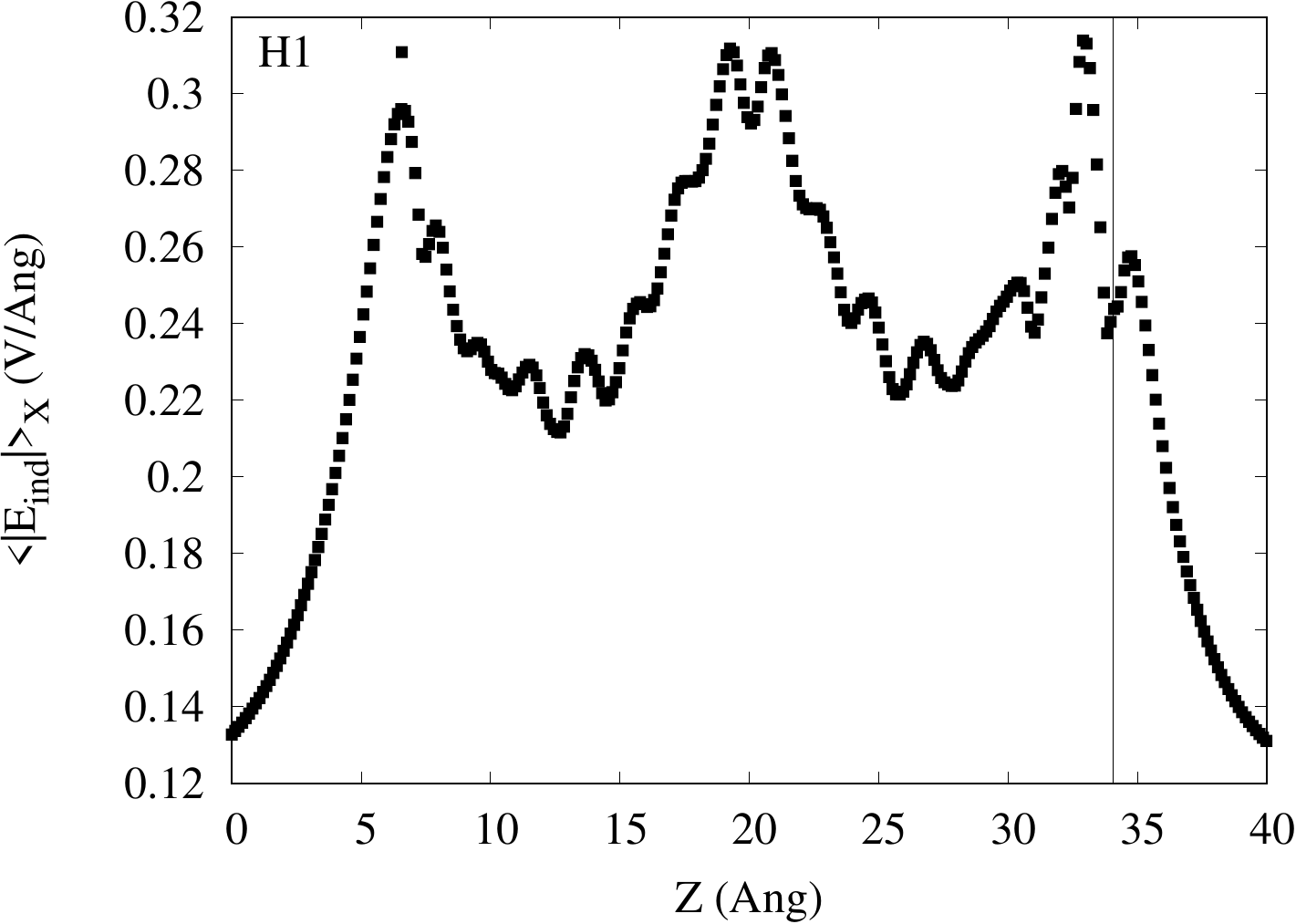} 
        \includegraphics[width=0.45\linewidth]{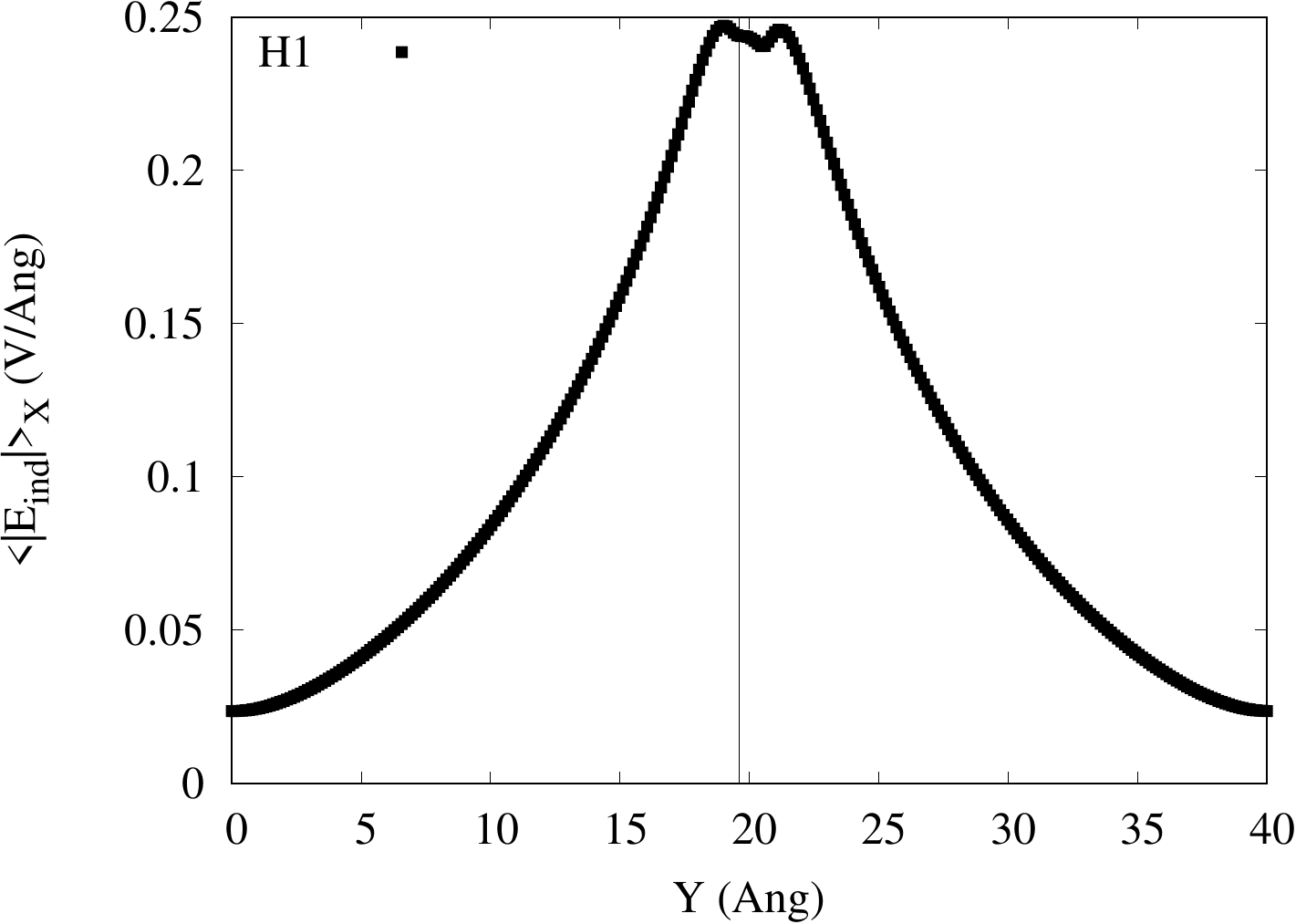}
        \includegraphics[width=0.45\linewidth]{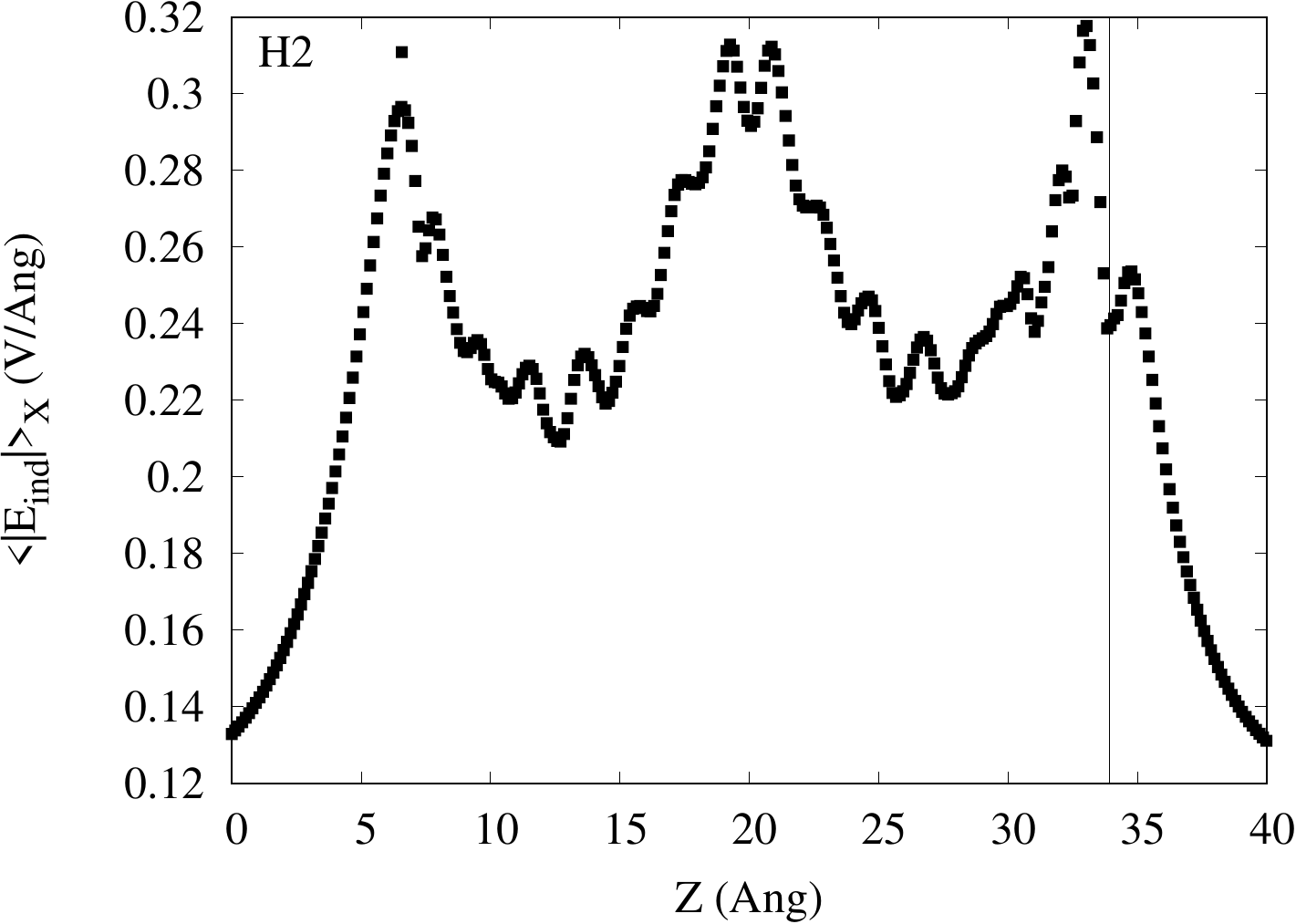} 
        \includegraphics[width=0.45\linewidth]{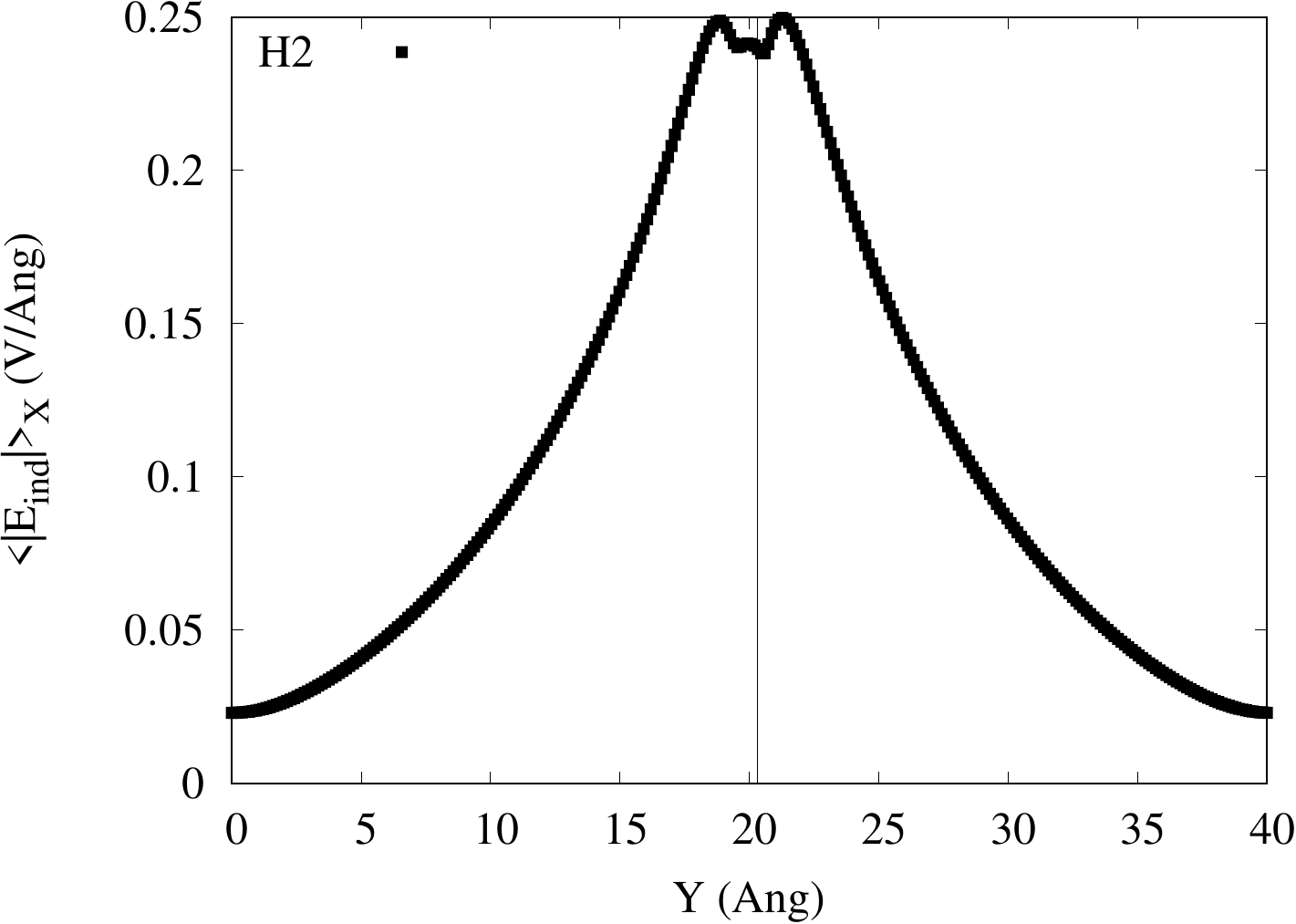}
		\caption{Same as Fig.~\ref{fig:field_2res} but for pulse-1 with frequency 2.48 eV at the time instant $t = 10$ fs. The molecule Cartesian coordinates are H1: (20.17012, 19.60066, 34.07398) and H2: (20.03598, 20.31432, 33.91182).}
		\label{fig:field_1res_1}
	\end{figure}

    \begin{figure}[h!]
 \centering
		\includegraphics[width=0.4\linewidth]{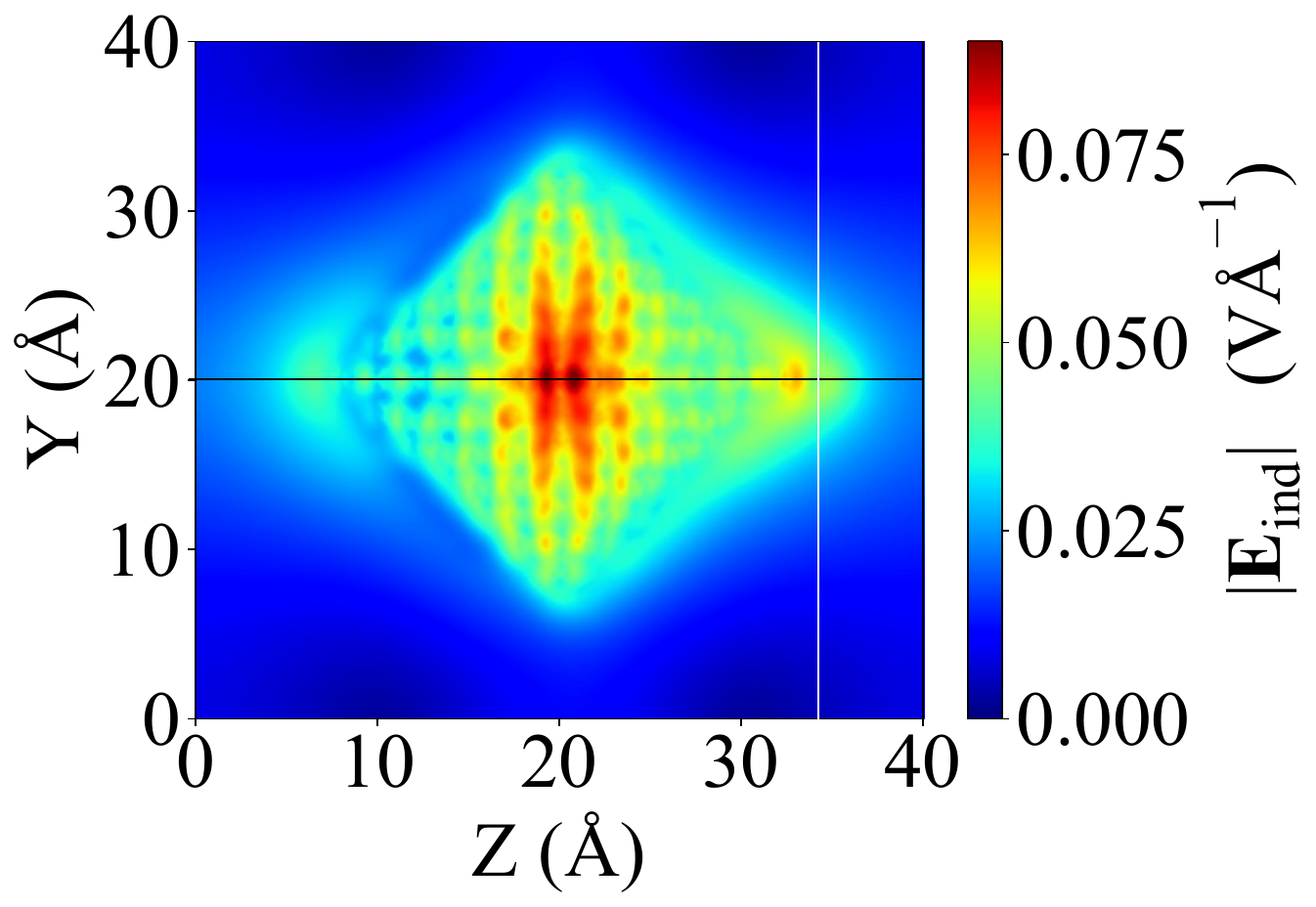}\\
        \includegraphics[width=0.45\linewidth]{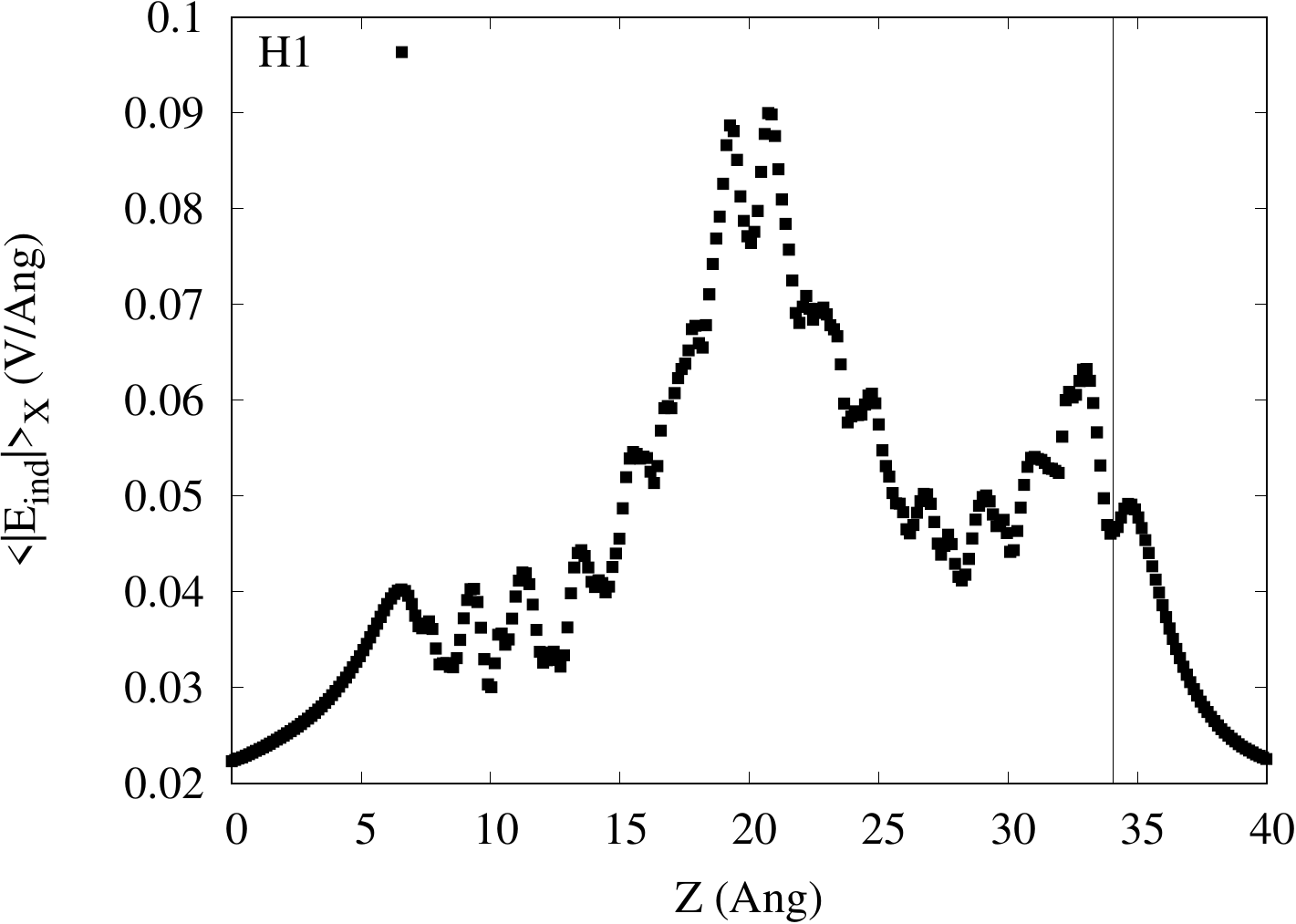} 
        \includegraphics[width=0.45\linewidth]{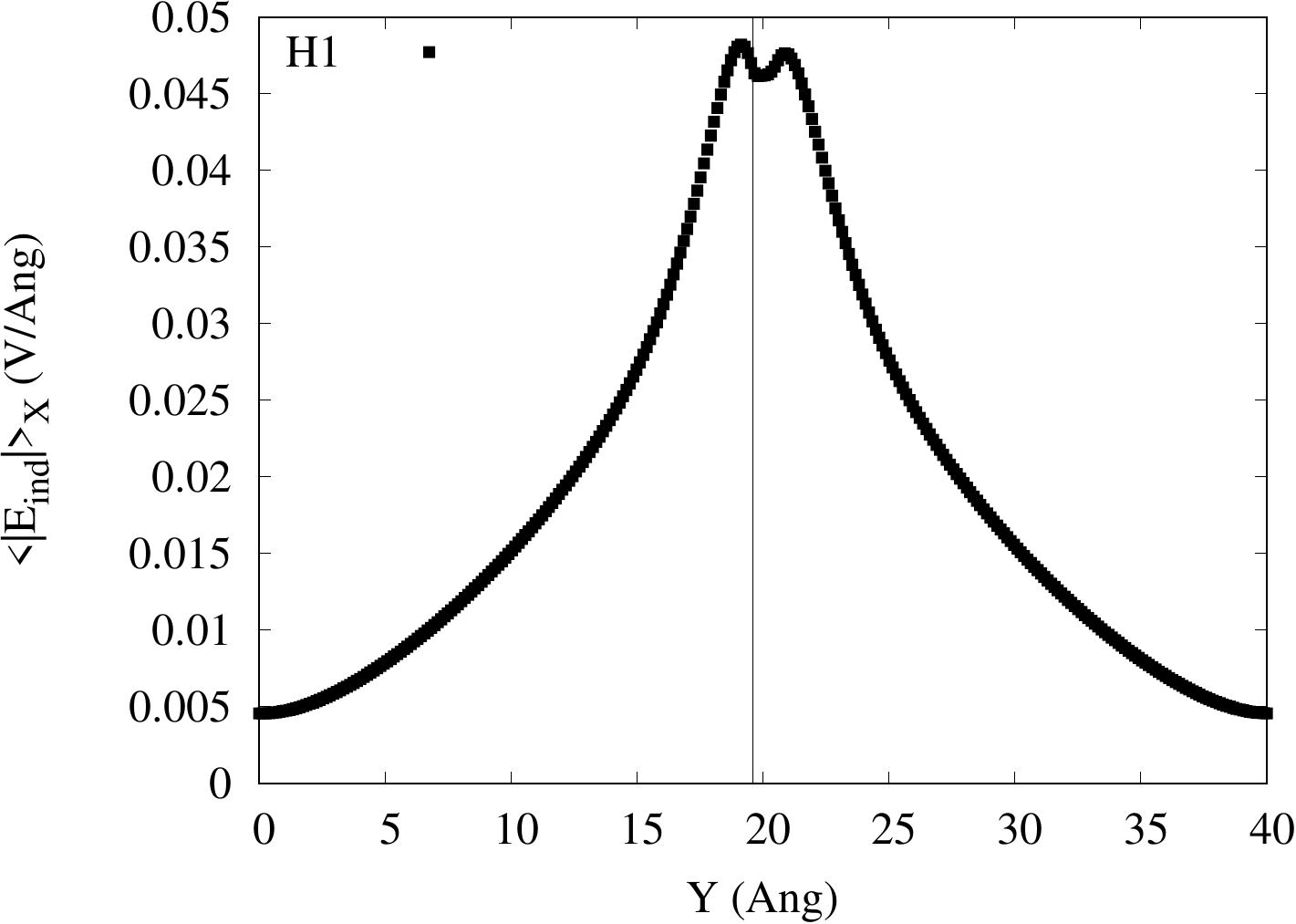}
        \includegraphics[width=0.45\linewidth]{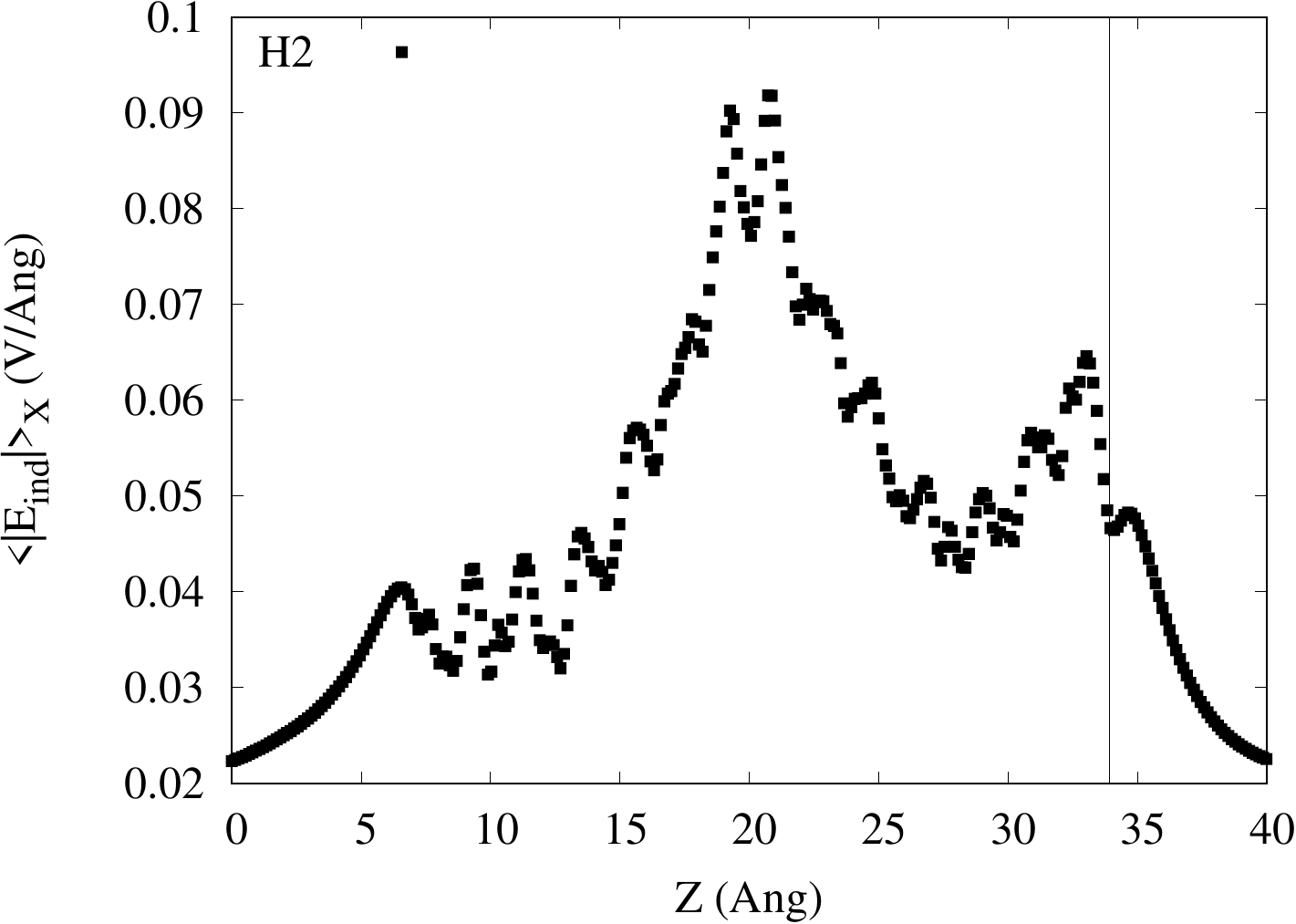} 
        \includegraphics[width=0.45\linewidth]{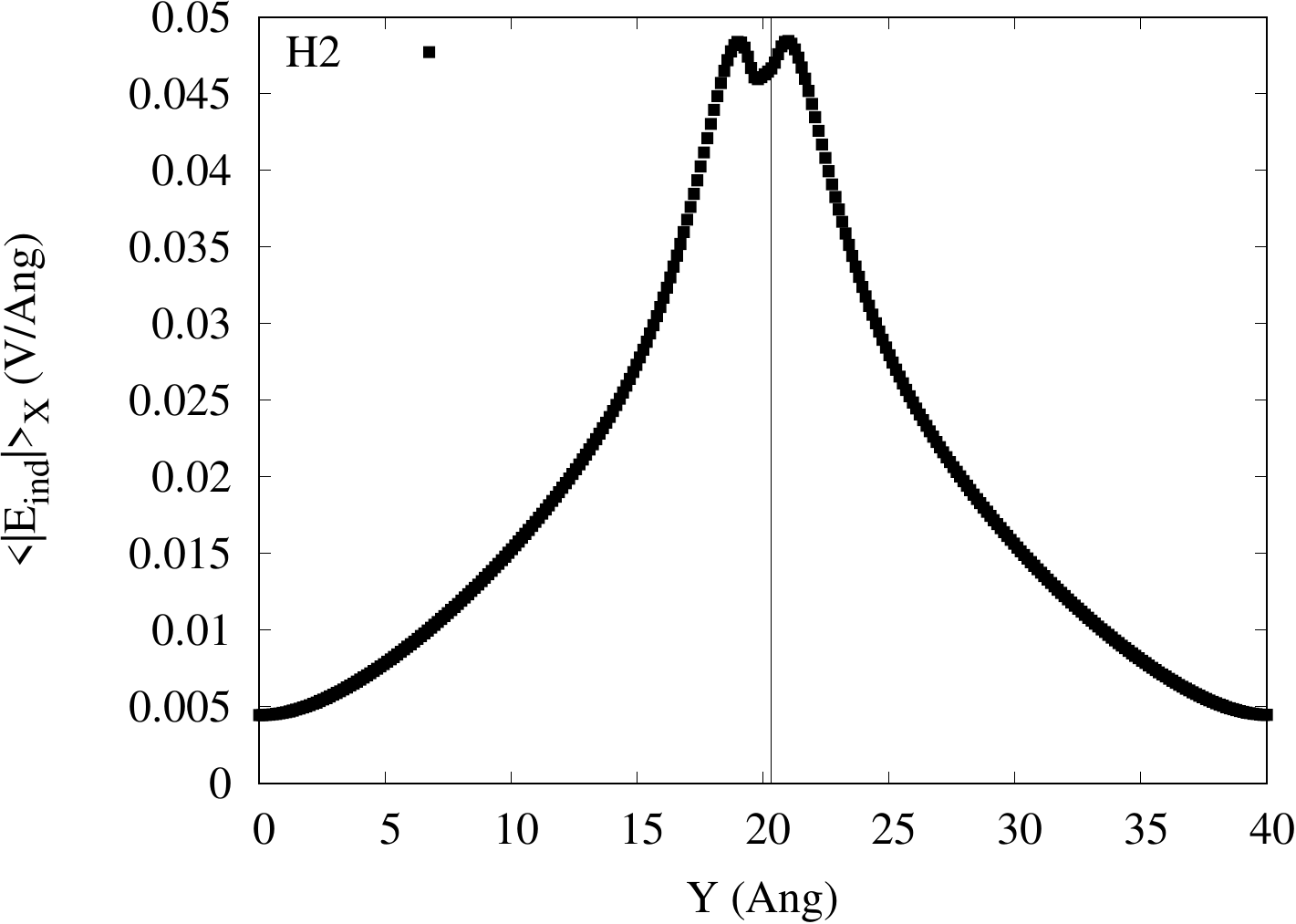}
		\caption{Same as Fig.~\ref{fig:field_2res} but for pulse-1 with frequency 8 eV at the time instant $t = 10$ fs. The molecule Cartesian coordinates are H1: (20.17003, 19.60123, 34.07358) and H2: (20.03605, 20.31386, 33.91217).}
		\label{fig:field_1nonres_1}
	\end{figure}